\documentclass[10pt, letterpaper, peerreview]{IEEEtran}

\usepackage{color}
\usepackage{colordvi}
\usepackage{fancybox,psfig}
\usepackage{graphics}
\usepackage{amsmath}
\usepackage{amssymb}
\usepackage{epsfig}
\usepackage{epic}

\newcommand{\real}{I\!\!R}

\newtheorem{theorem}{Theorem}[section]
\newtheorem{lemma}{Lemma}[section]
\newtheorem{example}{Example}[section]
\newtheorem{definition}[theorem]{Definition}
\newtheorem{corollary}{Corollary}[section]

\begin{document}

\title{\large \bf The necessity and sufficiency of anytime capacity for
stabilization of a linear system over a noisy communication link\\ Part I: scalar systems}

\author{Anant Sahai\footnote{A.~Sahai is with the Department of
Electrical Engineering and Computer Science, U.C.~Berkeley. Early
portions of this work appeared in his doctoral dissertation and a few
other results were presented at the 2004 Conference on Decision and
Control.} and Sanjoy Mitter\footnote{Department of Electrical Engineering
and Computer Science at the Massachusetts Institute of Technology.
Support for S.K.~Mitter was provided by the Army Research Office under
the MURI Grant: Data Fusion in Large Arrays of Microsensors
DAAD19-00-1-0466 and the Department of Defense MURI Grant: Complex
Adaptive Networks for Cooperative Control Subaward \#03-132 and the
National Science Foundation Grant CCR-0325774.} \\ {\small
sahai@eecs.berkeley.edu, mitter@mit.edu}}

        



\markboth{IEEE Transactions on Information Theory,~Vol.~??,
No.~??,~Month??~2005}{Shell \MakeLowercase{\textit{et al.}}: Bare
Demo of IEEEtran.cls for Journals}


\maketitle

\begin{abstract} 
We review how Shannon's classical notion of capacity is not enough to
characterize a noisy communication channel if the channel is intended
to be used as part of a feedback loop to stabilize an unstable scalar
linear system. While classical capacity is not enough, another sense
of capacity (parametrized by reliability) called ``anytime capacity''
is shown to be necessary for the stabilization of an unstable
process. The required rate is given by the log of the unstable system
gain and the required reliability comes from the sense of stability
desired. A consequence of this necessity result is a sequential
generalization of the Schalkwijk/Kailath scheme for communication over
the AWGN channel with feedback.

In cases of sufficiently rich information patterns between the encoder
and decoder, adequate anytime capacity is also shown to be sufficient
for there to exist a stabilizing controller. These sufficiency results
are then generalized to cases with noisy observations, delayed control
actions, and without any explicit feedback between the observer and
the controller. Both necessary and sufficient conditions are extended
to continuous time systems as well. We close with comments discussing
a hierarchy of difficulty for communication problems and how these
results establish where stabilization problems sit in that hierarchy.

\end{abstract}


\IEEEpeerreviewmaketitle

\section{Introduction}

For communication theorists, Shannon's classical channel capacity
theorems are not just beautiful mathematical results, they are useful
in practice as well. They let us summarize a diverse range of channels
by a single figure of merit: the capacity. For most non-interactive
point-to-point communication applications, the Shannon capacity of a
channel provides an upper bound on performance in terms of end-to-end
distortion through the distortion-rate function. As far as distortion
is concerned, all that matters is the channel capacity and the nature
of the source. Given enough tolerance for end-to-end delay, the source
can be encoded into bits and those bits can be reliably transported
across the noisy channel if the rate is less than the Shannon
capacity. As long as the source, distortion, and channel are
well-behaved\cite{Gallager, VembuVerdu}, there is asymptotically no
loss in separating the problems of source and channel coding. This
provides a justification for the layered architecture that lets
engineers isolate the problem of reliable communication from that of
using the communicated information. Recent advances in coding theory
have also made it possible to approach the capacity bounds very
closely in practical systems.

In order to extend our understanding of communication to interactive
settings, it is essential to have some model for interaction. Schulman
and others have studied interaction in the context of distributed
computation \cite{Schulman, Rajagopalan}. The interaction there is
between computational agents that have access to some private data and
wish to perform a global computation in a distributed way. The
computational agents can only communicate with each other through
noisy channels. In Schulman's formulation, capacity is not a question
of major interest since constant factor slowdowns are considered
acceptable.\footnote{Furthermore, such constant factor slowdowns
appear to be unavoidable when facing the very general class of
interactive computational problems.} Fundamentally, this is a
consequence of being able to design all the system dynamics. The rich
field of automatic control provides an interactive context to study
capacity requirements since the plant dynamics are given, rather than
something that can be designed. In control, we consider interaction
between an observer that gets to see the plant and a controller that
gets to control it. These two can be connected by a noisy channel.

Shannon himself had suggested looking to control problems for more
insight into reliable communication \cite{ShannonOriginalSource}.
\begin{quotation}
 ``$\ldots$ can be pursued further and is related to a duality between
 past and future\footnote{The differing roles of the past and future
 are made clear in \cite{OurUpperBoundPaper}.} and the notions of
 control and knowledge. Thus we may have knowledge of the past and
 cannot control it; we may control the future but have no knowledge of
 it.''
\end{quotation}

We are far from the first to attempt to bring together information and
control theory. In \cite{HoKastnerWong}, Ho, Kastner, and Wong drew
out a detailed diagram in which they summarized the then known
relationships among team theory, signaling, and information theory
from the perspective of distributed control. Rather than taking such a
broad perspective, we instead ask whether Shannon's classical capacity
is the appropriate characterization for communication channels arising
in distributed control systems. Our interest is in understanding the
fundamental relationship between problems of stabilization and
problems of communication.

Tatikonda's recent work on sequential rate distortion theory provides
an information-theoretic lower-bound on the achievable performance of
a control system over a channel. Because this bound is sometimes
infinite, it also implies that there is a fundamental rate of
information production, namely the sum of the logs of the unstable
eigenvalues of the plant, that is invariantly attached to an unstable
linear discrete-time process \cite{TatikondaThesis, Tatikonda1}. This
particular notion of rate was justified by showing how to stabilize
the system over a noiseless feedback link with capacity greater than
the intrinsic rate for the unstable process.\footnote{The sequential
rate-distortion bound is generally not attained even at higher rates
except in the case of perfectly matched channels.} Nair et
al.~extended this to cover the case of unbounded disturbances and
observation noise under suitable conditions \cite{NairPaper1,
NairPaper2}. In addition to noiseless channels, the results were
extended for almost-sure stabilization in the context of
undisturbed\footnote{In seminal work \cite{Tatikonda2}, there is no
persistent disturbance acting on the unstable plant.} control systems
with bounded initial conditions being stabilized over certain noisy
channels \cite{Tatikonda2}.

We had previously showed that it is possible to stabilize persistently
disturbed controlled Gauss-Markov processes over suitable
power-constrained AWGN (Additive White Gaussian Noise)
channels\cite{OurACC99Paper, OurMainLQGPaper} where it turns out that
Shannon capacity is tight and linear observers and controllers are
sufficient to achieve stabilization \cite{Bansal}. In contrast, we
showed that the Shannon capacity of the binary erasure channel (BEC)
is not sufficient to check stabilizability and introduced the anytime
capacity as a candidate figure of merit \cite{ACC00Paper}. Following
up on our treatment of the BEC case, Martins et al.~have studied more
general erasure-type models and have also incorporated bounded model
uncertainty in the plant \cite{Nuno}. There is also related work by
Elia that uses ideas from robust control to deal with communication
uncertainty in a mixed continuous/discrete context, but restricting to
linear operations \cite{EliaPaper1, EliaPaper2}. Basar and his
students have also considered such problems and have studied the
impact of a noisy channels on both the observations and the controls
\cite{YuksulBasar}. The area of control with communications
constraints continues to attract attention and the reader is directed
to the recent September 2004 issue of IEEE Transactions on Automatic
Control and the articles therein for a more comprehensive survey.

Many of the issues that arise in the control context also arise for
the conceptually simpler problem of merely estimating an unstable
open-loop process\footnote{The unstable open-loop processes discussed
here are first-order nonstationary autoregressive processes
\cite{Gray70}, of which an important special case is the Wiener
process considered by Berger \cite{BergerPaper}.}, across a noisy
channel. For this estimation problem in the limit of large, but
finite, end-to-end delays, we have proved a source coding theorem that
shows that the distortion-rate bound is achievable. Furthermore, it is
possible to characterize the information being produced by an unstable
process \cite{OurSourceCodingPaper}. It turns out that such processes
produce two qualitatively distinct types of information when it comes
to transport over a noisy channel. In addition to the classical
Shannon-type of information found in traditional rate-distortion
settings\footnote{In \cite{OurSourceCodingPaper}, we show how the
classical part of the information determines the shape of the
rate-distortion curve, while the unstable core is responsible for a
shift of this curve along the rate axis.}, there is an essential core
of information that captures the unstable nature of the source. While
classical Shannon reliability suffices for the classical information,
this unstable core requires {\em anytime} reliability for transport
across a noisy channel.\footnote{How to communicate such unstable
processes over noisy channels had been an open problem since Berger
had first developed a source-coding theorem for the Wiener process
\cite{BergerBook}. Berger had conjectured that it was impossible to
transport such processes over generic noisy channels with
asymptotically finite end-to-end distortion using traditional means.}
As also discussed in this paper, anytime reliability is a sense of
reliable transmission that lies between Shannon's classical
$\epsilon-$sense of reliable transmission and his zero-error
reliability \cite{ShannonZeroError}. In \cite{OurSourceCodingPaper},
we also review how the sense of anytime reliability is linked to
classical work on sequential tree codes with bounded delay
decoding.\footnote{Reference \cite{BorkarMitter} raised the
possibility of such a connection early on.}

\begin{figure}
\begin{center}
\setlength{\unitlength}{2300sp}%
\begingroup\makeatletter\ifx\SetFigFont\undefined%
\gdef\SetFigFont#1#2#3#4#5{%
  \reset@font\fontsize{#1}{#2pt}%
  \fontfamily{#3}\fontseries{#4}\fontshape{#5}%
  \selectfont}%
\fi\endgroup%

\begin{picture}(6624,1824)(889,-1873)
\thinlines
{\color[rgb]{0,0,0}\put(4801,-1861){\framebox(2700,1800){}}
}%
{\color[rgb]{0,0,0}\put(3601,-661){\vector( 1, 0){1200}}
}%
{\color[rgb]{0,0,0}\put(901,-1861){\framebox(2700,1800){}}
}%
{\color[rgb]{0,0,0}\put(4801,-1261){\vector(-1, 0){1200}}
}%
\put(6151,-736){\makebox(0,0)[b]{\smash{\SetFigFont{8}{14.4}{\rmdefault}{\mddefault}{\updefault}{\color[rgb]{0,0,0}Communication Problems}%
}}}
\put(6151,-961){\makebox(0,0)[b]{\smash{\SetFigFont{8}{14.4}{\rmdefault}{\mddefault}{\updefault}{\color[rgb]{0,0,0}over Noisy Channels}%
}}}
\put(6151,-1186){\makebox(0,0)[b]{\smash{\SetFigFont{8}{14.4}{\rmdefault}{\mddefault}{\updefault}{\color[rgb]{0,0,0}with Noiseless Feedback}%
}}}
\put(2251,-961){\makebox(0,0)[b]{\smash{\SetFigFont{8}{14.4}{\rmdefault}{\mddefault}{\updefault}{\color[rgb]{0,0,0}with}%
}}}
\put(2251,-1186){\makebox(0,0)[b]{\smash{\SetFigFont{8}{14.4}{\rmdefault}{\mddefault}{\updefault}{\color[rgb]{0,0,0}Noisy Feedback Channels}%
}}}
\put(2251,-736){\makebox(0,0)[b]{\smash{\SetFigFont{8}{14.4}{\rmdefault}{\mddefault}{\updefault}{\color[rgb]{0,0,0}Stabilization Problems}%
}}}
\end{picture}
\end{center}
\caption{The ``equivalence'' between stabilization over noisy feedback
  channels and reliable communication over noisy channels with
  feedback is the main result established in this
  paper. \label{fig:abstract_equivalence}}
\end{figure}
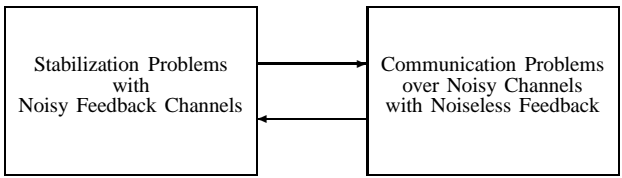

The new feature in control systems is their essential
interactivity. The information to be communicated is not a message
known in advance that is used by some completely separate
entity. Rather, it evolves through time and is used to control the
very process being encoded. This introduces two interesting
issues. First, causality is strictly enforced. The encoder and
controller must act in real time and so taking the limit of large
delays must be interpreted very carefully. Second, it is unclear what
the status of the controlled process is. If the controller succeeds in
stabilizing the process, it is no longer unstable. As explored in
Section~\ref{sec:externalvsinternal}, a purely external
non-interactive observer could treat the question of encoding the
controlled closed-loop system state using classical tools for the
encoding and communication of a stationary ergodic process. Despite
having to observe and encode the exact same closed-loop process, the
observer internal to the control system requires a channel as good as
that required to communicate the unstable open-loop process. This
seemingly paradoxical situation illustrates what can happen when the
encoding of information and its use are coupled together by
interactivity.

In this paper (Part I), the basic equivalence between feedback
stabilization and reliable communication is established. The scalar
problem (Figure~\ref{fig:problem}) is formally introduced in
Section~\ref{sec:setup} where classical capacity concepts are also
shown to be inadequate. In Section~\ref{sec:necessity}, it is shown
that adequate feedback anytime capacity is necessary for there to
exist an observer/controller pair able to stabilize the unstable
system across the noisy channel. This connection is also used to give
a sequential anytime version of the Schalkwijk/Kailath scheme for the
AWGN channel with noiseless feedback.

Section~\ref{sec:sufficiency} shows the sufficiency of feedback
anytime capacity for situations where the observer has noiseless
access to the channel outputs. In Section~\ref{sec:nofeedback}, these
sufficiency results are generalized to the case where the observer
only has noisy access to the plant state. Since the necessary and
sufficient conditions are tight in many cases, these results show the
asymptotic equivalence between the problem of control with ``noisy
feedback'' and the problem of reliable sequential communication with
noiseless feedback. In Section~\ref{sec:continuous}, these results are
further extended to the continuous time setting. Finally,
Section~\ref{sec:hierarchy} justifies why the problem of stabilization
of an unstable linear control system is ``universal'' in the same
sense that the Shannon formulation of reliable transmission of
messages over a noisy channel with (or without) feedback is
universal. This is done by introducing a hierarchy of communication
problems in which problems at a given level are equivalent to each
other in terms of which channels are good enough to solve
them. Problems high in the hierarchy are fundamentally more
challenging than the ones below them in terms of what they require
from the noisy channel.

In Part II, the necessity and sufficiency results are generalized to
the case of multivariable control systems on an unstable eigenvalue by
eigenvalue basis. The role of anytime capacity is played by a rate
region corresponding to a vector of anytime reliabilities. If there is
no explicit channel output feedback, the intrinsic delay of the
control system's input-output behavior plays an important role. It
shows that two systems with the same unstable eigenvalues can still
have potentially different channel requirements. These results
establish that in interactive settings, a single ``application'' can
fundamentally require different senses of reliability for its data
streams. No single number can adequately summarize the channel and any
layered communication architecture should allow applications to adjust
reliabilities on bitstreams.

There are many results in this paper. In order not to burden the
reader with repetitive details and unnecessarily lengthen this paper,
we have adopted a discursive style in some of the proofs. The reader
should not have any difficulty in filling in the omitted details.

\section{Problem definition and basic challenges} \label{sec:setup}

Section~\ref{sec:scalarproblem} formally introduces the control
problem of stabilizing an unstable scalar linear system driven by both
a control signal and a bounded disturbance. In
Section~\ref{sec:capacity}, classical notions of capacity are reviewed
along with how to stabilize an unstable system with a finite rate
noiseless channel. In Section~\ref{sec:inadequacy}, it is shown by
example that the classical concepts are inadequate when it comes to
evaluating a noisy channel for control purposes. Shannon's regular
capacity is too optimistic and zero-error capacity is too
pessimistic. Finally, Section~\ref{sec:externalvsinternal} shows that
the core issue of interactivity is different than merely requiring the
encoders and decoders to be delay-free.

\subsection{The control problem} \label{sec:scalarproblem}
\begin{equation} \label{eqn:discretesystem} 
X_{t+1} = \lambda X_{t} + U_{t} + W_{t}, \ \ t \geq 0
\end{equation} 
where $\{ X_{t} \}$ is a ${\real}$-valued state process.  $\{ U_{t}
\} $ is a ${\real}$-valued control process and $\{ W_{t} \}$ is a
bounded noise/disturbance process s.t.~$|W_t| \leq
\frac{\Omega}{2}$. This bound is assumed to hold with certainty. For
convenience, we also assume a known initial condition $X_0=0$. 

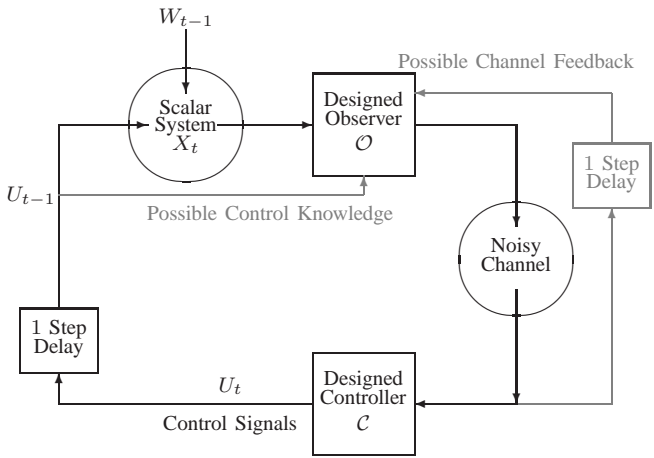
\begin{figure}
\begin{center}
\setlength{\unitlength}{2100sp}%
\begingroup\makeatletter\ifx\SetFigFont\undefined%
\gdef\SetFigFont#1#2#3#4#5{%
  \reset@font\fontsize{#1}{#2pt}%
  \fontfamily{#3}\fontseries{#4}\fontshape{#5}%
  \selectfont}%
\fi\endgroup%
\begin{picture}(7643,5307)(95,-4498)
\thinlines
\put(301,-3511){\framebox(900,750){}}
\put(751,-3061){\makebox(0,0)[b]{\smash{\SetFigFont{8}{14.4}{\rmdefault}{\mddefault}{\updefault}$1$ Step}}}
\put(751,-3286){\makebox(0,0)[b]{\smash{\SetFigFont{8}{14.4}{\rmdefault}{\mddefault}{\updefault}Delay}}}
\put(6150,-2169){\oval(1342,1342)}
\put(6151,-2311){\makebox(0,0)[b]{\smash{\SetFigFont{8}{14.4}{\rmdefault}{\mddefault}{\updefault}Channel}}}
\put(6151,-2086){\makebox(0,0)[b]{\smash{\SetFigFont{8}{14.4}{\rmdefault}{\mddefault}{\updefault}Noisy}}}
\put(4351,-361){\makebox(0,0)[b]{\smash{\SetFigFont{8}{14.4}{\rmdefault}{\mddefault}{\updefault}Designed}}}
\put(4351,-586){\makebox(0,0)[b]{\smash{\SetFigFont{8}{14.4}{\rmdefault}{\mddefault}{\updefault}Observer}}}
\put(4351,-3661){\makebox(0,0)[b]{\smash{\SetFigFont{8}{14.4}{\rmdefault}{\mddefault}{\updefault}Designed}}}
\put(4351,-3886){\makebox(0,0)[b]{\smash{\SetFigFont{8}{14.4}{\rmdefault}{\mddefault}{\updefault}Controller}}}
\put(2251,-586){\oval(1342,1342)}
\put(3751,-1186){\framebox(1200,1200){}}
\put(3751,-4486){\framebox(1200,1200){}}
\put(2626,-586){\vector( 1, 0){1125}}
\put(2251,539){\line( 0,-1){525}}
\put(2251, 14){\vector( 0,-1){225}}
\put(4951,-586){\line( 1, 0){1200}}
\put(6151,-586){\vector( 0,-1){1200}}
\put(3751,-3886){\line(-1, 0){3000}}
\put(751,-3886){\vector( 0, 1){375}}
\put(751,-2761){\line( 0, 1){2175}}
\put(751,-586){\vector( 1, 0){1050}}
{\color[gray]{0.5}
\put(3251,-1711){\makebox(0,0)[b]{\smash{\SetFigFont{8}{14.4}{\rmdefault}{\mddefault}{\updefault}Possible Control Knowledge}}}
\put(751,-1411){\line( 1, 0){3600}}
\put(4351,-1411){\vector( 0, 1){225}}
\put(6151, 89){\makebox(0,0)[b]{\smash{\SetFigFont{8}{14.4}{\rmdefault}{\mddefault}{\updefault}Possible Channel Feedback}}}
\put(6826,-1561){\framebox(900,750){}}
\put(7276,-1111){\makebox(0,0)[b]{\smash{\SetFigFont{8}{14.4}{\rmdefault}{\mddefault}{\updefault}$1$ Step}}}
\put(7276,-1336){\makebox(0,0)[b]{\smash{\SetFigFont{8}{14.4}{\rmdefault}{\mddefault}{\updefault}Delay}}}
\put(6151,-3886){\line( 1, 0){1125}}
\put(7276,-3886){\vector( 0, 1){2325}}
\put(7276,-811){\line( 0, 1){600}}
\put(7276,-211){\vector(-1, 0){2325}}
}
\put(2776,-3736){\makebox(0,0)[b]{\smash{\SetFigFont{8}{14.4}{\rmdefault}{\mddefault}{\updefault}$U_t$}}}
\put(2776,-4186){\makebox(0,0)[b]{\smash{\SetFigFont{8}{14.4}{\rmdefault}{\mddefault}{\updefault}Control Signals}}}
\put(4351,-886){\makebox(0,0)[b]{\smash{\SetFigFont{8}{14.4}{\rmdefault}{\mddefault}{\updefault}$\cal{O}$}}}
\put(4351,-4186){\makebox(0,0)[b]{\smash{\SetFigFont{8}{14.4}{\rmdefault}{\mddefault}{\updefault}$\cal{C}$}}}
\put(2251,-886){\makebox(0,0)[b]{\smash{\SetFigFont{8}{14.4}{\rmdefault}{\mddefault}{\updefault}$X_t$}}}
\put(2251,614){\makebox(0,0)[b]{\smash{\SetFigFont{8}{14.4}{\rmdefault}{\mddefault}{\updefault}$W_{t-1}$}}}
\put(426,-1486){\makebox(0,0)[b]{\smash{\SetFigFont{8}{14.4}{\rmdefault}{\mddefault}{\updefault}$U_{t-1}$}}}
\put(2251,-451){\makebox(0,0)[b]{\smash{\SetFigFont{8}{14.4}{\rmdefault}{\mddefault}{\updefault}Scalar}}}
\put(2251,-661){\makebox(0,0)[b]{\smash{\SetFigFont{8}{14.4}{\rmdefault}{\mddefault}{\updefault}System}}}
\put(6151,-3886){\vector(-1, 0){1200}}
\put(6151,-2536){\vector( 0,-1){1350}}
\end{picture}
\end{center}
\caption{Control over a noisy communication channel. The unstable
scalar system is persistently disturbed by $W_t$ and must be kept
stable in closed-loop through the actions of ${\cal O, C}$.}
\label{fig:problem}
\end{figure}

To make things interesting, consider $\lambda > 1$ so the open-loop
system is exponentially unstable. The distributed nature of the
problem (shown in Figure~\ref{fig:problem}) comes from having a noisy
communication channel in the feedback path. The observer/encoder
system $\cal O$ observes $X_t$ and generates inputs $a_t$ to the
channel. It may or may not have access to the control signals $U_t$ or
past channel outputs $B_{t-1}$ as well. The
decoder/controller\footnote{Because the decoder and controller are
both on the same side of the communication channel, they can be lumped
together into a single box.} system $\cal C$ observes channel outputs
$B_t$ and generates control signals $U_t$. Both $\cal O, C$ are
allowed to have unbounded memory and to be nonlinear in general.

\begin{definition} \label{def:stable}
A closed-loop dynamic system with state $X_t$ is {\em $f$-stable}
if ${\cal P}(|X_t| > m) < f(m)$ for all $t \geq 0$.
\end{definition}
\vspace{0.1in}

This definition requires the probability of a large state value to be
appropriately bounded. A looser sense of stability is given by:

\begin{definition} \label{def:momentstable}
A closed-loop dynamic system with state $X_t$ is {\em $\eta$-stable}
if there exists a constant $K$ s.t.~$E[|X_t|^\eta] \leq K$ for all
$t \geq 0$.
\end{definition}
\vspace{0.1in}

In both definitions, the bound is required to hold for all possible
sequences of bounded disturbances $\{W_t\}$ that satisfy the given
bound $\Omega$. We do not assume any specific probability model
governing the disturbances. Rather than having to specify a specific
target for the tail probability $f$, holding the $\eta$-moment within
bounds is a way of keeping large deviations rare. The larger $\eta$
is, the more strongly very large deviations are penalized. The
advantage of $\eta$-stability is that it allows constant factors to be
ignored while making sharp asymptotic statements. Furthermore,
Section~\ref{sec:scalar_implications} shows that for generic DMCs, no
sense stronger than $\eta$-stability is feasible.

{\em The goal in this paper is to find necessary and sufficient
conditions on the noisy channel for there to exist an observer ${\cal
O}$ and controller ${\cal C}$ so that the closed loop system shown in
Figure~\ref{fig:problem} is stable in the sense of definitions
\ref{def:stable} or \ref{def:momentstable}.} The problem is considered
under different information patterns corresponding to different
assumptions about what information is available at the observer ${\cal
O}$. The controller is always assumed to just have access to the
entire past history\footnote{In Section~\ref{sec:controllerlimits}, it
is shown that anything less than that can not work in general.} of
channel outputs. 

For discrete-time linear systems, the intrinsic rate of information
production (in units of bits per time) equals the sum of the
logarithms (base 2) of the unstable eigenvalues \cite{Tatikonda1}. In
the scalar case studied here, this is just $\log_2 \lambda$. This
means that it is generically\footnote{There are pathological cases
where it is possible to stabilize a system with less rate. These occur
when the driving disturbance is particularly structured instead of
just being unknown but bounded. An example is when the disturbance
only takes on values $\pm 1$ while $\lambda=4$. Clearly only one bit
per unit time is required even though $\log_2 \lambda = 2$.}
impossible to stabilize the system in any reasonable sense if the
feedback channel's Shannon classical capacity $C < \log_2 \lambda$.

\subsection{Classical notions of channels and capacity} \label{sec:capacity}
\begin{definition}
A {\em discrete time channel} is a probabilistic system with an
input. At every time step $t$, it takes an input $a_t \in {\cal A}$
and produces an output $b_t \in {\cal B}$ with
probability\footnote{This is a probability mass function in the case
of discrete alphabets $\cal B$, but is more generally an appropriate
probability measure over the output alphabet $\cal B$.}
$p(B_t|a_1^t,b_1^{t-1})$ where the notation $a_1^t$ is shorthand for
the sequence $a_1, a_2, \ldots, a_t$. In general, the current channel
output is allowed to depend on all inputs so far as well as on past
outputs.

The channel is {\em memoryless} if conditioned on $a_t$, $B_t$ is
independent of any other random variable in the system that occurs at
time $t$ or earlier. All that needs to be specified is $p(B_t|a_t)$.
\end{definition}
\vspace{0.1in}

The maximum rate achievable for a given sense of reliable
communication is called the associated capacity. Shannon's classical
reliability requires that after a suitably large end-to-end
delay\footnote{Traditionally, the community has used block-length for
a block code as the fundamental quantity rather than delay. It is easy
to see that doing encoding and decoding in blocks of size $n$
corresponds to a delay of between $n$ and $2n$ on the individual bits
being communicated.} $n$ that the average probability of error on each
bit is below a specified $\epsilon$. {\em Shannon classical capacity}
$C$ can also be calculated in the case of memoryless channels by
solving an optimization problem:
$$C = \sup_{{\cal P}(A)} I(A;B)$$ where the maximization is over the
input probability distribution and $I(A;B)$ represents the mutual
information through the channel \cite{Gallager}. This is referred to as
a single letter characterization of channel capacity for memoryless
channels. Similar formulae exist using limits in cases of channels
with memory.  There is another sense of reliability and its associated
capacity $C_0$ called {\em zero-error capacity} which requires the
probability of error to be exactly zero with sufficiently large
$n$. It does not have a simple single-letter
characterization \cite{ShannonZeroError}.

\begin{example} \label{example:noiseless}
Consider a system (\ref{eqn:discretesystem}) with $\Omega = 1$ and
$\lambda=\frac{3}{2}$. Suppose that the memoryless communication
channel is a noiseless one bit channel. So ${\cal A} = {\cal B} =
\{0,1\}$ and $p(B_t = 1| a_t = 1) = p(B_t = 0| a_t = 0) = 1$ while
$p(B_t = 1 | a_t = 0) = p(B_t = 0 | a_t = 1 ) = 0$. This channel has
$C_0 = C = 1 > \log_2 \frac{3}{2}$.

Use a memoryless observer 
$${\cal O}(x) = \left\{ \begin{array}{ll}
                0 & \mbox{if }x \leq 0 \\
                1 & \mbox{if }x > 0
			\end{array} \right.$$
and memoryless controller 
$${\cal C}(B) = \left\{ \begin{array}{ll}
                +\frac{3}{2} & \mbox{if }B = 0 \\
                -\frac{3}{2} & \mbox{if }B = 1
			\end{array} \right.$$

Assume that the closed loop system state is within the interval $[-2,
+2]$. If it is positive, then it is in the interval $[0, +2]$. At the
next time, $\frac{3}{2}X + W$ would be in the interval $[-\frac{1}{2},
\frac{7}{2}]$. The applied control of $-\frac{3}{2}$ shifts the state
back to within the interval $[-2, + 2]$. The same argument holds by
symmetry on the negative side. Since it starts at $0$, by induction it
will stay within $[-2,+2]$ forever. As a consequence, the second
moment will stay less than $4$ for all time, and all the other moments
will be similarly bounded.
\end{example}
\vspace{0.1in}

In addition to the Shannon and zero-error senses of reliability,
information theory has various reliability functions. Such reliability
functions (or error exponents) are traditionally considered an
internal matter for channel coding and were viewed as mathematically
tractable proxies for the issue of implementation complexity
\cite{Gallager}. Reliability functions study how fast the probability
of error goes to zero as the relevant system parameter is
increased. Thus, the reliability functions for block-codes are given
in terms of the block length, reliability functions for convolutional
codes in terms of the constraint length\cite{ForneyML}, and
reliability functions for variable-length codes in terms of the
expected block length \cite{burnashev}. With the rise of sparse code
constructions and iterative decoding, the prominence of error
exponents in channel coding has diminished since the computational
burden is not superlinear in the block-length.

For memoryless channels, the presence or absence of feedback does not
alter the classical Shannon capacity \cite{Gallager}. More
surprisingly, for symmetric DMCs, the fixed block coding reliability
functions also do not change with feedback, at least in the high rate
regime \cite{DobrushinReliability}. From a control perspective, this
is the first indication that neither Shannon's capacity nor
block-coding reliability functions are the perfect fit for control
applications.

\subsection{Counterexample showing classical concepts are inadequate} \label{sec:inadequacy}

We use erasure channels to construct a counterexample showing the
inadequacy of the Shannon classical capacity in characterizing
channels for control. While both erasure and AWGN channels are easy to
deal with, it turns out that AWGN channels can not be used for a
counterexample since they can be treated in the classical LQG
framework \cite{Bansal}. The deeper reason for why AWGN channels do
not provide a counterexample is given in
Section~\ref{sec:awgncasewithfeedback}.

\subsubsection{Erasure channels}
The packet erasure channel models situations where errors can be
reliably detected at the receiver. In the model, sometimes the packet
being sent does not make it through with probability $\delta$, but
otherwise it makes it through correctly. Explicitly:

\begin{definition}
The {\em $L$-bit packet erasure channel} is a memoryless channel with
${\cal A} = \{0,1\}^L$, ${\cal B} = \{0,1\}^L \cup \{ \emptyset \}$
and $p(x|x) = 1-\delta$ while $p(\emptyset|x) = \delta$.
\end{definition}
\vspace{0.1in}

It is well known that the Shannon capacity of the packet erasure
channel is $(1-\delta)L$ bits per channel use regardless of whether
the encoder has feedback or not \cite{Gallager}. Furthermore, because
a long string of erasures is always possible, the zero-error capacity
$C_0$ of this channel is $0$. There are also variable-length packet
erasure channels where the packet-length is something the encoder can
choose. See \cite{AllertonPacket2004} for a discussion of such
channels.

To construct a simple counterexample, consider a further abstraction:
\begin{definition}
The {\em real packet erasure channel} has ${\cal A} = {\cal B} = \real
$ and $p(x|x) = 1-\delta$ while $p(0|x) = \delta$. 
\end{definition}
\vspace{0.1in}

This model has also been explored in the context of Kalman filtering
with lossy observations \cite{MassimoKalmanRef, AndreaKalmanRef}. It
has infinite classical capacity since a single real number can carry
arbitrarily many bits within its binary expansion, while the
zero-error capacity remains $0$.

\subsubsection{The inadequacy of Shannon capacity}

Consider the problem from example \ref{example:noiseless}, except over
the real erasure channel instead of the one bit noiseless channel. The
goal is for the second moment to be bounded ($\eta=2$) and recall that
$\lambda=\frac{3}{2}$. Let $\delta = \frac{1}{2}$ so that there is a
$50\%$ chance of any real number being erased. Assume the bounded
disturbance $W_t$, assume that it is zero-mean and iid with variance
$\sigma^2$. By assuming an explicit probability model for the
disturbance, the problem is only made easier as compared to the
arbitrarily-varying but bounded model introduced earlier.

In this case, the optimal control is obvious --- set $a_t = X_t$ as
the channel input and use $U_t = -\lambda B_t$ as the control. With
every successful reception, the system state is reset to the initial
condition of zero. For an arbitrary time $t$, the time since it was
last reset is distributed like a geometric-$\frac{1}{2}$ random
variable. Thus the second moment is:

\begin{eqnarray*} 
E[|X_{t+1}|^2] & > & \sum_{i=0}^{t} \frac{1}{2}(\frac{1}{2})^i
E[ (\sum_{j=0}^i (\frac{3}{2})^j W_{t-j})^2 ] \\
& = & \sum_{i=0}^{t} \frac{1}{2}(\frac{1}{2})^i
\sum_{j=0}^i \sum_{k=0}^i (\frac{3}{2})^{j+k} E[ W_{t-j} W_{t-k}] \\
& = & \sum_{i=0}^{t} (\frac{1}{2})^{i+1}
\sum_{j=0}^i (\frac{9}{4})^{j} \sigma^2 \\
& = & \frac{4\sigma^2}{5} \sum_{i=0}^{t} 
\left((\frac{9}{8})^{i+1} - (\frac{1}{2})^{i+1} \right)
\end{eqnarray*}
This diverges as $t \rightarrow \infty$ since $\frac{9}{8} > 1$.

Notice that the root of the problem is that
$(\frac{3}{2})^2(\frac{1}{2}) > 1$. Intuitively, the system is
exploding faster than the noisy channel is able to give
reliability. This causes the second moment to diverge. In contrast,
the first moment $E[|X_t|]$ is bounded for all $t$ since
$(\frac{3}{2})(\frac{1}{2}) < 1$.

The adequacy of the channel depends on which moment is required to be
bounded. {\em Thus no single-number characterization like classical
capacity can give the figure-of-merit needed to evaluate a channel for
control applications.}

\subsection{Non-interactive observation of a closed-loop process}
\label{sec:externalvsinternal}

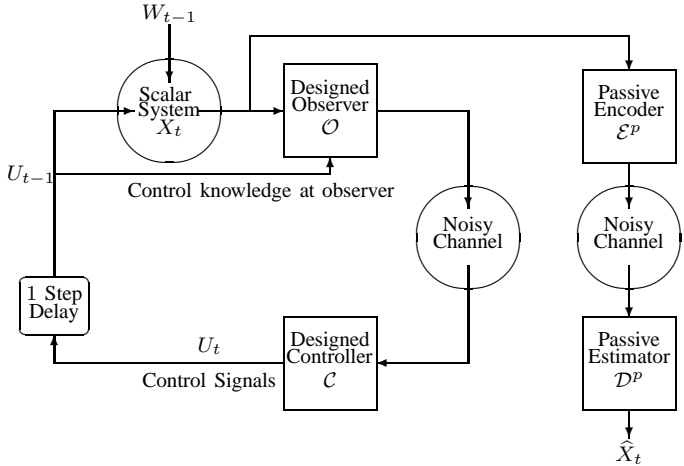
\begin{figure}
\begin{center}
\setlength{\unitlength}{1900sp}%
\begingroup\makeatletter\ifx\SetFigFont\undefined%
\gdef\SetFigFont#1#2#3#4#5{%
  \reset@font\fontsize{#1}{#2pt}%
  \fontfamily{#3}\fontseries{#4}\fontshape{#5}%
  \selectfont}%
\fi\endgroup%
\begin{picture}(8640,5989)(289,-5219)
\thinlines
{\color[rgb]{0,0,0}\put(406,-3406){\oval(210,210)[bl]}
\put(406,-2866){\oval(210,210)[tl]}
\put(1096,-3406){\oval(210,210)[br]}
\put(1096,-2866){\oval(210,210)[tr]}
\put(406,-3511){\line( 1, 0){690}}
\put(406,-2761){\line( 1, 0){690}}
\put(301,-3406){\line( 0, 1){540}}
\put(1201,-3406){\line( 0, 1){540}}
}%
\put(751,-3061){\makebox(0,0)[b]{\smash{\SetFigFont{8}{14.4}{\rmdefault}{\mddefault}{\updefault}{\color[rgb]{0,0,0}$1$ Step}%
}}}
\put(751,-3286){\makebox(0,0)[b]{\smash{\SetFigFont{8}{14.4}{\rmdefault}{\mddefault}{\updefault}{\color[rgb]{0,0,0}Delay}%
}}}
{\color[rgb]{0,0,0}\put(2251,-586){\oval(1342,1342)}
}%
{\color[rgb]{0,0,0}\put(8250,-2244){\oval(1342,1342)}
}%
{\color[rgb]{0,0,0}\put(6150,-2244){\oval(1342,1342)}
}%
{\color[rgb]{0,0,0}\put(3751,-4486){\framebox(1200,1200){}}
}%
{\color[rgb]{0,0,0}\put(2626,-586){\vector( 1, 0){1125}}
}%
{\color[rgb]{0,0,0}\put(2251,539){\line( 0,-1){525}}
\put(2251, 14){\vector( 0,-1){225}}
}%
{\color[rgb]{0,0,0}\put(4951,-586){\line( 1, 0){1200}}
\put(6151,-586){\vector( 0,-1){1275}}
}%
{\color[rgb]{0,0,0}\put(3751,-3886){\line(-1, 0){3000}}
\put(751,-3886){\vector( 0, 1){375}}
}%
{\color[rgb]{0,0,0}\put(751,-2761){\line( 0, 1){2175}}
\put(751,-586){\vector( 1, 0){1050}}
}%
{\color[rgb]{0,0,0}\put(751,-1411){\line( 1, 0){3600}}
\put(4351,-1411){\vector( 0, 1){225}}
}%
{\color[rgb]{0,0,0}\put(7651,-4486){\framebox(1200,1200){}}
}%
{\color[rgb]{0,0,0}\put(7651,-1261){\framebox(1200,1200){}}
}%
{\color[rgb]{0,0,0}\put(3751,-1186){\framebox(1200,1200){}}
}%
\put(2776,-3736){\makebox(0,0)[b]{\smash{\SetFigFont{8}{14.4}{\rmdefault}{\mddefault}{\updefault}{\color[rgb]{0,0,0}$U_t$}%
}}}
\put(2776,-4186){\makebox(0,0)[b]{\smash{\SetFigFont{8}{14.4}{\rmdefault}{\mddefault}{\updefault}{\color[rgb]{0,0,0}Control Signals}%
}}}
\put(2251,-451){\makebox(0,0)[b]{\smash{\SetFigFont{8}{14.4}{\rmdefault}{\mddefault}{\updefault}{\color[rgb]{0,0,0}Scalar }%
}}}
\put(2251,-661){\makebox(0,0)[b]{\smash{\SetFigFont{8}{14.4}{\rmdefault}{\mddefault}{\updefault}{\color[rgb]{0,0,0}System}%
}}}
\put(3451,-1711){\makebox(0,0)[b]{\smash{\SetFigFont{8}{14.4}{\rmdefault}{\mddefault}{\updefault}{\color[rgb]{0,0,0}Control knowledge at observer}%
}}}
\put(451,-1486){\makebox(0,0)[b]{\smash{\SetFigFont{8}{14.4}{\rmdefault}{\mddefault}{\updefault}{\color[rgb]{0,0,0}$U_{t-1}$}%
}}}
\put(2251,-886){\makebox(0,0)[b]{\smash{\SetFigFont{8}{14.4}{\rmdefault}{\mddefault}{\updefault}{\color[rgb]{0,0,0}$X_t$}%
}}}
\put(2251,614){\makebox(0,0)[b]{\smash{\SetFigFont{8}{14.4}{\rmdefault}{\mddefault}{\updefault}{\color[rgb]{0,0,0}$W_{t-1}$}%
}}}
\put(4351,-3661){\makebox(0,0)[b]{\smash{\SetFigFont{8}{14.4}{\rmdefault}{\mddefault}{\updefault}{\color[rgb]{0,0,0}Designed}%
}}}
\put(4351,-3886){\makebox(0,0)[b]{\smash{\SetFigFont{8}{14.4}{\rmdefault}{\mddefault}{\updefault}{\color[rgb]{0,0,0}Controller}%
}}}
\put(8251,-2386){\makebox(0,0)[b]{\smash{\SetFigFont{8}{14.4}{\rmdefault}{\mddefault}{\updefault}{\color[rgb]{0,0,0}Channel}%
}}}
\put(8251,-2161){\makebox(0,0)[b]{\smash{\SetFigFont{8}{14.4}{\rmdefault}{\mddefault}{\updefault}{\color[rgb]{0,0,0}Noisy}%
}}}
\put(8251,-436){\makebox(0,0)[b]{\smash{\SetFigFont{8}{14.4}{\rmdefault}{\mddefault}{\updefault}{\color[rgb]{0,0,0}Passive}%
}}}
\put(8251,-3661){\makebox(0,0)[b]{\smash{\SetFigFont{8}{14.4}{\rmdefault}{\mddefault}{\updefault}{\color[rgb]{0,0,0}Passive}%
}}}
\put(8251,-3886){\makebox(0,0)[b]{\smash{\SetFigFont{8}{14.4}{\rmdefault}{\mddefault}{\updefault}{\color[rgb]{0,0,0}Estimator}%
}}}
\put(8251,-661){\makebox(0,0)[b]{\smash{\SetFigFont{8}{14.4}{\rmdefault}{\mddefault}{\updefault}{\color[rgb]{0,0,0}Encoder}%
}}}
\put(8251,-961){\makebox(0,0)[b]{\smash{\SetFigFont{8}{14.4}{\rmdefault}{\mddefault}{\updefault}{\color[rgb]{0,0,0}${\cal E}^p$}%
}}}
\put(8251,-4186){\makebox(0,0)[b]{\smash{\SetFigFont{8}{14.4}{\rmdefault}{\mddefault}{\updefault}{\color[rgb]{0,0,0}${\cal D}^p$}%
}}}
\put(4351,-4186){\makebox(0,0)[b]{\smash{\SetFigFont{8}{14.4}{\rmdefault}{\mddefault}{\updefault}{\color[rgb]{0,0,0}${\cal C}$}%
}}}
\put(4351,-361){\makebox(0,0)[b]{\smash{\SetFigFont{8}{14.4}{\rmdefault}{\mddefault}{\updefault}{\color[rgb]{0,0,0}Designed}%
}}}
\put(4351,-586){\makebox(0,0)[b]{\smash{\SetFigFont{8}{14.4}{\rmdefault}{\mddefault}{\updefault}{\color[rgb]{0,0,0}Observer}%
}}}
\put(4351,-886){\makebox(0,0)[b]{\smash{\SetFigFont{8}{14.4}{\rmdefault}{\mddefault}{\updefault}{\color[rgb]{0,0,0}${\cal O}$}%
}}}
\put(6151,-2386){\makebox(0,0)[b]{\smash{\SetFigFont{8}{14.4}{\rmdefault}{\mddefault}{\updefault}{\color[rgb]{0,0,0}Channel}%
}}}
\put(6151,-2161){\makebox(0,0)[b]{\smash{\SetFigFont{8}{14.4}{\rmdefault}{\mddefault}{\updefault}{\color[rgb]{0,0,0}Noisy}%
}}}
{\color[rgb]{0,0,0}\put(6151,-2611){\line( 0,-1){1275}}
\put(6151,-3886){\vector(-1, 0){1200}}
}%
{\color[rgb]{0,0,0}\put(3301,-586){\line( 0, 1){975}}
\put(3301,389){\line( 1, 0){4950}}
\put(8251,389){\vector( 0,-1){450}}
}%
{\color[rgb]{0,0,0}\put(8251,-1261){\vector( 0,-1){600}}
}%
{\color[rgb]{0,0,0}\put(8251,-2611){\vector( 0,-1){675}}
}%
{\color[rgb]{0,0,0}\put(8251,-4486){\vector( 0,-1){375}}
}%
\put(8251,-5161){\makebox(0,0)[b]{\smash{\SetFigFont{8}{14.4}{\rmdefault}{\mddefault}{\updefault}{\color[rgb]{0,0,0}$\widehat{X}_t$}%
}}}
\end{picture}
\end{center}
\caption{The control system with an additional passive joint
source-channel encoder ${\cal E}^p$ watching the closed loop state
$X_t$ and communicating it to a passive estimator ${\cal D}^p$. The
controller ${\cal C}$ implicitly needs a good causal estimate for
$X_t$ and the passive estimator ${\cal D}^p$ explicitly needs the same
thing. Which requires the better channel?}
\label{fig:externalvsinternal}
\end{figure}

Consider the system shown in Figure~\ref{fig:externalvsinternal}. In
this, there is an additional passive joint source-channel encoder
${\cal E}^p$ watching the closed loop state $X_t$ and communicating it
to a passive estimator ${\cal D}^p$ through a second independent noisy
channel. Both the passive and internal observers have access to the
same plant state and we can also require the passive encoder and
decoder to be causal --- no end-to-end delay is permitted. At first
glance, it certainly appears that the communication situations are
symmetric. If anything, the internal observer is better off since it
also has access to the control signals while the passive observer is
denied access to them.

Suppose that the closed-loop process (\ref{eqn:discretesystem}) had
already been stabilized by the observer and controller system of
\ref{example:noiseless}, so that the second moment $E[X_t^2] \leq K$
for all $t$. Suppose that the noisy channel facing the passive encoder
is the real $\frac{1}{2}$-erasure channel of the previous section. It
is interesting to consider how well the passive observer does at
estimating this process.

The optimal encoding rule is clear, set $a_t = X_t$. It is certainly
feasible to use $\widehat{X}_t = B_t$ itself as the estimator for the
process. This passive observation system clearly achieves
$E[(\widehat{X}_t - X_t)^2] \leq \frac{K}{2} < K$ since the
probability of non-erasure is $\frac{1}{2}$. The causal decoding rule
is able to achieve a finite end-to-end squared error distortion over
this noisy channel in a causal and memoryless way.

This example makes it clear that the challenge here is arising from
interactivity, not simply being forced to be delay-free. The passive
external encoder and decoder do not have to face the unstable nature
of the source while the internal observer and controller do. An error
made while estimating $\widehat{X}_t$ by the passive decoder has no
consequence for the next state $X_{t+1}$ while a similar error by the
controller does. 

\section{Anytime capacity and its necessity} \label{sec:necessity}
Anytime reliability is introduced and related to classical notions of
reliability in \cite{OurSourceCodingPaper}. Here, the focus is on the
maximum rate achievable for a given sense of reliability rather than
the maximum reliability possible at a given rate. The two are of
course related since fundamentally there is an underlying region of
feasible rate/reliability pairs.

Since the open-loop system state has the potential to grow
exponentially, the controller's knowledge of the past must become
certain at a fast rate in order to prevent a bad decision made in the
past from continuing to corrupt the future. When viewed in the context
of reliably communicating bits from an encoder to a decoder, this
suggests that the estimates of the bits at the decoder must become
increasingly reliable with time. The sense of anytime reliability is
made precise in
Section~\ref{sec:anytimecapacity}. Section~\ref{sec:scalarnecessity}
then establishes the key result of this paper relating the problem of
stabilization to the reliable communication of messages in the anytime
sense. Finally, some consequences of this connection are studied in
Section~\ref{sec:scalar_implications}. Among these consequences is a
sequential generalization of the Schalkwijk/Kailath scheme for
communication over an AWGN channel that achieves a doubly-exponential
convergence to zero of the probability of bit error universally over
all delays simultaneously.

\subsection{Anytime reliability and capacity} \label{sec:anytimecapacity}

The entire message is not assumed to be known ahead of time. Rather,
it is made available gradually as time evolves. For simplicity of
notation, let $M_i$ be the $R$ bit message that the channel encoder
gets at time $i$. At the channel decoder, no target delay is assumed
--- {\em i.e.} the channel decoder does not necessarily know when the
message $i$ will be needed by the application. A past message may even
be needed more than once by the application. Consequently, the anytime
decoder produces estimates $\widehat{M}_{i}(t)$ which are the best
estimates for message $i$ at time $t$ based on all the channel outputs
received so far. If the application is using the past messages with a
delay $d$, the relevant probability of error is ${\cal
P}(\widehat{M}_1^{t-d}(t) \neq M_1^{t-d})$. This corresponds to an
uncorrected error anywhere in the distant past ({\em ie} on messages
$M_1, M_2, \ldots, M_{t-d}$) beyond $d$ channel uses ago.

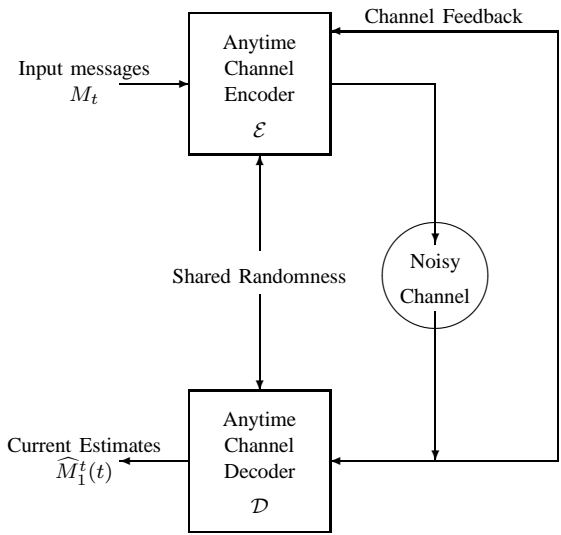
\begin{figure}
\begin{center}
\setlength{\unitlength}{2900sp}%
\begingroup\makeatletter\ifx\SetFigFont\undefined%
\gdef\SetFigFont#1#2#3#4#5{%
  \reset@font\fontsize{#1}{#2pt}%
  \fontfamily{#3}\fontseries{#4}\fontshape{#5}%
  \selectfont}%
\fi\endgroup%
\begin{picture}(4062,4497)(751,-3298)
{\color[rgb]{0,0,0}\thinlines
\put(3751,-1096){\circle{1500}}
}%
{\color[rgb]{0,0,0}\put(1651,-2686){\vector(-1, 0){600}}
}%
{\color[rgb]{0,0,0}\put(2251,-1261){\vector( 0,-1){825}}
}%
{\color[rgb]{0,0,0}\put(2251,-886){\vector( 0, 1){825}}
}%
{\color[rgb]{0,0,0}\put(1051,539){\vector( 1, 0){600}}
}%
{\color[rgb]{0,0,0}\put(2851,539){\line( 1, 0){900}}
\put(3751,539){\vector( 0,-1){1370}}
}%
{\color[rgb]{0,0,0}\put(1651,-61){\framebox(1200,1200){}}
}%
{\color[rgb]{0,0,0}\put(1651,-3286){\framebox(1200,1200){}}
}%
\put(2251,-1163){\makebox(0,0)[b]{\smash{\SetFigFont{8}{7.2}{\rmdefault}{\mddefault}{\updefault}{\color[rgb]{0,0,0}Shared Randomness}%
}}}
{\color[rgb]{0,0,0}\put(3751,-1411){\vector( 0,-1){1275}}
}%
{\color[rgb]{0,0,0}\put(3751,-2686){\line( 1, 0){1050}}
\put(4801,-2686){\line( 0, 1){3675}}
\put(4801,989){\vector(-1, 0){1950}}
}%
{\color[rgb]{0,0,0}\put(3751,-2686){\vector(-1, 0){900}}
}%
\put(2251,839){\makebox(0,0)[b]{\smash{\SetFigFont{8}{7.2}{\rmdefault}{\mddefault}{\updefault}{\color[rgb]{0,0,0}Anytime}%
}}}
\put(2251,614){\makebox(0,0)[b]{\smash{\SetFigFont{8}{7.2}{\rmdefault}{\mddefault}{\updefault}{\color[rgb]{0,0,0}Channel}%
}}}
\put(2251,389){\makebox(0,0)[b]{\smash{\SetFigFont{8}{7.2}{\rmdefault}{\mddefault}{\updefault}{\color[rgb]{0,0,0}Encoder}%
}}}
\put(751,-2611){\makebox(0,0)[b]{\smash{\SetFigFont{8}{7.2}{\rmdefault}{\mddefault}{\updefault}{\color[rgb]{0,0,0}Current Estimates}%
}}}
\put(3751,-1036){\makebox(0,0)[b]{\smash{\SetFigFont{8}{7.2}{\rmdefault}{\mddefault}{\updefault}{\color[rgb]{0,0,0}Noisy}%
}}}
\put(3751,-1336){\makebox(0,0)[b]{\smash{\SetFigFont{8}{7.2}{\rmdefault}{\mddefault}{\updefault}{\color[rgb]{0,0,0}Channel}%
}}}
\put(3826,1064){\makebox(0,0)[b]{\smash{\SetFigFont{8}{7.2}{\rmdefault}{\mddefault}{\updefault}{\color[rgb]{0,0,0}Channel Feedback}%
}}}
\put(2251,-2386){\makebox(0,0)[b]{\smash{\SetFigFont{8}{7.2}{\rmdefault}{\mddefault}{\updefault}{\color[rgb]{0,0,0}Anytime}%
}}}
\put(2251,-2611){\makebox(0,0)[b]{\smash{\SetFigFont{8}{7.2}{\rmdefault}{\mddefault}{\updefault}{\color[rgb]{0,0,0}Channel}%
}}}
\put(2251,-2836){\makebox(0,0)[b]{\smash{\SetFigFont{8}{7.2}{\rmdefault}{\mddefault}{\updefault}{\color[rgb]{0,0,0}Decoder}%
}}}
\put(2251,-3136){\makebox(0,0)[b]{\smash{\SetFigFont{8}{7.2}{\rmdefault}{\mddefault}{\updefault}{\color[rgb]{0,0,0}$\cal D$}%
}}}
\put(2251, 89){\makebox(0,0)[b]{\smash{\SetFigFont{8}{7.2}{\rmdefault}{\mddefault}{\updefault}{\color[rgb]{0,0,0}$\cal E$}%
}}}
\put(751,614){\makebox(0,0)[b]{\smash{\SetFigFont{8}{7.2}{\rmdefault}{\mddefault}{\updefault}{\color[rgb]{0,0,0}Input messages}%
}}}
\put(751,389){\makebox(0,0)[b]{\smash{\SetFigFont{8}{7.2}{\rmdefault}{\mddefault}{\updefault}{\color[rgb]{0,0,0}$M_t$}%
}}}
\put(751,-2836){\makebox(0,0)[b]{\smash{\SetFigFont{8}{7.2}{\rmdefault}{\mddefault}{\updefault}{\color[rgb]{0,0,0}$\widehat{M}_1^{t}(t)$}%
}}}
\end{picture}
\end{center}
\caption{The problem of communicating messages in an anytime
fashion. Both the encoder ${\cal E}$ and decoder ${\cal D}$ are causal
maps and the decoder in principle provides updated estimates for {\em
all} past messages. These estimates must converge to the
true message values appropriately rapidly with increasing delay.}
\label{fig:anytimecom}
\end{figure}

\begin{definition} \label{def:anytime}
As illustrated in figure \ref{fig:anytimecom}, a rate $R$ {\em
communication system} over a noisy channel is an encoder ${\cal E}$
and decoder ${\cal D}$ pair such that:
\begin{itemize}
  \item $R$-bit message $M_i$ enters\footnote{In what follows,
    messages are considered to be composed of bits for simplicity of
    exposition. The $i$-th bit arrives at the encoder at time
    $\frac{i}{R}$ and thus $M_i$ is composed of the bits $S_{\lfloor
    (i-1)R\rfloor +1}^{\lfloor iR \rfloor}$.} the encoder at discrete
    time $i$
  \item The encoder produces a channel input at integer times based on
        all information that it has seen so far. For encoders with
        access to feedback with delay $1+\theta$, this also includes
        the past channel outputs $B_1^{t-1-\theta}$. 
  \item The decoder produces updated channel estimates $\widehat{M}_i(t)$
        for all $i \leq t$ based on all channel outputs observed till
        time $t$.
\end{itemize}

A rate $R$ sequential communication system achieves {\em anytime
reliability} $\alpha$ if there exists a constant $K$ such that:
\begin{equation} \label{eqn:anytime_req}
 {\cal P}(\widehat{M}_1^i(t) \neq M_1^i) \leq K 2^{-\alpha (t-i)}
\end{equation}
holds for every $i,t$. The probability is taken over the channel
noise, the $R$ bit messages $M_i$, and all of the common randomness
available in the system. 

If (\ref{eqn:anytime_req}) holds for every possible realization of the
messages $M$, then the system is said to achieve {\em uniform anytime
reliability} $\alpha$.

Communication systems that achieve {\em anytime reliability} are
called {\em anytime codes} and similarly for {\em uniform anytime
codes}. 
\end{definition}
\vspace{0.1in}

We could alternatively have bounded the probability of error by
$2^{-\alpha(d-\log_2 K)}$ and interpreted $\log_2 K$ as the minimum
delay imposed by the communication system. 

\begin{definition} \label{def:anytimecapacity}
The {\em $\alpha$-anytime capacity} $C_{\mbox{any}}(\alpha)$ of a
channel is the least upper bound of the rates $R$ (in bits) at which the
channel can be used to construct a rate $R$ communication system
that achieves uniform anytime reliability $\alpha$. 

{\em Feedback anytime capacity} is used to refer to the anytime
capacity when the encoder has access to noiseless feedback of the
channel outputs with unit delay.
\end{definition}
\vspace{0.1in}

The requirement for exponential decay in the probability of error with
delay is reminiscent of the block-coding reliability functions $E(R)$
of a channel given in \cite{Gallager}. There is one crucial
difference. With standard error exponents, both the encoder and
decoder vary with blocklength or delay $n$. Here, the encoding is
required to be fixed and the decoder in principle has to work at all
delays since it must produce updated estimates of the message $M_i$ at
all times $t > i$.

This additional requirement is why it is called ``anytime''
capacity. The decoding process can be queried for a given bit at any
time and the answer is required to be increasingly accurate the longer
we wait. The anytime reliability $\alpha$ specifies the exponential
rate at which the quality of the answers must improve. The anytime
sense of reliable transmission lies between that represented by
classical zero-error capacity $C_0$ (probability of error becomes zero
at a large but finite delay) and classical capacity $C$ (probability
of error becomes something small at a large but finite delay).  It is
clear that $\forall \alpha, C_0 \leq C_{\mbox{any}}(\alpha) \leq C$.

By using a random coding argument over infinite tree codes, it is
possible to show the existence of anytime codes without using feedback
between the encoder and decoder for all rates less than the Shannon
capacity. This shows:
$$C_{\mbox{any}}(E_r(R)) \geq R $$ where $E_r(R)$ is Gallager's random
coding error exponent calculated in base $2$ and $R$ is the rate in
bits \cite{SahaiThesis, OurSourceCodingPaper}. Since feedback plays an
essential role in control, it turns out that we are interested in the
anytime capacity with feedback. It is interesting to note that in many
cases for which the block-coding error exponents are not increased
with feedback, the anytime reliabilities are increased considerably
\cite{OurUpperBoundPaper}.

\subsection{Necessity of anytime capacity} \label{sec:scalarnecessity}

Anytime reliability and capacity are defined in terms of digital
messages that must be reliably communicated from point to
point. Stability is a notion involving the analog value of the state
of a plant in interaction with a controller over a noisy feedback
channel. At first glance, these two problems appear to have nothing in
common except the noisy channel. Even on that point there is a
difference. The observer/encoder ${\cal O}$ in the control system may
have no explicit access to the noisy output of the channel. It can
appear to be using the noisy channel {\em without feedback}. Despite
this, it turns out that the relevant digital communication problem
involves access to the noisy channel {\em with noiseless channel
feedback} coming back to the message encoder.
\begin{theorem} \label{thm:scalar_control_necessity}
For a given noisy channel and $\eta > 0$, if there exists an observer
${\cal O}$ and controller ${\cal C}$ for the unstable scalar system
that achieves $E[|X_t|^\eta] < K$ for all sequences of bounded driving
noise $|W_t| \leq \frac{\Omega}{2}$, then the channel's feedback
anytime capacity $C_{\mbox{any}}(\eta \log_2 \lambda) \geq \log_2
\lambda$ bits per channel use.
\end{theorem}
\vspace{0.1in}

The proof of this spans the next few sections. Assume that there is an
observer/controller pair $({\cal O, C})$ that can $\eta$-stabilize an
unstable system with a particular $\lambda$ and are robust to all
bounded disturbances of size $\Omega$. The goal is to use the pair to
construct a rate $R < \log_2 \lambda$ anytime encoder and decoder for
the channel with noiseless feedback, thereby reducing\footnote{In
traditional rate-distortion theory, this ``necessity'' direction is
shown by going through the mutual information characterizations of
both the rate-distortion function and the channel capacity
function. In the case of stabilization, mutual information is not
discriminating enough and so the reduction of anytime reliable
communication to stabilization must be done directly.} the problem of
anytime communication to a problem of stabilization.

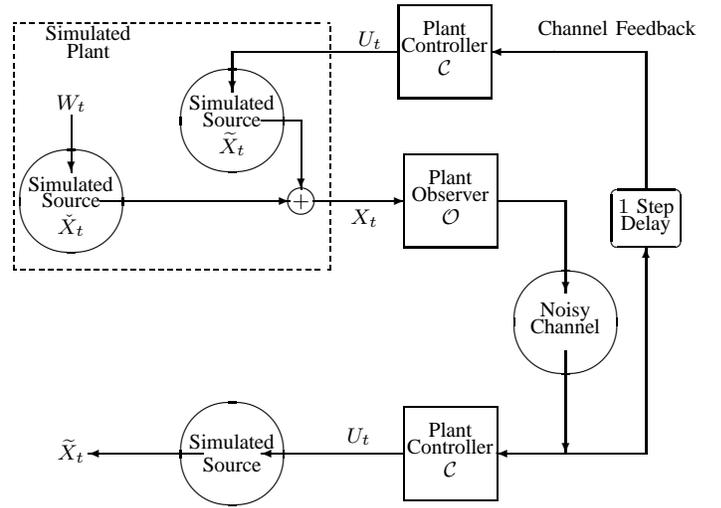
\begin{figure}
\begin{center}
\setlength{\unitlength}{1900sp}%
\begingroup\makeatletter\ifx\SetFigFont\undefined%
\gdef\SetFigFont#1#2#3#4#5{%
  \reset@font\fontsize{#1}{#2pt}%
  \fontfamily{#3}\fontseries{#4}\fontshape{#5}%
  \selectfont}%
\fi\endgroup%
\begin{picture}(8724,6541)(64,-6140)
\thinlines
{\color[rgb]{0,0,0}\put(7981,-2656){\oval(210,210)[bl]}
\put(7981,-2116){\oval(210,210)[tl]}
\put(8671,-2656){\oval(210,210)[br]}
\put(8671,-2116){\oval(210,210)[tr]}
\put(7981,-2761){\line( 1, 0){690}}
\put(7981,-2011){\line( 1, 0){690}}
\put(7876,-2656){\line( 0, 1){540}}
\put(8776,-2656){\line( 0, 1){540}}
}%
\put(8326,-2311){\makebox(0,0)[b]{\smash{\SetFigFont{8}{14.4}{\rmdefault}{\mddefault}{\updefault}{\color[rgb]{0,0,0}$1$ Step}%
}}}
\put(8326,-2536){\makebox(0,0)[b]{\smash{\SetFigFont{8}{14.4}{\rmdefault}{\mddefault}{\updefault}{\color[rgb]{0,0,0}Delay}%
}}}
{\color[rgb]{0,0,0}\put(2926,-5461){\oval(1342,1342)}
}%
{\color[rgb]{0,0,0}\put(826,-2161){\oval(1342,1342)}
}%
{\color[rgb]{0,0,0}\put(7275,-3744){\oval(1342,1342)}
}%
{\color[rgb]{0,0,0}\put(2926,-1111){\oval(1342,1342)}
}%
{\color[rgb]{0,0,0}\put(3826,-2161){\circle{336}}
}%
{\color[rgb]{0,0,0}\put(7276,-5461){\line( 1, 0){1050}}
\put(8326,-5461){\vector( 0, 1){2700}}
}%
{\color[rgb]{0,0,0}\put(8326,-2011){\line( 0, 1){1800}}
\put(8326,-211){\vector(-1, 0){2025}}
}%
{\color[rgb]{0,0,0}\put(6376,-2161){\line( 1, 0){900}}
\put(7276,-2161){\vector( 0,-1){1200}}
}%
{\color[rgb]{0,0,0}\put(3976,-2161){\vector( 1, 0){1200}}
}%
{\color[rgb]{0,0,0}\put(5101,-211){\line(-1, 0){2175}}
\put(2926,-211){\line( 0,-1){300}}
\put(2926,-511){\vector( 0,-1){225}}
}%
{\color[rgb]{0,0,0}\put(1201,-2161){\vector( 1, 0){2475}}
}%
{\color[rgb]{0,0,0}\put(3301,-1111){\line( 1, 0){525}}
\put(3826,-1111){\vector( 0,-1){900}}
}%
{\color[rgb]{0,0,0}\put(2551,-5461){\vector(-1, 0){1500}}
}%
{\color[rgb]{0,0,0}\put( 76,-3061){\dashbox{60}(4125,3225){}}
}%
{\color[rgb]{0,0,0}\put(826,-1036){\line( 0,-1){525}}
\put(826,-1561){\vector( 0,-1){225}}
}%
{\color[rgb]{0,0,0}\put(5176,-2761){\framebox(1200,1200){}}
}%
{\color[rgb]{0,0,0}\put(5176,-6061){\framebox(1200,1200){}}
}%
{\color[rgb]{0,0,0}\put(5101,-811){\framebox(1200,1200){}}
}%
{\color[rgb]{0,0,0}\put(6376,-5461){\makebox(1.6667,11.6667){\SetFigFont{5}{6}{\rmdefault}{\mddefault}{\updefault}.}}
}%
{\color[rgb]{0,0,0}\put(5176,-5461){\vector(-1, 0){1875}}
}%
{\color[rgb]{0,0,0}\put(7126,-211){\makebox(1.6667,11.6667){\SetFigFont{5}{6}{\rmdefault}{\mddefault}{\updefault}.}}
}%
\put(4576,-5311){\makebox(0,0)[b]{\smash{\SetFigFont{8}{14.4}{\rmdefault}{\mddefault}{\updefault}{\color[rgb]{0,0,0}$U_t$}%
}}}
\put(2926,-5686){\makebox(0,0)[b]{\smash{\SetFigFont{8}{14.4}{\rmdefault}{\mddefault}{\updefault}{\color[rgb]{0,0,0}Source}%
}}}
\put(826,-2236){\makebox(0,0)[b]{\smash{\SetFigFont{8}{14.4}{\rmdefault}{\mddefault}{\updefault}{\color[rgb]{0,0,0}Source}%
}}}
\put(826,-961){\makebox(0,0)[b]{\smash{\SetFigFont{8}{14.4}{\rmdefault}{\mddefault}{\updefault}{\color[rgb]{0,0,0}$W_{t}$}%
}}}
\put(7951, 14){\makebox(0,0)[b]{\smash{\SetFigFont{8}{14.4}{\rmdefault}{\mddefault}{\updefault}{\color[rgb]{0,0,0}Channel Feedback}%
}}}
\put(5776,-2461){\makebox(0,0)[b]{\smash{\SetFigFont{8}{14.4}{\rmdefault}{\mddefault}{\updefault}{\color[rgb]{0,0,0}$\cal{O}$}%
}}}
\put(5776,-5761){\makebox(0,0)[b]{\smash{\SetFigFont{8}{14.4}{\rmdefault}{\mddefault}{\updefault}{\color[rgb]{0,0,0}$\cal{C}$}%
}}}
\put(5776,-5461){\makebox(0,0)[b]{\smash{\SetFigFont{8}{14.4}{\rmdefault}{\mddefault}{\updefault}{\color[rgb]{0,0,0}Controller}%
}}}
\put(5776,-2161){\makebox(0,0)[b]{\smash{\SetFigFont{8}{14.4}{\rmdefault}{\mddefault}{\updefault}{\color[rgb]{0,0,0}Observer}%
}}}
\put(5776,-1936){\makebox(0,0)[b]{\smash{\SetFigFont{8}{14.4}{\rmdefault}{\mddefault}{\updefault}{\color[rgb]{0,0,0}Plant}%
}}}
\put(5776,-5236){\makebox(0,0)[b]{\smash{\SetFigFont{8}{14.4}{\rmdefault}{\mddefault}{\updefault}{\color[rgb]{0,0,0}Plant}%
}}}
\put(5701,-511){\makebox(0,0)[b]{\smash{\SetFigFont{8}{14.4}{\rmdefault}{\mddefault}{\updefault}{\color[rgb]{0,0,0}$\cal{C}$}%
}}}
\put(5701,-211){\makebox(0,0)[b]{\smash{\SetFigFont{8}{14.4}{\rmdefault}{\mddefault}{\updefault}{\color[rgb]{0,0,0}Controller}%
}}}
\put(5701, 14){\makebox(0,0)[b]{\smash{\SetFigFont{8}{14.4}{\rmdefault}{\mddefault}{\updefault}{\color[rgb]{0,0,0}Plant}%
}}}
\put(7276,-3886){\makebox(0,0)[b]{\smash{\SetFigFont{8}{14.4}{\rmdefault}{\mddefault}{\updefault}{\color[rgb]{0,0,0}Channel}%
}}}
\put(7276,-3661){\makebox(0,0)[b]{\smash{\SetFigFont{8}{14.4}{\rmdefault}{\mddefault}{\updefault}{\color[rgb]{0,0,0}Noisy}%
}}}
\put(1051,-286){\makebox(0,0)[b]{\smash{\SetFigFont{8}{14.4}{\rmdefault}{\mddefault}{\updefault}{\color[rgb]{0,0,0}Plant}%
}}}
\put(826,-2026){\makebox(0,0)[b]{\smash{\SetFigFont{8}{14.4}{\rmdefault}{\mddefault}{\updefault}{\color[rgb]{0,0,0}Simulated}%
}}}
\put(826,-2581){\makebox(0,0)[b]{\smash{\SetFigFont{8}{14.4}{\rmdefault}{\mddefault}{\updefault}{\color[rgb]{0,0,0}$\check{X}_t$}%
}}}
\put(2926,-961){\makebox(0,0)[b]{\smash{\SetFigFont{8}{14.4}{\rmdefault}{\mddefault}{\updefault}{\color[rgb]{0,0,0}Simulated}%
}}}
\put(2926,-1186){\makebox(0,0)[b]{\smash{\SetFigFont{8}{14.4}{\rmdefault}{\mddefault}{\updefault}{\color[rgb]{0,0,0}Source}%
}}}
\put(4726,-136){\makebox(0,0)[b]{\smash{\SetFigFont{8}{14.4}{\rmdefault}{\mddefault}{\updefault}{\color[rgb]{0,0,0}$U_{t}$}%
}}}
\put(2926,-1531){\makebox(0,0)[b]{\smash{\SetFigFont{8}{14.4}{\rmdefault}{\mddefault}{\updefault}{\color[rgb]{0,0,0}$\widetilde{X}_{t}$}%
}}}
\put(4651,-2461){\makebox(0,0)[b]{\smash{\SetFigFont{8}{14.4}{\rmdefault}{\mddefault}{\updefault}{\color[rgb]{0,0,0}$X_t$}%
}}}
\put(826,-5536){\makebox(0,0)[b]{\smash{\SetFigFont{8}{14.4}{\rmdefault}{\mddefault}{\updefault}{\color[rgb]{0,0,0}$\widetilde{X}_t$}%
}}}
\put(1051,-61){\makebox(0,0)[b]{\smash{\SetFigFont{8}{14.4}{\rmdefault}{\mddefault}{\updefault}{\color[rgb]{0,0,0}Simulated}%
}}}
\put(2926,-5386){\makebox(0,0)[b]{\smash{\SetFigFont{8}{14.4}{\rmdefault}{\mddefault}{\updefault}{\color[rgb]{0,0,0}Simulated}%
}}}
\put(3826,-2236){\makebox(0,0)[b]{\smash{\SetFigFont{8}{14.4}{\rmdefault}{\mddefault}{\updefault}{\color[rgb]{0,0,0}$+$}%
}}}
{\color[rgb]{0,0,0}\put(7276,-4111){\vector( 0,-1){1350}}
}%
{\color[rgb]{0,0,0}\put(7276,-5461){\vector(-1, 0){900}}
}%
\end{picture}
\end{center}
\caption{The construction of a feedback anytime code from a control
system. The messages are used to generate the $\{W_t\}$ inputs which
are causally combined to generate $\{\check{X}_t\}$ within the
encoder. The channel outputs are used to generate control signals at
both the encoder and decoder. Since the simulated plant is stable,
$-\widetilde{X}$ and $\check{X}$ are close to each other. The past
message bits are estimated from the $\widetilde{X}$ at the decoder.}
\label{fig:control_estimation_equivalence}
\end{figure}

The heart of the construction is illustrated in figure
\ref{fig:control_estimation_equivalence}. The ``black-box'' observer
and controller are wrapped around a simulated plant mimicking
(\ref{eqn:discretesystem}). Since the $\{U_t\}$ must be generated by
the black-box controller ${\cal C}$ and the $\lambda$ is prespecified,
the disturbances $\{W_t\}$ must be used to carry the message. So, the
encoder must embed the messages $\{M_t\}$ into an appropriate sequence
$\{W_t\}$, taking care to stay within the $\Omega$ size limit.

While both the observer and controller can be simulated at the encoder
thanks to the noiseless channel output feedback, at the decoder only
the channel outputs are available. Consequently, these channel outputs
are connected to a copy of the black-box controller ${\cal C}$,
thereby giving access to the controls $\{U_t\}$ at the decoder. To
extract the messages from these control signals, they are first
causally preprocessed through a simulated copy of the unstable plant,
except with no disturbance input. All past messages are then estimated
from the current state of this simulated plant.

The key is to think of the
simulated plant state as the sum of the states of two different
unstable LTI systems. The first, with state denoted $\widetilde{X}_t$, is
driven entirely by the controls and starts in state $0$.
\begin{equation}\label{eqn:tildeX}
 \widetilde{X}_{t+1} = \lambda\widetilde{X}_t + U_t
\end{equation}
$\widetilde{X}$ is available at both the decoder and the encoder due
to the presence of noiseless feedback.\footnote{If the controller is
randomized, then the randomness is required to be common and shared
between the encoder and decoder.}  The other, with state denoted
$\check{X}_t$, is driven entirely by a simulated driving noise that is
generated from the data stream to be communicated.
\begin{equation}\label{eqn:checkX}
 \check{X}_{t+1} = \lambda\check{X}_t + W_t
\end{equation}
The sum $X_t = (\widetilde{X}_t + \check{X}_t)$ behaves exactly like it
was coming from (\ref{eqn:discretesystem}) and is fed to the observer
which uses it to generate inputs for the noisy channel.

The fact that the original observer/controller pair stabilized the
original system implies that $|X_t|=|\check{X} - (-\widetilde{X}_t)|$
is small and hence $-\widetilde{X}_t$ stays close to $\check{X}_t$.


\subsubsection{Encoding data into the state}

As long as the bound $\Omega$ is satisfied, the encoder is free to
choose any disturbance\footnote{In \cite{OurSourceCodingPaper}, a
similar strategy is followed assuming a specific density for the iid
disturbance $W_t$. In that context, it is important to choose a
simulated disturbance sequence that behaves stochastically like
$W_t$. This is accomplished by using common randomness shared between
the encoder and decoder to dither the kind of disturbances produced
here into ones with the desired density.} for the simulated plant. The
choice will be determined by the data rate $R$ and the specific
messages to be sent. Rather than working with general messages $M_i$,
consider a bitstream $S_i$ with bit $i$ becoming available at time
$\frac{i}{R}$. Everything generalizes naturally to non-binary
alphabets for the messages, but the notation is cleaner in the binary
case with $S_i = \pm 1$.

$\check{X}_t$ is the part of $X_t$ driven only by the $\{W_t\}$. 

\begin{eqnarray*}
\check{X}_{t} 
& = & \lambda \check{X}_{t-1} + W_{t-1} \\
& = & \sum_{i=0}^{t-1} \lambda^i W_{t-1-i} \\
& = & \lambda^{t-1} \sum_{j=0}^{t-1} \lambda^{-j} W_{j}
\end{eqnarray*}

This looks like the representation of a fractional number in base
$\lambda$ which is then multiplied by $\lambda^{t-1}$. This is
exploited in the encoding by choosing the bounded disturbance sequence
so that:\footnote{For a rough understanding, ignore the $\epsilon_1$
and suppose that the message were encoded in binary. It is intuitive
that any good estimate of the $\check{X}_t$ state is going to agree
with $\check{X}_t$ in all the high order bits. Since the system is
unstable, all the encoded bits eventually become high-order bits
as time goes on. So no bit error could persist for too long and still
keep the estimate close to $\check{X}_t$. The $\epsilon_1$ in the
encoding is a technical device to make this reasoning hold uniformly
for all bit strings, rather than merely ``typical'' ones. This is
important since we are aiming for exponentially small bounds and so
cannot neglect rare events.}
\begin{equation} \label{eqn:scalarcantorencoding}
\check{X}_{t} = \gamma \lambda^t \sum_{k=0}^{\lfloor Rt \rfloor}
(2+\epsilon_1)^{-k} S_k 
\end{equation}
where $S_k$ is the $k$-th bit\footnote{For the next section, it is
convenient to have the disturbances balanced around zero and so we
choose to represent the bit $S_i$ as $+1$ or $-1$ rather than the
usual 1 or 0.}  of data that the anytime encoder has to send and
$\lfloor Rt \rfloor$ is just the total number of bits that are
available by time $t$. $\gamma, \epsilon_1$ are constants to be
specified.

To see that (\ref{eqn:scalarcantorencoding}) is always possible to
achieve by appropriate choice of $W$, use
induction. (\ref{eqn:scalarcantorencoding}) clearly holds for
$t=0$. Now assume that it holds for time $t$ and consider time $t+1$:
\begin{eqnarray*} 
\check{X}_{t+1} 
& = & \lambda \check{X}_t + W_t \\
& = & \gamma \lambda^{t+1} (\sum_{k=0}^{\lfloor Rt \rfloor}
(2+\epsilon_1)^{-k} S_k ) + W_t
\end{eqnarray*}
So setting 
\begin{equation} \label{eqn:scalardisturbance} 
W_t = \gamma \lambda^{t+1}\sum_{k=\lfloor Rt \rfloor+1}^{\lfloor 
R(t+1) \rfloor} (2+\epsilon_1)^{-k} S_k
\end{equation}
gives the desired result. Manipulate (\ref{eqn:scalardisturbance}) to
get $W_t =$
\begin{eqnarray*}
 & & \gamma \lambda^{t+1} (2+\epsilon_1)^{-\lfloor Rt \rfloor}
\sum_{j=1}^{\lfloor R(t+1) \rfloor - \lfloor Rt \rfloor} 
(2+\epsilon_1)^{-j} S_{\lfloor Rt \rfloor + j} \\
& = & \gamma \lambda \frac{(2+\epsilon_1)^{Rt - (\lfloor Rt \rfloor)}}
{\lambda^{-t(1-R \frac{\log_2(2+\epsilon_1)}{\log_2 \lambda})}}
\sum_{j=1}^{\lfloor R(t+1) \rfloor - \lfloor Rt \rfloor} 
(2+\epsilon_1)^{-j} S_{\lfloor Rt \rfloor + j}
\end{eqnarray*}

To keep this bounded, choose 
\begin{equation} \label{eqn:epsilonchoice}
\epsilon_1 = 2^{\frac{\log_2 \lambda}{R}} - 2
\end{equation}
which is strictly positive if $R < \log_2 \lambda$. Applying that
substitution gives $|W_t| = $
\begin{eqnarray*}
& &  |\gamma \lambda (2+\epsilon_1)^{Rt - (\lfloor Rt \rfloor)}
\sum_{j=1}^{\lfloor R(t+1) \rfloor - \lfloor Rt \rfloor} 
(2+\epsilon_1)^{-j} S_{\lfloor Rt \rfloor + j} |\\
& < & |\gamma \lambda (2+\epsilon_1)| \\
& = & |\gamma \lambda^{1 + \frac{1}{R}}|
\end{eqnarray*}
So by choosing 
\begin{equation} \label{eqn:gammachoice}
\gamma = \frac{\Omega}{2 \lambda^{1 + \frac{1}{R}}}
\end{equation}
the simulated disturbance is guaranteed to stay within the specified
bounds.

\subsubsection{Extracting data bits from the state estimate}
\label{sec:extractionalgorithm}

\begin{lemma} \label{lem:erroreventbound}
 Given a channel with access to noiseless feedback, for any rate $R <
 \log_2 \lambda$, it is possible to encode bits into the simulated
 scalar plant so that the uncontrolled process behaves like
 (\ref{eqn:scalarcantorencoding}) by using disturbances given in
 (\ref{eqn:scalardisturbance}) and the formulas
 (\ref{eqn:epsilonchoice}) and (\ref{eqn:gammachoice}). At the output
 end of the noisy channel, it is possible to extract estimates
 $\widehat{S}_i(t)$ for the $i$-th bit sent for which the error event
 \begin{equation} \label{eqn:erroreventbound}
   \{\omega | \exists i\leq j, \widehat{S}_i(t) \neq S_i(t) \} \subseteq 
   \{\omega | |X_t| \geq \lambda^{t-\frac{j}{R}} \left(\frac{ \gamma \epsilon_1}{1+\epsilon_1}\right) \}
 \end{equation} and thus:
 \begin{equation} \label{eqn:errorprobbound}
 {\cal P}(\widehat{S}_1^j(t) \neq S_1^j(t)) \leq  {\cal P}(|X_t| \geq \lambda^{t-\frac{j}{R}} \left(\frac{ \gamma \epsilon_1}{1+\epsilon_1}\right))
 \end{equation}
\end{lemma}
{\em Proof:} Here $\omega$ is used to denote members of the underlying
sample space.\footnote{If the bits to be sent are deterministic, this
is the sample space giving channel noise realizations.}

The decoder has $-\widetilde{X}_t = \check{X}_t - X_t$ which is close
to $\check{X}$ since $X_t$ is small. To see how to extract bits from
$-\widetilde{X}_t$, first consider how to recursively extract those
bits from $\check{X}_t$.

Starting with the first bit, notice that the set of all possible
$\check{X}_t$ that have $S_0 = +1$ is separated from the set of all
possible $\check{X}_t$ that have $S_0 = -1$ by a gap of
\begin{eqnarray*}
 & & 
 \gamma \lambda^t \left( (1 - \sum_{k=1}^{\lfloor Rt \rfloor}
 (2+\epsilon_1)^{-k}) - (-1 + \sum_{k=1}^{\lfloor Rt
 \rfloor} (2+\epsilon_1)^{-k}) \right) \\
 & > & 
 \gamma \lambda^t 2 (1 - \sum_{k=1}^{\infty}
 (2+\epsilon_1)^{-k}) \\
 & = & 
 \gamma \lambda^t 2 (1 - \frac{1}{1+\epsilon_1}) \\
 & = & 
 \lambda^t \left(\frac{2 \epsilon_1 \gamma}{1+\epsilon_1}\right)
\end{eqnarray*}

\begin{figure}[htbp]
\begin{center}
\mbox{\epsfxsize=3.3in \epsfysize=0.6in \epsfbox{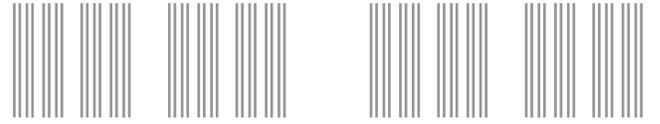}}
\end{center}
\caption{The data bits are used to sequentially refine a point on a
Cantor set. Its natural tree structure allows bits to be encoded
sequentially. The Cantor set also has finite gaps between all points
corresponding to bit sequences that first differ in a particular bit
position. These gaps allow the uniformly reliable extraction of bit
values from noisy observations.}
\label{fig:cantorset}
\end{figure}

Notice that this worst-case gap\footnote{The typical gap is larger and
so the probability of error is actually lower than this bound says it
is.} is a positive number that is growing exponentially in $t$. If the
first $i-1$ bits are the same, then both sides can be scaled by
$(2+\epsilon_1)^i = \lambda^{\frac{i}{R}}$ to get the same expressions
above and so by induction, it quickly follows that the minimum gap
between the encoded state corresponding to two sequences of bits that
first differ in bit position $i$ is given by $\mbox{gap}_i(t) =$
\begin{equation} \label{eqn:gapbound}
 \inf_{\bar{S} : \bar{S}_i \neq S_i} |\check{X}_t(S) -
 \check{X}_t(\bar{S})| > 
 \left\{\begin{array}{ll} \lambda^{t-\frac{i}{R}} 
 \left(\frac{2 \gamma  \epsilon_1}{1+\epsilon_1}\right) & \mbox{if }i
 \leq \lfloor Rt \rfloor \\
 0 & \mbox{otherwise}	  
        \end{array}\right.
\end{equation}
Because the gaps are all positive, (\ref{eqn:gapbound}) shows that it
is always possible to perfectly extract the data bits from
$\check{X}_t$ by using an iterative procedure.\footnote{This is a
minor twist on the procedure followed by serial A/D converters.} To
extract bit information from an input $I_t$:
 \begin{enumerate}
 \item Initialize threshold $T_0 = 0$ and counter $i=0$. 
 \item Compare input $I_t$ to $T_i$. If $I_t \geq
 T_i$, set $\widehat{S}_i(t) = +1$. If $I_t <
 T_i$, set $\widehat{S}_i(t) = -1$. 
 \item Increment counter $i$ and update threshold $T_i = \gamma \lambda^t
 \sum_{k=0}^{i-1} (2+\epsilon_1)^{-k} \widehat{S}_k$
 \item Goto step 2 as long as $i \leq \lfloor Rt \rfloor$
 \end{enumerate}

Since the gaps given by (\ref{eqn:gapbound}) are always positive, the
procedure works perfectly if applied to input $I_t = \check{X}_t$. At 
the decoder, apply the procedure to $I_t = -\widetilde{X}_t$ instead.

With this, (\ref{eqn:erroreventbound}) is easy to verify by looking at
the complementary event $\{\omega | |X_t| < \frac{
\lambda^{t-\frac{j}{R}} \gamma \epsilon_1}{1+\epsilon_1} \}$. The
bound (\ref{eqn:gapbound}) thus implies that we are less than halfway
across the minimum gap for bit $j$ at time $t$. Consequently, there is
no error in the step 2 comparison of the procedure at
iterations $i \leq j$. \hfill $\Box$ \vspace{0.15in}

\subsubsection{Probability of error for bounded moment and other
  senses of stability}

{\em Proof of Theorem \ref{thm:scalar_control_necessity}:} Using
Markov's inequality: 
\begin{eqnarray*}
{\cal P}(|X_t|> m) & = &
{\cal P}(|X_t|^\eta > m^\eta) \\
& \leq & E[|X_t|^\eta] m^{-\eta} \\
& < & K m^{-\eta}
\end{eqnarray*}
Combining with Lemma~\ref{lem:erroreventbound}, gives:
\begin{eqnarray*}
{\cal P}(\widehat{S}_1^i(t) \neq S_1^i(t)) 
& \leq & {\cal P}(|X_t|
   \geq \lambda^{t-\frac{i}{R}} \left(\frac{\gamma \epsilon_1}{1+\epsilon_1}\right)) \\
& < & K (\frac{1}{\gamma} + \frac{1}{\gamma \epsilon_1})^{\eta} 
      \lambda^{-\eta(t-\frac{i}{R})} \\
& = & (K (\frac{1}{\gamma} + \frac{1}{\gamma \epsilon_1})^{\eta})
      2^{-(\eta \log_2 \lambda)(t-\frac{i}{R})}
\end{eqnarray*}
Since $t-\frac{i}{R}$ represents the delay between the time that bit
$i$ was ready to be sent and the decoding time, the theorem is
proved. \hfill $\Box$ \vspace{0.15in}

All that was needed from the bounded moment sense of stability was
some bound on the probability that $X_t$ took on large values. Thus,
the proof above immediately generalizes to other senses of stochastic
stability if we suitably generalize the sense of anytime capacity to
allow for other bounds on the probability of error with delay.

\begin{definition} \label{def:ganytime}
A rate $R$ communication system achieves {\em $g-$anytime reliability}
given by a function $g(d)$ if
$${\cal P}(\widehat{M}_1^{t-d}(t) \neq  M_1^{t-d}(t)) < g(d)$$

$g(d)$ is assumed to be $1$ for all negative values of $d$. 

The {\em $g-$anytime capacity} $C_{\mbox{g-any}}(g)$ of a noisy
channel is the least upper bound of the rates $R$ at which the channel
can be used to construct a sequential communication system that
achieves $g-$anytime reliability given by the function $g(d)$.
\end{definition}
\vspace{0.1in}

Notice that for $\alpha$-anytime capacity, $g(d) = K 2^{-\alpha d}$
for some $K$. 

\begin{theorem} \label{thm:general_scalar_control_necessity}
For a given noisy channel and decreasing function $f(m)$, if there
exists an observer ${\cal O}$ and controller ${\cal C}$ for the
unstable scalar system that achieves ${\cal P}(|X_t| > m) < f(m)$ for
all sequences of bounded driving noise $|W_t| \leq \frac{\Omega}{2}$,
then $C_{\mbox{g-any}}(g) \geq \log_2 \lambda$ for the noisy channel
considered with the encoder having access to noiseless feedback and
$g(d)$ having the form $g(d) = f(K \lambda^d)$ for some constant $K$.
\end{theorem}
{\em Proof:} For any rate $R < \log_2 \lambda$, 
\begin{eqnarray*}
{\cal P}(\widehat{S}_1^i(t) \neq S_1^i(t)) 
& \leq & {\cal P}(|X_t|
   \geq \frac{ \lambda^{t-\frac{i}{R}} \gamma \epsilon_1}{1+\epsilon_1}) \\
& = & f(\frac{\gamma \epsilon_1}{1+\epsilon_1} \lambda^{t-\frac{i}{R}}) 
\end{eqnarray*}
Since the delay $d = t - \frac{i}{R}$, the theorem is proved. \hfill $\Box$
\vspace{0.15in}

\subsection{Implications} \label{sec:scalar_implications}

At this point, it is interesting to consider a few implications of
Theorem~\ref{thm:general_scalar_control_necessity}. 

\subsubsection{Weaker senses of stability than $\eta$-moment}

There are senses of stability weaker than specifying a specific
$\eta$-th moment or a specific tail decay target $f(m)$. An example is
given by the requirement $\lim_{m \rightarrow \infty} {\cal P}(|X_t| >
m) = 0$ uniformly for all $t$. This can be explored by taking the
limit of $C_{\mbox{any}}(\alpha)$ as $\alpha \downarrow 0$. We have
shown elsewhere\cite{SahaiThesis, OurSourceCodingPaper} that:
$$\lim_{\alpha \downarrow 0} C_{\mbox{any}}(\alpha) = C$$ where $C$ is
the Shannon classical capacity. This holds for all discrete memoryless
channels since the $\alpha$-anytime reliability goes to zero at
Shannon capacity but is $>0$ for all lower rates even without feedback
being available at the encoder. Thus, classical Shannon capacity is
the natural candidate for the relevant figure of merit.

To see why Shannon capacity can not be beaten, it is useful to
consider an even more lax sense of stability. Suppose the requirement
were only that $\lim_{m \rightarrow \infty} {\cal P}(|X_t| > m) =
10^{-5} > 0$ uniformly for all $t$. This imposes the constraint that
the probability of a large state stays below $10^{-5}$ for all
time. Theorem~\ref{thm:general_scalar_control_necessity} would thus
only requires the probability of decoding error to be less than
$10^{-5}$. However, Wolfowitz' strong converse to the coding
theorem\cite{Gallager} implies that since the block-length in this
case is effectively going to infinity, the Shannon capacity of the
noisy channel still must satisfy $C \geq \log_2 \lambda$. Adding a
finite tolerance for unboundedly large states does not get around the
need to be able to communicate $\log_2 \lambda$ bits reliably.

\subsubsection{Stronger senses of stability than $\eta$-moment}

Having $f$ decrease only as a power law might not be suitable for
certain applications. Unfortunately, this is all that can be hoped for
in generic situations. Consider a DMC with no zero entries in its
transition matrix. Define $\rho = \min_{i,j} p(i,j)$. For such a
channel, with or without feedback, the probability of error after $d$
time steps is lower bounded by $\rho^d$ since that lower bounds the
probability of all channel output sequences of length $d$. This
implies that the probability of error can drop no more than
exponentially in $d$ for such DMCs. Tighter upper-bounds on anytime
reliability with feedback are available in \cite{TuncThesis} and
\cite{OurUpperBoundPaper}.

Theorem~\ref{thm:general_scalar_control_necessity} therefore implies
that the only $f$-senses of stability which are possible over such
channels are those for which: 
\begin{eqnarray*}
f(K \lambda^d) & \geq & \rho^d \\
f(m) & \geq & \rho^{\frac{\log_2(\frac{m}{K})}{\log_2 \lambda}} \\
f(m) & \geq & K' m^{-\frac{\log_2 \frac{1}{\rho}}{\log_2 \lambda}}
\end{eqnarray*}
which is a power law. This rules out the ``risk sensitive'' sense of
stability in which $f$ is required to decrease exponentially. In the
context of Theorem~\ref{thm:scalar_control_necessity}, this also
implies that there is an $\eta$ beyond which all moments must be
infinite!  

\begin{corollary} \label{cor:powerlawrequired}
If any unstable process is controlled over a discrete memoryless
channel with no feedback zero-error capacity, then the resulting state
can have at best a power-law bound (Pareto distribution) on its tail.
\end{corollary}
\vspace{0.1in}

This is very much related to how sequential decoding must have
computational effort distributions with at best a Pareto
distribution\cite{JacobsBerlekamp}. In both cases, the result follows
from the interaction of two exponentials. The difference is that the
computational search effort distributions assumed a particular
structure on the decoding algorithm while the bound here is
fundamental to the stabilization problem regardless of the observers
or controllers.

Thus {\em for DMCs and a given $\lambda$, we are either limited to a
power-law tail for the controlled state because of an anytime
reliability that is at most singly exponential in delay or it is
possible to hold the state inside a finite box since there is adequate
feedback zero-error capacity.} Nothing in between can happen with a
DMC.

\subsubsection{Limiting the controller effort or memory} \label{sec:controllerlimits}
If there was a hard limit on actuator effort ($|U| \leq {\cal U}$ for
some ${\cal U} > 0$), then the only way to maintain stability is to
also have a hard limit on how big the state $X$ can get.
Theorem~\ref{thm:general_scalar_control_necessity} immediately
gives a fundamental requirement for feedback zero-error capacity $\geq
\log_2 \lambda$ since $g(d) = 0$ for sufficiently large $d$.

Similarly, consider limited-memory time-invariant controllers which
only have access to the past $k$ channel outputs. If the channel has a
finite output alphabet and no randomization is permitted at the
controller, limited memory immediately translates into only a finite
number of possible control inputs. Since there must be a largest one,
it reduces to the case of having a hard limit on actuator effort. 

We conjecture that even with randomization and time-variation, finite
memory at the controller implies that the channel must have feedback
zero-error capacity $\geq \log_2 \lambda$. Intuitively, if the channel
has zero-error capacity $< \log_2 \lambda$, it can misbehave for
arbitrarily long times and build up a huge ``backlog'' of uncertainty
that can not be resolved at the controller. With finite memory, the
controller has no way of knowing what uncertainty it is actually
facing and so is unable to properly interpret the channel outputs to
devise the proper control signals.


\subsubsection{The AWGN case with an average input power constraint}
\label{sec:awgncasewithfeedback}

The tight relationship between control and communication established
in Theorem~\ref{thm:general_scalar_control_necessity} allows the
construction of sequential codes for noisy channels with noiseless
feedback if we know how to stabilize linear plants over such
channels. Consider the problem of stabilizing an unstable plant driven
by finite variance driving noise over an AWGN channel. A linear
observer and controller strategy achieve mean-square stability for
such systems since the problem fits into the standard LQG
framework \cite{OurMainLQGPaper}. 

By looking more closely at the actual tail probabilities achieved by
the linear observer/controller strategy, we obtain a natural anytime
generalization of Schalkwijk and Kailath's
scheme\cite{Kailath66,Schalkwijk66} for communicating over the power
constrained additive white Gaussian noise channel with noiseless
feedback. Its properties are summarized in Figure~\ref{fig:skcompare},
but the highlight is that it achieves doubly exponential reliability
with delay, universally over all sufficiently long delays.

\begin{theorem} \label{thm:sequential_code_awgn_feedback}
It is possible to communicate bits reliably across a discrete-time
average-power constrained AWGN channel with noiseless feedback at any
rate $R < \frac{1}{2} \log_2(1 + \frac{P}{\sigma^2})$ while achieving
a $g-$anytime reliability of at least
\begin{equation} \label{eqn:double_exponential}
g(d) = 2 e^{- K (4^{R d} - O(2^{Rd}))}
\end{equation}
for some constant $K$ that depends only on the rate $R$, power
constraint $P$, and channel noise power $\sigma^2$.
\end{theorem}
{\em Proof:} To avoid having to drag $\sigma^2$ around, just normalize
units so as to consider power constraint $P' = \frac{P}{\sigma^2}$ and
a channel with iid unit variance noise $N_t$. Choose the $\lambda$ for
the simulated (\ref{eqn:discretesystem}) so that $R < \log_2 \lambda <
\frac{1}{2}\log_2(1+P')$.

The observer/encoder used is a linear map:
\begin{equation} \label{eqn:awgn_observer}
a_t = \beta X_t
\end{equation}
so the channel output $B_t = \beta X_t + N_t$. Use a linear controller:
\begin{equation} \label{eqn:awgn_controller}
U_t = -\lambda \phi B_t
\end{equation}
giving the closed-loop system:
\begin{equation} \label{eqn:discretesystemawgn} 
X_{t+1} = \lambda (1 - \beta \phi) X_{t} + W_{t} - \lambda\phi N_t
\end{equation} 
where the $\beta,\phi$ are constants to be chosen. For the closed-loop
system to be stable:
\begin{equation} \label{eqn:stabilityconditionawgn}
0 < \lambda(1-\beta \phi) < 1
\end{equation}
Thus $\beta \phi \in (1 - \frac{1}{\lambda}, 1)$. Assuming
(\ref{eqn:stabilityconditionawgn}) holds and temporarily setting the
$W_t = 0$ for analysis, it is clear that the closed-loop $X_t$ is
Gaussian with a growing variance asymptotically tending to
\begin{equation} \label{eqn:sigmaxawgn}
\sigma_x^2 = \frac{\lambda^2 \phi^2}{1 - \lambda^2(1-\beta \phi)^2}
\end{equation}
The channel input power satisfies:
$$E[a_t^2] \leq \frac{\lambda^2 (\beta\phi)^2}{1 - \lambda^2(1-\beta
\phi)^2}$$ 
Since $\lambda^2 < 1 + P'$, define $P'' = \lambda^2 - 1 < P'$ and substitute to get:
\begin{equation} \label{eqn:inputpower}
E[a_t^2] \leq \frac{(P'' + 1)(\beta\phi)^2}{1 - (P'' +
  1)(1-\beta\phi)^2}
\end{equation}
By setting $\beta\phi = \frac{P''}{P'' + 1}$, the left hand side of
(\ref{eqn:inputpower}) is identically $P''$ as desired. All that
remains is to verify the stability condition
(\ref{eqn:stabilityconditionawgn}):
\begin{eqnarray*}
\lambda(1-\beta \phi) & = & 
\frac{\lambda}{P'' + 1} \\
& = & 
\frac{\sqrt{P'' + 1}}{P'' + 1} \\
& = & 
\frac{1}{\sqrt{P'' + 1}} \\
& < & 1
\end{eqnarray*}
So the closed loop system is stable and the channel noise alone
results in an average input power of at most $P'' < P'$. 

Rather than optimizing the choice of $\beta$ and $\phi$ to get the
best tradeoff point, just set $\beta = 1$ and $\phi = \frac{P''}{P'' +
1}$ for simplicity. In that case, $\sigma_x^2 = P''$.

Now consider the impact of the $W_t$ alone on the closed-loop control
system. These are going through a stable system and so by expanding
the recursion (\ref{eqn:discretesystemawgn}) and setting $N_t = 0$,
\begin{eqnarray*}
|X^{w}_t| & \leq & 
\sum_{i=0}^\infty \left(\lambda(1-\beta\phi)\right)^i \frac{\Omega}{2} \\
& = & 
\sum_{i=0}^\infty \left(\frac{1}{\sqrt{P'' + 1}}\right)^i \frac{\Omega}{2} \\
& = & 
\frac{\Omega}{2 (1 - \frac{1}{\sqrt{P'' + 1}})} \\
& = & 
\frac{\Omega \sqrt{P'' + 1}}{2(\sqrt{P'' + 1} - 1)}
\end{eqnarray*}
which is a constant that can be made as small as desired by choice of
$\Omega$. Assume that the data stream $S$ to be transmitted is
independent of the channel noise $N$. Then, the total average input
power is bounded by:
\begin{eqnarray*} 
\sigma_x^2 + \beta^2 |X^{w}_t|^2 
& \leq & 
P'' + (\frac{\Omega \sqrt{P'' + 1}}{2 (\sqrt{P'' + 1} - 1)})^2 \\
& \leq & 
P'' + \Omega^2 \frac{P'' + 1}{4(P'' + 2(1 - \sqrt{P'' + 1}))}
\end{eqnarray*}
Since $P'' < P'$, we can choose an $\Omega$ small enough so that the
channel input satisfies the average power constraint regardless of the
message bits to be sent.

All that remains is to see what $f(m)$ this control system meets for
such arbitrary, but bounded, disturbances. $X_t$ is asymptotically the
sum of a Gaussian with zero mean and variance $P''$ together with the
closed-loop impact of the disturbance $X^w(t)$. Since the total impact
of the disturbance part is bounded:
\begin{eqnarray*}
{\cal P}(|X_t| > m) & \leq & 
{\cal P}(|N_{\sigma_x^2}| 
  > m - \frac{\Omega \sqrt{P'' + 1}}{2(\sqrt{P'' + 1}  - 1)}) \\
& = & 
{\cal P}(|N| > \frac{1}{\sqrt{P''}}
(m - \frac{\Omega \sqrt{P'' + 1}}{2(\sqrt{P'' + 1}  - 1)})) \\
& \leq & 2 e^{-\frac{1}{2P''} 
(m - \frac{\Omega \sqrt{P'' + 1}}{2(\sqrt{P'' + 1}  - 1)})^2}
\end{eqnarray*}
Ignoring the details of the constants, this gives an $f(m) = 2
e^{-K_1(m-K_2)^2} = 2 e^{-K_1(m^2 -2K_2m -K_3)}$. Applying
Theorem~\ref{thm:general_scalar_control_necessity} immediately gives
(\ref{eqn:double_exponential}) since $\lambda^d > 2^{R d}$. \hfill
$\Box$
\vspace{0.15in}

\begin{figure}
\begin{center}
\begin{tabular}{|l | c | c |}
\hline
Scheme: & Schalkwijk \cite{Schalkwijk66} & Theorem~\ref{thm:sequential_code_awgn_feedback} \\
\hline
{\em Message}: & known in advance & streams in \\ 
\hline
{\em Delay}: & prespecified & {\bf universal}  \\
\hline
Error exponent: & double-exponential & double-exponential \\
\hline
Channel: & known AWGN & known AWGN \\
\hline
Constraint: & average power & average power \\
\hline 
Noiseless & required & required \\
Feedback: &          &          \\
\hline
{\em Initialization}: & $2^{nR}$-PAM & none \\
                      & + ML feedback &   \\
\hline
{\em Ongoing}: & MMSE feedback & MMSE feedback  \\
               &               & + small $R$-PAM \\ 
               &               & perturbations \\
\hline
{\em Channel input:} & Gaussian & Perturbed Gaussian \\
\hline 
Decoding: & Minimum distance & Minimum distance \\
\hline
Equivalent  & unstable & unstable \\
Plant:      & $R < \log_2 \lambda < C$ & $R < \log_2 \lambda < C$ \\
\hline 
{\em Initial condition}: & bounded & zero \\
\hline 
{\em Disturbance}: & zero & bounded \\
\hline
{\em Stability sense}: & almost-sure \cite{EliaPaper1} & exponential tail \\
\hline
\end{tabular}
\end{center}
\caption{Quick comparison of the Schalkwijk/Kailath scheme to the
  anytime generalization in this paper.}
\label{fig:skcompare}
\end{figure}

Since the convergence is double exponential, it is faster than any
exponential and hence
$$C_{\mbox{any}}(\alpha) = \frac{1}{2} \log_2 (1 +
\frac{P}{\sigma^2})$$ for all $\alpha > 0$ on the AWGN channel. If the
additive channel noise were not Gaussian, but had bounded support with
the same variance, then this proof immediately reveals that the
zero-error capacity of such a bounded noise channel with feedback
satisfies: $C_0 \geq \frac{1}{2} \log_2 (1 + \frac{P}{\sigma^2})$.

In the Gaussian case, it is not immediately clear whether there are
ideas analogous to those in \cite{IntermittentFeedback} that can be
used to further boost the $g$-anytime reliability beyond double
exponential. It is clear that if it were possible, it would require
nonlinear control strategies.

The AWGN case is merely one
example. Theorem~\ref{thm:general_scalar_control_necessity} gives a
way to lower-bound the anytime capacity for channels with feedback in
cases where the optimal control behavior is easy to see. The finite
moments of the closed-loop state reveal what anytime reliability is
being achieved. Often, there is a simple upper-bound that matches up
with the lower-bound thereby giving the anytime capacity itself.  The
BEC case discussed in \cite{ACC00Paper,SahaiThesis,OurUpperBoundPaper}
is such an example. In addition,
Theorem~\ref{thm:general_scalar_control_necessity} gives us the
ability to mix and match communication and control tools to study a
problem. This is exploited in \cite{AllertonPacket2004,
AllertonAWGN2004} to understand the feedback anytime capacity of
constrained packet erasure channels and the power constrained
AWGN+erasure channel. In \cite{GEISIT05}, these results are extended
to the Gilbert-Eliot channel with feedback. It is also exploited in
\cite{TuncThesis} to lower bound the anytime reliability achieved by a
particular code for the BSC with feedback.

\section{The sufficiency of anytime capacity} \label{sec:sufficiency}

\subsection{Overview}

When characterizing a noisy channel for control, the choice of
information pattern\cite{WitsenhausenPatterns} can be critical
\cite{OurMainLQGPaper}. The sufficiency result is first established
for cases with an explicit noiseless feedback path from the channel
outputs back to the observer. Section~\ref{sec:almostsure} takes a
quick look at the simpler problem of almost-sure stabilization when
the system is undisturbed and all the uncertainty comes from either
the channel or the initial condition. Then, in
Section~\ref{sec:blockedtime}, the impact of viewing time in blocks of
size $n$ and only acting on the slower time-scale is
examined. Finally, Sections \ref{sec:quantizedcontrols} and
\ref{sec:noisyobservations} give models for boundedly noisy or
quantized controls and/or observations and show that such bounded
noise can be tolerated.

To prove the sufficiency theorem addressing the situation illustrated
in figure \ref{fig:problem}, we need to design an observer/controller
pair that deals with the analog plant and communicates across the
channel by using an anytime communication system. The anytime
communication system works with noiseless feedback from the channel
output available at the bit encoder and is considered a ``black box.''

\begin{theorem} \label{thm:general_scalar_control_sufficiency}
For a given noisy channel, if there exists an anytime encoder/decoder
pair with access to noiseless feedback that achieves
$C_{\mbox{g-any}}(g) \geq \log_2 \lambda$, then it is possible to
stabilize an unstable scalar plant with parameter $\lambda$ that is
driven by bounded driving noise through the noisy channel by using an
observer that has noiseless access to the noisy channel
outputs. Furthermore, there exists a constant $K$ so that ${\cal
P}(|X_t| > m) \leq g(K + \log_\lambda m)$.
\end{theorem}
\vspace{0.1in}

To prove this theorem, explicit constructions are given for the
observer and controller in the next sections.

\subsection{Observer} \label{sec:observerwithmemory}

Since the observer has access to the channel outputs, it can run a
copy of the controller and hence has access to the control signals
$U_t$. Since $W_t = X_{t+1}- \lambda X_t - U_t$, and the observer
receives $X_t$ from the plant, the observer also effectively has
access to the $W_t$. However, it is not sufficient to merely encode
the $W_t$ independently to some precision.\footnote{This is because
the unstable plant will eventually blow up even tiny uncorrected
discrepancies between the encoded and actual $W_t$.} Instead, the
observer will act as though it is working with a virtual controller
through a noiseless channel of finite rate $R$ in the manner of
example \ref{example:noiseless}. The resulting bits will be sent
through the anytime code.

The observer is constructed to keep the state uncertainty at the
virtual controller inside a box of size $\Delta$ by using bits at the
rate $R$. It does this by simulating a virtual process $\bar{X}_t$
governed by:

\begin{equation}\label{eqn:barupdate}
\bar{X}_{t+1} = \lambda \bar{X}_{t} + W_{t} + \bar{U}_t
\end{equation}
where the $\bar{U}_t$ represent the computed actions of the virtual
controller. This gives rise to a virtual counterpart of $\widetilde{X}_t$ 

\begin{equation} \label{eqn:XbarU}
X^{\bar{U}}_{t+1} = \lambda X^{\bar{U}}_{t} + \bar{U}_t
\end{equation}

which satisfies the relationship $\bar{X}_t = \check{X}_t +
X^{\bar{U}}_{t}$. Because $\bar{X}_t$ will be kept within a box, it is
known that $-X^{\bar{U}}_{t}$ is close to $\check{X}_t$. The actual
controller will pick controls designed to keep $\widetilde{X}_t$ close
to $X^{\bar{U}}_{t}$.

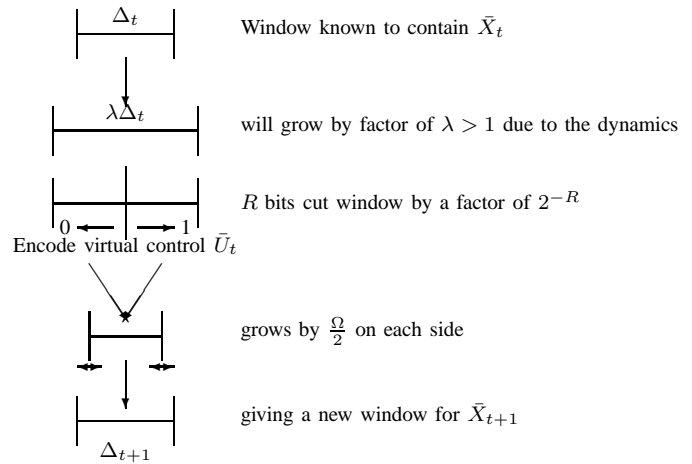
\begin{figure}
\begin{center}
\setlength{\unitlength}{2000sp}%
\begingroup\makeatletter\ifx\SetFigFont\undefined%
\gdef\SetFigFont#1#2#3#4#5{%
  \reset@font\fontsize{#1}{#2pt}%
  \fontfamily{#3}\fontseries{#4}\fontshape{#5}%
  \selectfont}%
\fi\endgroup%
\begin{picture}(6858,5620)(868,-5069)
\put(3226,239){\makebox(0,0)[lb]{\smash{\SetFigFont{8}{14.4}{\rmdefault}{\mddefault}{\updefault}{\color[rgb]{0,0,0}Window known to contain $\bar{X}_t$}%
}}}
\put(3226,-1936){\makebox(0,0)[lb]{\smash{\SetFigFont{8}{14.4}{\rmdefault}{\mddefault}{\updefault}{\color[rgb]{0,0,0}$R$ bits cut window by a factor of $2^{-R}$}%
}}}
\put(1051,-2236){\makebox(0,0)[b]{\smash{\SetFigFont{8}{14.4}{\rmdefault}{\mddefault}{\updefault}{\color[rgb]{0,0,0}0}%
}}}
\put(2551,-2236){\makebox(0,0)[b]{\smash{\SetFigFont{8}{14.4}{\rmdefault}{\mddefault}{\updefault}{\color[rgb]{0,0,0}1}%
}}}
\thinlines
{\color[rgb]{0,0,0}\put(1201,239){\line( 1, 0){1200}}
}%
{\color[rgb]{0,0,0}\put(1201,539){\line( 0,-1){600}}
}%
{\color[rgb]{0,0,0}\put(2401,539){\line( 0,-1){600}}
}%
{\color[rgb]{0,0,0}\put(1801,-61){\vector( 0,-1){600}}
}%
{\color[rgb]{0,0,0}\put(2701,-661){\line( 0,-1){600}}
}%
{\color[rgb]{0,0,0}\put(901,-661){\line( 0,-1){600}}
}%
\thicklines
{\color[rgb]{0,0,0}\put(901,-961){\line( 1, 0){1800}}
}%
\thinlines
{\color[rgb]{0,0,0}\put(1801,-1411){\line( 0,-1){900}}
}%
\thicklines
{\color[rgb]{0,0,0}\put(901,-1861){\line( 1, 0){1800}}
}%
\thinlines
{\color[rgb]{0,0,0}\put(1951,-2161){\vector( 1, 0){450}}
}%
{\color[rgb]{0,0,0}\put(1651,-2161){\vector(-1, 0){450}}
}%
{\color[rgb]{0,0,0}\put(1351,-2611){\vector( 2,-3){484.615}}
}%
{\color[rgb]{0,0,0}\put(2251,-2611){\vector(-2,-3){484.615}}
}%
{\color[rgb]{0,0,0}\put(901,-1561){\line( 0,-1){600}}
}%
{\color[rgb]{0,0,0}\put(2701,-1561){\line( 0,-1){600}}
}%
{\color[rgb]{0,0,0}\put(1351,-3511){\line( 1, 0){900}}
}%
{\color[rgb]{0,0,0}\put(1351,-3211){\line( 0,-1){600}}
}%
{\color[rgb]{0,0,0}\put(2251,-3211){\line( 0,-1){600}}
}%
{\color[rgb]{0,0,0}\put(1801,-3811){\vector( 0,-1){600}}
}%
{\color[rgb]{0,0,0}\put(1201,-3886){\vector(-1, 0){  0}}
\put(1201,-3886){\vector( 1, 0){300}}
}%
{\color[rgb]{0,0,0}\put(2101,-3886){\vector(-1, 0){  0}}
\put(2101,-3886){\vector( 1, 0){300}}
}%
{\color[rgb]{0,0,0}\put(1201,-4561){\line( 1, 0){1200}}
}%
{\color[rgb]{0,0,0}\put(2401,-4261){\line( 0,-1){600}}
}%
{\color[rgb]{0,0,0}\put(1201,-4261){\line( 0,-1){600}}
}%
\put(3226,-3511){\makebox(0,0)[lb]{\smash{\SetFigFont{8}{14.4}{\rmdefault}{\mddefault}{\updefault}{\color[rgb]{0,0,0}grows by $\frac{\Omega}{2}$ on each side}%
}}}
\put(3226,-4561){\makebox(0,0)[lb]{\smash{\SetFigFont{8}{14.4}{\rmdefault}{\mddefault}{\updefault}{\color[rgb]{0,0,0}giving a new window for $\bar{X}_{t+1}$}%
}}}
\put(3226,-961){\makebox(0,0)[lb]{\smash{\SetFigFont{8}{14.4}{\rmdefault}{\mddefault}{\updefault}{\color[rgb]{0,0,0}will grow by factor of $\lambda>1$ due to the dynamics }%
}}}
\put(1801,-5011){\makebox(0,0)[b]{\smash{\SetFigFont{8}{14.4}{\rmdefault}{\mddefault}{\updefault}{\color[rgb]{0,0,0}$\Delta_{t+1}$}%
}}}
\put(1801,-2461){\makebox(0,0)[b]{\smash{\SetFigFont{8}{14.4}{\rmdefault}{\mddefault}{\updefault}{\color[rgb]{0,0,0}Encode virtual control $\bar{U}_t$}%
}}}
\put(1801,-811){\makebox(0,0)[b]{\smash{\SetFigFont{8}{14.4}{\rmdefault}{\mddefault}{\updefault}{\color[rgb]{0,0,0}$\lambda\Delta_t$}%
}}}
\put(1801,389){\makebox(0,0)[b]{\smash{\SetFigFont{8}{14.4}{\rmdefault}{\mddefault}{\updefault}{\color[rgb]{0,0,0}$\Delta_t$}%
}}}
\end{picture}
\end{center}
\caption{Virtual controller for R=1. How the virtual state
$\bar{X}$ evolves.}
\label{fig:causalmarkovcode}
\end{figure}
 
Because of the rate constraint, the virtual control $\bar{U}_t$ takes
on one of $2^{\lfloor R(t+1) \rfloor - \lfloor Rt \rfloor}$
values. For simplicity of exposition, we ignore the integer effects
and consider it to be one of $2^R$ values\footnote{For the details of
how to deal with fractional $R$, please see the causal source code
discussion in \cite{SahaiThesis}.} and proceed by induction. Assume
that $\bar{X}_t$ is known to lie within
$[-\frac{\Delta}{2},\frac{\Delta}{2}]$. Then $\lambda\bar{X}_t$ will
lie within $[-\frac{\lambda \Delta}{2},\frac{\lambda \Delta}{2}]$. By
choosing $2^R$ control values uniformly spaced within that interval,
it is guaranteed that $\lambda \bar{X}_t + \bar{U}_t$ will lie within
$[-\frac{\lambda \Delta}{2^{R+1}},\frac{\lambda
\Delta}{2^{R+1}}]$. Finally, the state will be disturbed by $W_t$ and
so $\bar{X}_{t+1}$ will be known to lie within $[-\frac{\lambda
\Delta}{2^{R+1}} - \frac{\Omega}{2}, \frac{\lambda \Delta}{2^{R+1}} +
\frac{\Omega}{2}]$.

Since the initial condition has no uncertainty, induction will be
complete if
\begin{equation} \label{eqn:inductioncondition}
\frac{\lambda}{2^{R}} \Delta + \Omega \leq \Delta
\end{equation}
To get the minimum $\Delta$ required as a function of $R$, we can
solve for (\ref{eqn:inductioncondition}) being an equality. This
occurs\footnote{In reality, the uncertainty approaches this from below
since the system starts at the known initial condition $0$.} when
$\Delta = \frac{\Omega}{1 - \lambda 2^{-R}}$ for every case where $R >
\log_2 \lambda$. Since the slope $\frac{\lambda}{2^R}$ on the left
hand side of (\ref{eqn:inductioncondition}) is less than $1$, any
larger $\Delta$ also works. 

Since they arose from dividing the uncertainty window to $2^R$
disjoint segments, it is clear that the virtual controls $\bar{U}_t$
can be encoded causally using $R$ bits per unit time. These bits are
sent to the anytime encoder for transport over the noisy
channel.

\subsection{Controller} \label{sec:sufficiency_controller_basic}

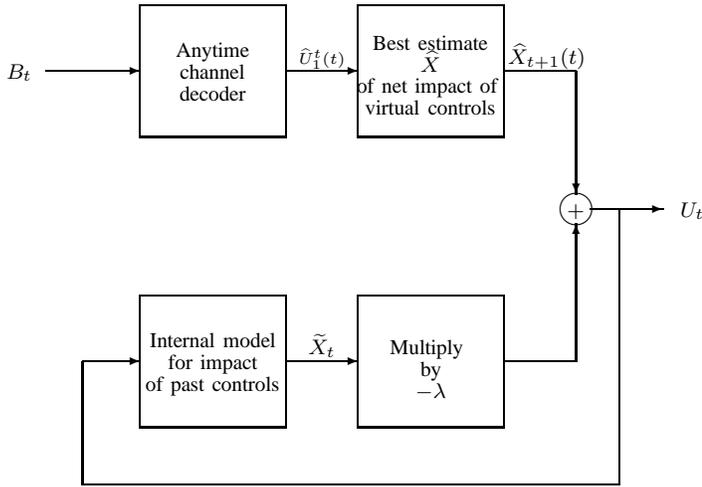
\begin{figure}
\begin{center}
\setlength{\unitlength}{2400sp}%
\begingroup\makeatletter\ifx\SetFigFont\undefined%
\gdef\SetFigFont#1#2#3#4#5{%
  \reset@font\fontsize{#1}{#2pt}%
  \fontfamily{#3}\fontseries{#4}\fontshape{#5}%
  \selectfont}%
\fi\endgroup%
\begin{picture}(6975,4974)(376,-4873)
{\color[rgb]{0,0,0}\thinlines
\put(6151,-2011){\circle{336}}
}%
{\color[rgb]{0,0,0}\put(676,-586){\vector( 1, 0){975}}
}%
{\color[rgb]{0,0,0}\put(3151,-586){\vector( 1, 0){750}}
}%
{\color[rgb]{0,0,0}\put(3901,-1261){\framebox(1500,1350){}}
}%
{\color[rgb]{0,0,0}\put(1651,-1261){\framebox(1500,1350){}}
}%
{\color[rgb]{0,0,0}\put(5401,-586){\line( 1, 0){750}}
\put(6151,-586){\vector( 0,-1){1275}}
}%
{\color[rgb]{0,0,0}\put(1651,-4261){\framebox(1500,1350){}}
}%
{\color[rgb]{0,0,0}\put(3151,-3586){\vector( 1, 0){750}}
}%
{\color[rgb]{0,0,0}\put(3901,-4261){\framebox(1500,1350){}}
}%
{\color[rgb]{0,0,0}\put(5401,-3586){\line( 1, 0){750}}
\put(6151,-3586){\vector( 0, 1){1425}}
}%
{\color[rgb]{0,0,0}\put(6301,-2011){\vector( 1, 0){750}}
}%
{\color[rgb]{0,0,0}\put(6601,-2011){\line( 0,-1){2850}}
\put(6601,-4861){\line(-1, 0){5550}}
\put(1051,-4861){\line( 0, 1){1275}}
\put(1051,-3586){\vector( 1, 0){600}}
}%
\put(2401,-436){\makebox(0,0)[b]{\smash{\SetFigFont{8}{7.2}{\rmdefault}{\mddefault}{\updefault}{\color[rgb]{0,0,0}Anytime}%
}}}
\put(2401,-661){\makebox(0,0)[b]{\smash{\SetFigFont{8}{7.2}{\rmdefault}{\mddefault}{\updefault}{\color[rgb]{0,0,0}channel}%
}}}
\put(2401,-886){\makebox(0,0)[b]{\smash{\SetFigFont{8}{7.2}{\rmdefault}{\mddefault}{\updefault}{\color[rgb]{0,0,0}decoder}%
}}}
\put(2401,-3436){\makebox(0,0)[b]{\smash{\SetFigFont{8}{7.2}{\rmdefault}{\mddefault}{\updefault}{\color[rgb]{0,0,0}Internal model}%
}}}
\put(2401,-3661){\makebox(0,0)[b]{\smash{\SetFigFont{8}{7.2}{\rmdefault}{\mddefault}{\updefault}{\color[rgb]{0,0,0}for impact}%
}}}
\put(2401,-3886){\makebox(0,0)[b]{\smash{\SetFigFont{8}{7.2}{\rmdefault}{\mddefault}{\updefault}{\color[rgb]{0,0,0}of past controls}%
}}}
\put(4651,-361){\makebox(0,0)[b]{\smash{\SetFigFont{8}{7.2}{\rmdefault}{\mddefault}{\updefault}{\color[rgb]{0,0,0}Best estimate}%
}}}
\put(4651,-586){\makebox(0,0)[b]{\smash{\SetFigFont{8}{7.2}{\rmdefault}{\mddefault}{\updefault}{\color[rgb]{0,0,0}$\widehat{X}$}%
}}}
\put(4651,-811){\makebox(0,0)[b]{\smash{\SetFigFont{8}{7.2}{\rmdefault}{\mddefault}{\updefault}{\color[rgb]{0,0,0}of net impact of }%
}}}
\put(4651,-1036){\makebox(0,0)[b]{\smash{\SetFigFont{8}{7.2}{\rmdefault}{\mddefault}{\updefault}{\color[rgb]{0,0,0}virtual controls}%
}}}
\put(4651,-3511){\makebox(0,0)[b]{\smash{\SetFigFont{8}{7.2}{\rmdefault}{\mddefault}{\updefault}{\color[rgb]{0,0,0}Multiply }%
}}}
\put(4651,-3736){\makebox(0,0)[b]{\smash{\SetFigFont{8}{7.2}{\rmdefault}{\mddefault}{\updefault}{\color[rgb]{0,0,0}by}%
}}}
\put(4651,-3961){\makebox(0,0)[b]{\smash{\SetFigFont{8}{7.2}{\rmdefault}{\mddefault}{\updefault}{\color[rgb]{0,0,0}$-\lambda$}%
}}}
\put(6151,-2086){\makebox(0,0)[b]{\smash{\SetFigFont{8}{7.2}{\rmdefault}{\mddefault}{\updefault}{\color[rgb]{0,0,0}$+$}%
}}}
\put(3526,-3511){\makebox(0,0)[b]{\smash{\SetFigFont{8}{7.2}{\rmdefault}{\mddefault}{\updefault}{\color[rgb]{0,0,0}$\widetilde{X}_t$}%
}}}
\put(5851,-511){\makebox(0,0)[b]{\smash{\SetFigFont{8}{7.2}{\rmdefault}{\mddefault}{\updefault}{\color[rgb]{0,0,0}$\widehat{X}_{t+1}(t)$}%
}}}
\put(3526,-511){\makebox(0,0)[b]{\smash{\SetFigFont{6}{7.2}{\rmdefault}{\mddefault}{\updefault}{\color[rgb]{0,0,0}$\widehat{U}_1^{t}(t)$}%
}}}
\put(400,-661){\makebox(0,0)[b]{\smash{\SetFigFont{8}{7.2}{\rmdefault}{\mddefault}{\updefault}{\color[rgb]{0,0,0}$B_t$}%
}}}
\put(7351,-2086){\makebox(0,0)[b]{\smash{\SetFigFont{8}{7.2}{\rmdefault}{\mddefault}{\updefault}{\color[rgb]{0,0,0}$U_t$}%
}}}
\end{picture}
\end{center}
\caption{The controller remembers what it did in the past and uses the
  anytime decoder to get an updated sense of where the observer wants
  it to go. It then applies a control designed to correct for any past
  errors and move the state to be close to the virtual state
  controlled by the observer.} \label{fig:basic_controller}
\end{figure}

The controller uses the updated bit estimates from the anytime decoder
to choose a control to attempt to make the true state $X_t$ stay close
to the virtual state $\bar{X}_t$. It does this by having a pair of
internal models as shown in figure \ref{fig:basic_controller}.

The first, $\widetilde{X}_t$ from (\ref{eqn:tildeX}), models the
unstable system driven only by the actual controls. The second is its
best estimate $\widehat{X}_{t}$, based on the current bit estimates
from the anytime decoder, of where the unstable system should be
driven only by the virtual controls $\bar{U}_t$. Of course, the
controller does not have the exact virtual controls, only its best
estimates $\widehat{U}_1^t(t)$ for them.
\begin{equation}\label{eqn:trackingX}
 \widehat{X}_{t+1}(t) = \sum_{i=0}^{t} \lambda^{i} \widehat{U}_{t-i}(t)
\end{equation}
This is not given in recursive form since all of the past estimates
for the virtual controls are subject to re-estimation at the current
time $t$. The control $U_t$ is chosen to make $\widetilde{X}_{t+1}$ =
$\widehat{X}_{t+1}(t)$.
\begin{equation}\label{eqn:actualcontrol}
 U_t = \widehat{X}_{t+1}(t) - \lambda\widetilde{X}_{t}
\end{equation}

\subsection{Evaluating stability} \label{sec:sufficiencyerrorproof}
{\em Proof of Theorem \ref{thm:general_scalar_control_sufficiency}:}
With controls given by (\ref{eqn:actualcontrol}), the true state $X_t$
can be written as:  
\begin{eqnarray*}
X_t & = & \check{X}_t + \widetilde{X}_t = \check{X}_t + \widehat{X}_{t}(t-1) \\
& = & \sum_{i=0}^{t-1} \lambda^i (W_{t-i} + \widehat{U}_{t-i}(t-1))
\end{eqnarray*}
Notice that the actual state $X_t$ differs from the virtual state
$\bar{X}_t$ only due to errors in virtual control estimation due to
channel noise. If there were no errors in the prefix $\widehat{U}_1^{t-d}$
and arbitrarily bad errors for $\widehat{U}_{t-d+1}^t$, then we could
start at $\bar{X}_{t-d}$ and see how much the errors could have
propagated since then:
$$
X_t = \lambda^{d}\bar{X}_{t-d} + \sum_{i=0}^{d-1} \lambda^i (W_{t-i} + \widehat{U}_{t-i}(t-1))
$$
Comparing this with $\bar{X}_t$, and noticing that the maximum
possible difference between two virtual controls is $\lambda\Delta$
gives:  
\begin{eqnarray*}
|X_t - \bar{X}_t| 
& = & |\sum_{i=0}^{d-1} \lambda^i (\bar{U}_{t-i} - \widehat{U}_{t-i}(t-1))| \\
& \leq & \sum_{i=0}^{d-1} \lambda^i |\bar{U}_{t-i} - \widehat{U}_{t-i}(t-1)| \\
& \leq & \sum_{i=0}^{d-1} \lambda^{i+1}\Delta \\
 & < & \Delta \lambda^{d} \sum_{i=0}^{\infty} \lambda^{-i} \\
& = & \lambda^{d} \frac{\Delta}{1 - \lambda^{-1}}
\end{eqnarray*}
Since $|\bar{X}_t| \leq \frac{\Delta}{2}$, if we know that there were
no errors in the prefix of estimated virtual controls until $d$ time
steps ago, then
\begin{equation}\label{eqn:controlledstatebound}
\{\widehat{U}_0^{t-d}(t-1) = \bar{U}_0^{t-d}\}
\Rightarrow
\{|X_t| < \lambda^{d} \frac{2 \Delta}{1 - \lambda^{-1}}\}
\end{equation}

(\ref{eqn:controlledstatebound}) immediately gives:
\begin{eqnarray*}
& & {\cal P}(|X_t| \geq m ) \\
& = & 
{\cal P}(|X_t| \geq \lambda^{\frac{\log_2 m}{\log_2 \lambda}} \lambda^{\frac{\log_2(1-\lambda^{-1}) - \log_2 (2\Delta)}{\log_2 \lambda}} \left(\frac{2 \Delta}{1 - \lambda^{-1}}\right)) \\
& \leq & 
{\cal P}(|X_t| \geq \lambda^{ \left\lfloor \frac{\log_2 m + \log_2(1-\lambda^{-1}) - \log_2
    (2\Delta)}{\log_2 \lambda} \right\rfloor}
    \left( \frac{2 \Delta}{1 -  \lambda^{-1}} \right)) \\
& \leq & 
g(\left\lfloor \frac{\log_2 m + \log_2(1-\lambda^{-1}) - \log_2
    (2\Delta)}{\log_2 \lambda} \right\rfloor) \\
& \leq & 
g(K'' + \frac{\log_2 m}{\log_2 \lambda})
\end{eqnarray*}
where $g$ bounds the probability of error for the $g-$anytime
code and $K''$ is some constant. \hfill $\Box$ \vspace{0.15in}

Specializing to the case of $\alpha$-anytime capacity, it is clear that:
\begin{eqnarray*}
{\cal P}(|X_t| \geq m) 
& \leq & K''' 2^{-\alpha \frac{\log_2 m}{\log_2 \lambda}} \\
& = & K''' m^{- \frac{\alpha}{\log_2 \lambda}} 
\end{eqnarray*}
which gives a power-law bound on the tail. If the goal is a finite
$\eta$-th moment,
\begin{eqnarray*}
E[|X_t|^\eta]  & = &
\int_{0}^\infty {\cal P}(|X_t|^\eta \geq m) dm \\
& = &
\int_{0}^\infty {\cal P}(|X_t| \geq m^{\frac{1}{\eta}}) dm \\
& \leq & 1 + 
K''' \int_{1}^\infty m^{- \frac{\alpha}{\eta \log_2 \lambda}} dm 
\end{eqnarray*}
As long as $\alpha > \eta \log_2 \lambda$, the integral above converges and
hence the controlled process has a bounded $\eta$-moment. 

\begin{theorem} \label{thm:moment_scalar_control_sufficiency}
It is possible to control an unstable scalar process driven by a
bounded disturbance over a noisy channel so that the $\eta$-moment of
$|X_t|$ stays finite for all time if the channel has feedback anytime
capacity $C_{\mbox{any}}(\alpha) > \log_2 \lambda$ for some $\alpha >
\eta \log_2 \lambda$ and the observer is allowed to observe the noisy
channel outputs and the state exactly.
\end{theorem}
\vspace{0.1in}

Aside from the usual gap between $>$ and $\geq$, this shows that the
necessity condition in Theorem~\ref{thm:scalar_control_necessity} is
tight. Since there are no assumptions on the disturbance process
except for its boundedness, the sufficiency theorems here
automatically cover the case of stochastic disturbances having any
sort of memory structure as long as they remain bounded in support.

\subsection{Almost-sure stability} \label{sec:almostsure}
Control theorists are sometimes interested in an even simpler problem
for which there is no disturbance (i.e.~$W_t = 0$ for all $t$) but the
initial condition $X_0$ is unknown to within some bound $\Omega$. For
this problem, the goal is ensuring that the state $X_t$ tends to zero
almost surely. This short section constructively shows that any
sufficiency result for $\eta$-stability also extends to almost-sure
stabilization. To do this, we consider the system:
\begin{equation} \label{eqn:almostsuresystem}
X'_{t+1} = \lambda' X'_t + U'_t + W'_t
\end{equation}
and use it to prove a key lemma:
\begin{lemma} \label{lem:almostsuremapping}
If it is possible to $\eta'$-stabilize a persistently disturbed system
from (\ref{eqn:almostsuresystem}) when driven by any driving noise
$W'$ bounded by $\Omega$, then there exists a time-varying observer
with noiseless access to the state and a time-varying controller so
that any undisturbed system (\ref{eqn:discretesystem}) with initial
condition $|X_0| \leq \frac{\Omega}{2}$, $W_t = 0$, and $0 < \lambda <
\lambda'$ can be stabilized in the sense that there exists a $K$ so
that:
\begin{equation} \label{eqn:goingtozero}
E[|X_t|^{\eta'}] \leq K (\frac{\lambda}{\lambda'})^{\eta' t} 
\end{equation}
\end{lemma}
{\em Proof:} Since $W_t = 0$ for $t>0$, it is immediately clear that
the system of (\ref{eqn:almostsuresystem}) can be related to the
original system of (\ref{eqn:discretesystem}) by the following scaling
relationships: 
\begin{eqnarray*}
W'_0 & = & X_0 \\
W'_t & = & 0 \mbox{ if } t>0 \\
X'_0 & = & 0 \\
X'_t & = & (\frac{\lambda'}{\lambda})^{t-1} X_{t-1} \mbox{ if }t>0 \\
U'_t & = & (\frac{\lambda'}{\lambda})^{t-1} U_{t-1}
\end{eqnarray*}
It is possible to use an observer/controller design for the system of
(\ref{eqn:almostsuresystem}) to construct one for the original system
(\ref{eqn:discretesystem}) through the same mapping. The input to the
observer constructed with $X'$ in mind will just be
$(\frac{\lambda'}{\lambda})^{t} X_t$ and the controls $U'$ just need
to be scaled down by a factor $(\frac{\lambda}{\lambda'})^{t}$ so that
they will properly apply to the $X_t$ system.

Since (\ref{eqn:almostsuresystem}) can be $\eta'$-stabilized, there
exists a $K'$ so that for all $t \geq 0$,
\begin{eqnarray*}
K' & \geq & E[|X'_t|^{\eta'}] \\
& = & E[(\frac{\lambda'}{\lambda})^{(t-1)\eta'} |X_{t-1}|^{\eta'}] \\
& = & (\frac{\lambda'}{\lambda})^{\eta'(t-1)}E[|X_{t-1}|^{\eta'}]
\end{eqnarray*}
which immediately yields (\ref{eqn:goingtozero}). \hfill $\Box$
\vspace{0.1in}

Lemma~\ref{lem:almostsuremapping} can be used to get almost-sure
stability by noticing that:
\begin{eqnarray*}
E[ \sum_{t=0}^\infty |X_t|^{\eta'}]
& = & \sum_{t=0}^\infty E[ |X_t|^{\eta'}] \\
& \leq &\sum_{t=0}^\infty K (\frac{\lambda}{\lambda'})^{\eta' t} \\
& \leq & \frac{K}{1 - (\frac{\lambda}{\lambda'})^{\eta'}}
\end{eqnarray*}
which is bounded. It immediately follows that:
\begin{eqnarray*}
\lim_{t \rightarrow \infty} |X_t|^{\eta'} & = & 0 \mbox{ almost
  surely} \\
\lim_{t \rightarrow \infty} X_t & = & 0 \mbox{ almost surely}
\end{eqnarray*}
which is summarized in the following theorem:
\begin{theorem} \label{thm:almostsureresult}
If it is possible to $\eta'$-stabilize a persistently disturbed system
from (\ref{eqn:almostsuresystem}) when driven by any driving noise
$W'$ bounded by $\Omega$, then there exists a time-varying observer
with noiseless access to the state and a time-varying controller so
that any undisturbed system (\ref{eqn:discretesystem}) with initial
condition $|X_0| \leq \frac{\Omega}{2}$, $W_t = 0$, and $0 < \lambda <
\lambda'$ can be stabilized in the almost-sure\footnote{Here, the
probability is over the channel's noisy actions and any randomness
present at the observer and controller. The convergence holds for
every possible initial condition and so it does not matter if the
initial condition is included in the probability model.} sense:
$$\lim_{t \rightarrow \infty} X_t = 0 \mbox{ almost surely}$$
\end{theorem}
\vspace{0.1in}

The important thing to notice about Lemma~\ref{lem:almostsuremapping}
and Theorem~\ref{thm:almostsureresult} is that they do not depend on
the detailed structure of the original problem except for the need to
observe the state perfectly at the encoder and to be able to apply
controls with perfect precision. It is clear that if either the state
observation or the control application was limited in precision, then
there would be no way to drive the state to zero almost surely. 

Theorem~\ref{thm:almostsureresult} is used in
Section~\ref{sec:nofeedback} to get
Corollary~\ref{cor:almost_sure_sufficiency} which says that for
almost-sure stabilization of an undisturbed plant across a discrete
memoryless channel (DMC), Shannon capacity larger than $\log_2
\lambda$ suffices regardless of the information pattern.

\subsection{Time in blocks and delayed observations} \label{sec:blockedtime}

In the discussion so far, time has operated at the same scale for 
channel uses, system dynamics, plant observations, and control
application. Furthermore, the only structural delay in the system was
the one-step-delay across the noisy channel needed to allow
the interconnection of the controller, observer, channel, and plant to
make sense. It is interesting to consider different parts of the
system operating at slightly different time scales and to see the
impact of fixed and known delays in the system.

\subsubsection{Observing and controlling the plant on a slower time
  scale} \label{sec:subblocktime}

In the control context, it is natural to consider cases where the
plant evolves on a slower time scale than communication. Formally,
suppose that time is grouped into blocks of size $n$ and the observer
is restricted to only encode the value of $X_t$ at times that are
integer multiples of $n$. Similarly, suppose that the controller only
takes an action\footnote{The controller can take ``no action'' by
setting $U_t = 0$.} immediately before the observer will sample the
state. The effective system dynamics change to
\begin{equation} \label{eqn:scalartimeblocks}
X_{n(k+1)} = \lambda^n X_{nk} + U_{n(k+1)-1} + W'_k
\end{equation}
where $W'_{k} = \sum_{j=0}^{n-1} \lambda^{n-1-j} W_{nk + j}$. Observe
that $|W'_k|$ is known to be bounded within an interval of size
$\Omega' < \lambda^{n} \frac{\Omega}{\lambda - 1}$. Essentially,
everything has just scaled up by a factor of $\lambda^n$. Thus all the
results above continue to hold above for a system described by
(\ref{eqn:scalartimeblocks}) at times which are integer multiples of
$n$. The rate must be larger than $\log_2 \lambda^n = n \log_2
\lambda$ bits per $n$ time steps which translates to $\log_2 \lambda$
bits per time step. The anytime reliability $\alpha > \eta \log_2
\lambda^n = n (\eta \log_2 \lambda)$ for delay measured in units of
$n$ time-steps translates into $\alpha > \eta \log_2 \lambda$ for
delay measured in unit time steps. This is the same as it was for the
system described by (\ref{eqn:discretesystem}).

The only remaining question is what happens to the state at times
within the blocks since no controls are being applied while the state
continues to grow on its own. At such times, the state has just grown
by a factor of at most $\lambda^n$ with an additive term of at most
$\lambda^{n} \frac{\Omega}{\lambda - 1}$.
\begin{eqnarray*}
& & E[(\lambda^n(X_{nk} + \frac{\Omega}{\lambda - 1}))^\eta] \\
& = & 
\lambda^{\eta n} E[(X_{nk} + \frac{\Omega}{\lambda - 1})^\eta] \\
& \leq & 
\lambda^{\eta n}  E[\left(2 \max(|X_{nk}|,\frac{\Omega}{\lambda - 1})\right)^\eta] \\
& = & \lambda^{\eta n} 2^\eta \int_0^\infty 
{\cal P}(\max(|X_{nk}|^\eta,(\frac{\Omega}{\lambda-1})^\eta) \geq \tau) d\tau \\
& = & \lambda^{\eta n} 2^\eta \left((\frac{\Omega}{\lambda-1})^\eta +
\int_{(\frac{\Omega}{\lambda-1})^\eta}^\infty {\cal P}(|X_{nk}|^\eta \geq
\tau) d\tau\right) \\
& < & \lambda^{\eta n} 2^\eta \left((\frac{\Omega}{\lambda-1})^\eta +
\int_{0}^\infty {\cal P}(|X_{nk}|^\eta \geq \tau) d\tau\right) \\
& = & \lambda^{\eta n} 2^\eta \left((\frac{\Omega}{\lambda-1})^\eta + E[|X_{nk}|^\eta]\right)
\end{eqnarray*}
which is finite since the original is finite. Thus:
\begin{theorem} \label{thm:time_block_impact}
If for all $\Omega > 0$, it is possible to stabilize a particular
unstable scalar system with gain $\lambda^n$ and arbitrary disturbance
signal bounded by $\Omega$ when we are allowed $n$ uses of a
particular channel between when the control-system evolves, then for
any $\Omega > 0$ it is also possible to stabilize an unstable scalar
system with gain $\lambda$ that evolves on the same time scale as the
channel using an observer restricted to only observe the system every
$n$ time steps.
\end{theorem}
\vspace{0.1in}

By simple application of Theorem~\ref{thm:time_block_impact}, it is
known that Theorem~\ref{thm:moment_scalar_control_sufficiency} and
similarly Theorem~\ref{thm:scalar_control_necessity} continue to hold
even if the observers/controllers only get access to the analog system
at timesteps that are integer multiples of some $n$. This is used when
considering noisy observations in Section~\ref{sec:noisyobservations}
and in the context of vector-valued states in Part II.

\subsubsection{Known fixed delays} \label{sec:feedbackdelay}

Similarly, we can study cases where the assumed ``round trip delay''
is larger than one. Suppose the control signal applied at time $t$
depends only on channel outputs up to time $t-v$ for some $v > 0$.

It is easy to see that while this sort of deterministic delay does
degrade performance, it does not change stability. The proof of
Theorem~\ref{thm:general_scalar_control_sufficiency} goes through as
before. Specifically, in
Section~\ref{sec:sufficiency_controller_basic}, (\ref{eqn:trackingX})
will change to: 
\begin{equation} \label{eqn:delayedTrackingX}
 \widehat{X}_{t+1}(t) = \sum_{i=0}^{t} \lambda^{i} \widehat{U}_{t-i}(t-v)
\end{equation}
Everything else proceeds as before, just that in place of $d$ for the
probability of error we will have $d + v$. Specifically, in place of
(\ref{eqn:controlledstatebound}), we now know only that:
\begin{eqnarray}\label{eqn:delayedcontrolledstatebound}
|X_t| 
& < & \lambda^{d+v} \frac{2 \Delta}{1 - \lambda^{-1}} \nonumber\\
& = & \lambda^{d} \frac{2 \Delta \lambda^v}{1 - \lambda^{-1}}
\end{eqnarray}

This is just a change in the constant factor and results in a smaller
(more negative) constant $K$ to deal with the larger uncertainty. This
change of constant does not make a bounded $\eta$-moment become
unbounded. The result is summarized in the following theorem:

\begin{theorem} \label{thm:delayed_sufficiency}
Theorems~\ref{thm:general_scalar_control_sufficiency} and
\ref{thm:moment_scalar_control_sufficiency}, continue to hold if the
control signal $U_t$ is required to depend only on the channel outputs
up through time $t-v$ where $v \geq 0$. Only the constants grow
larger. 
\end{theorem}
\vspace{0.1in}

\subsection{Noisy or quantized controls} \label{sec:quantizedcontrols}

The control signals $U_t$ may not be able to be set by the controller
to infinite precision. The applied control $U_t$ at the plant might be
different from the intended control $U^i_t$ generated at the
controller. This section considers the case of $\Gamma_c$-precise
controls where the difference is bounded so $|U_t - U^i_t| \leq
\frac{\Gamma_c}{2}$ for some constant $\Gamma_c$ to reflect the noise
at the controller. It is easy to see that the plant dynamics now
effectively change from \ref{eqn:discretesystem} to:
$$X_{t+1} = \lambda X_{t} + U^i_{t} + \left(W_{t} + (U_t -
U^i_{t})\right)$$ where the term $\left(W_{t} + (U_t -
U^i_{t})\right)$ can be considered the new bounded disturbance for the
system. So in place of $\Omega$, we simply use the new bound
$\Omega+\Gamma_c$. Thus, all the previous results continue to hold in
the case of boundedly noisy control signals.

\begin{theorem} \label{thm:control_noise_impact}
If for all $\Omega > 0$, it is possible to stabilize a particular
unstable scalar system with arbitrary disturbance signal bounded by
$\Omega$ given the ability to apply precise control signals, then for
all $\Gamma_c > 0$ and $\Omega > 0$, it remains possible to stabilize
the same unstable scalar system with arbitrary disturbance signal
bounded by $\Omega$ given the ability to apply only $\Gamma_c$-precise
control signals.
\end{theorem}
\vspace{0.1in}
\subsection{Noisy or quantized observations} \label{sec:noisyobservations}

The observer of Section~\ref{sec:observerwithmemory} has exact
knowledge of the state $X_t$. Suppose that the observation is instead
$X_{noisy}(t) = X_t + N_t$ where $N_t$ is known to be within a bound
$(\frac{-\Gamma}{2},\frac{+\Gamma}{2})$. For example, this models
situations where the input to the encoder has already been quantized
to some resolution.\footnote{The quantization is assumed to be coarse,
but with infinite dynamic range. Section~\ref{sec:scalar_implications}
tells us that finite dynamic range will impose the
requirement of zero-error capacity on the link.}

The observer needs to ensure that the virtual state $\bar{X}$ is
within an interval of size $\Delta$. To do this, just choose a large
enough $\Delta > 2 \Gamma$ so that $X_{noisy}(t)$ and $X_t$ both pick
out the same interval for the state. As figure \ref{fig:overlaps}
illustrates, this is not quite enough since the intervals used in
Section~\ref{sec:observerwithmemory} are partitions of the real
line. Meanwhile, each observation of $X_{noisy}(t)$ gives rise to an
uncertainty window for $X_t \in (X_{noisy}(t) - \frac{\Gamma}{2},
X_{noisy}(t) + \frac{\Gamma}{2})$ that might straddle a boundary of
the partition.\footnote{This will not arise for statically quantized
states since those will have fixed boundaries. In that case, nothing
needs to be done except ensuring that the partitions respect those
boundaries.} Doubling the number of intervals and having them overlap
by half ensures that the uncertainty window can always fit inside a
single interval. Such a doubling increases the data rate by at most an
additional bit. To amortize this additional bit,
Theorem~\ref{thm:time_block_impact} from Section~\ref{sec:blockedtime}
is used and time is considered in blocks of size $n$. Then, the
required rate for achievability with blocked time is $R > 1 + \log_2
\lambda^n$ bits per $n$ time-steps or $R > \frac{1}{n} + \log_2
\lambda$ bits per time step. Since $n$ can be large enough, $R >
\log_2 \lambda$ is good enough. Delayed control actions also causes no
new concerns. Thus, we get the following corollary to
Theorems~\ref{thm:moment_scalar_control_sufficiency} 
and \ref{thm:delayed_sufficiency}:

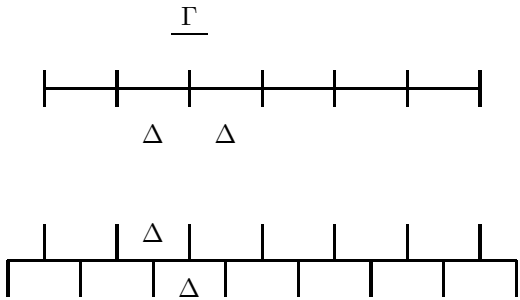
\begin{figure}
\begin{center}
\setlength{\unitlength}{3000sp}%
\begingroup\makeatletter\ifx\SetFigFont\undefined%
\gdef\SetFigFont#1#2#3#4#5{%
  \reset@font\fontsize{#1}{#2pt}%
  \fontfamily{#3}\fontseries{#4}\fontshape{#5}%
  \selectfont}%
\fi\endgroup%
\begin{picture}(4244,2431)(1329,-1985)
\thinlines
{\color[rgb]{0,0,0}\put(2701,239){\line( 1, 0){300}}
}%
\thicklines
{\color[rgb]{0,0,0}\put(1351,-1636){\line( 1, 0){4200}}
}%
{\color[rgb]{0,0,0}\put(1651,-1336){\line( 0,-1){300}}
}%
{\color[rgb]{0,0,0}\put(2251,-1336){\line( 0,-1){300}}
}%
{\color[rgb]{0,0,0}\put(2851,-1336){\line( 0,-1){300}}
}%
{\color[rgb]{0,0,0}\put(3451,-1336){\line( 0,-1){300}}
}%
{\color[rgb]{0,0,0}\put(4051,-1336){\line( 0,-1){300}}
}%
{\color[rgb]{0,0,0}\put(4651,-1336){\line( 0,-1){300}}
}%
{\color[rgb]{0,0,0}\put(5251,-1336){\line( 0,-1){300}}
}%
{\color[rgb]{0,0,0}\put(1351,-1636){\line( 0,-1){300}}
}%
{\color[rgb]{0,0,0}\put(1951,-1636){\line( 0,-1){300}}
}%
{\color[rgb]{0,0,0}\put(2551,-1636){\line( 0,-1){300}}
}%
{\color[rgb]{0,0,0}\put(3151,-1636){\line( 0,-1){300}}
}%
{\color[rgb]{0,0,0}\put(3751,-1636){\line( 0,-1){300}}
}%
{\color[rgb]{0,0,0}\put(4351,-1636){\line( 0,-1){300}}
}%
{\color[rgb]{0,0,0}\put(4951,-1636){\line( 0,-1){300}}
}%
{\color[rgb]{0,0,0}\put(5551,-1636){\line( 0,-1){300}}
}%
{\color[rgb]{0,0,0}\put(1651,-211){\line( 1, 0){3600}}
}%
{\color[rgb]{0,0,0}\put(1651,-61){\line( 0,-1){300}}
}%
{\color[rgb]{0,0,0}\put(2251,-61){\line( 0,-1){300}}
}%
{\color[rgb]{0,0,0}\put(2851,-61){\line( 0,-1){300}}
}%
{\color[rgb]{0,0,0}\put(3451,-61){\line( 0,-1){300}}
}%
{\color[rgb]{0,0,0}\put(4051,-61){\line( 0,-1){300}}
}%
{\color[rgb]{0,0,0}\put(4651,-61){\line( 0,-1){300}}
}%
{\color[rgb]{0,0,0}\put(5251,-61){\line( 0,-1){300}}
}%
\put(2851,314){\makebox(0,0)[b]{\smash{\SetFigFont{10}{14.4}{\rmdefault}{\mddefault}{\updefault}{\color[rgb]{0,0,0}$\Gamma$}%
}}}
\put(2551,-1486){\makebox(0,0)[b]{\smash{\SetFigFont{10}{14.4}{\rmdefault}{\mddefault}{\updefault}{\color[rgb]{0,0,0}$\Delta$}%
}}}
\put(2851,-1936){\makebox(0,0)[b]{\smash{\SetFigFont{10}{14.4}{\rmdefault}{\mddefault}{\updefault}{\color[rgb]{0,0,0}$\Delta$}%
}}}
\put(2551,-661){\makebox(0,0)[b]{\smash{\SetFigFont{10}{14.4}{\rmdefault}{\mddefault}{\updefault}{\color[rgb]{0,0,0}$\Delta$}%
}}}
\put(3151,-661){\makebox(0,0)[b]{\smash{\SetFigFont{10}{14.4}{\rmdefault}{\mddefault}{\updefault}{\color[rgb]{0,0,0}$\Delta$}%
}}}
\end{picture}
\end{center}
\caption{With noisy observations, no strict partition of the line can
  adequately capture the uncertainty since it can straddle the
  boundary of two regions. By doubling the number of bins, it is
  guaranteed that the uncertainty arising from observation noise can
  be contained inside a single bin.}
  \label{fig:overlaps}
\end{figure}

\begin{corollary} \label{cor:moment_scalarsufficiency_with_noise}
It is possible to control an unstable scalar process driven by a
bounded disturbance over a noisy channel so that the $\eta$-moment of
$|X_t|$ stays finite for all time if the channel has feedback anytime
capacity $C_{\mbox{any}}(\alpha) > \log_2 \lambda$ for some $\alpha >
\eta \log_2 \lambda$ and the observer is allowed to observe the noisy
channel outputs exactly and has a boundedly noisy view of the state.

This is true even if the control $U_t$ is only allowed to depend on
channel outputs up through time $t-v$ where $v \geq 0$. 
\end{corollary}

\section{Relaxing feedback} \label{sec:nofeedback}
In this section, we relax the (unrealistic) assumption that the
observer can observe the outputs of the noisy channel directly. This
change of information pattern has the potential to make the problem
more difficult. In distributed control, this was first brought out in
\cite{Witsenhausen68} by the famous Witsenhausen counterexample. This
showed that even in the case of LQG problems, nonlinear solutions can
be optimal when the information patterns are not classical. This same
example also showed how the ``control'' signals can start to play a
dual role --- simultaneously being used for control and to communicate
missing information from one party to another
\cite{AreaExamPaper}. Information theory also has experience with
the new challenges that arise in distributed problems of source and
channel coding \cite{CoverThomas}.

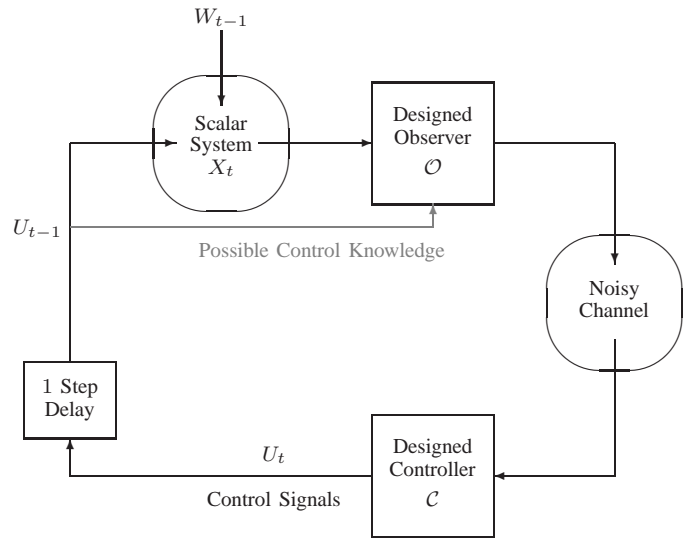
\begin{figure}
\begin{center}
\setlength{\unitlength}{2500sp}%
\begingroup\makeatletter\ifx\SetFigFont\undefined%
\gdef\SetFigFont#1#2#3#4#5{%
  \reset@font\fontsize{#1}{#2pt}%
  \fontfamily{#3}\fontseries{#4}\fontshape{#5}%
  \selectfont}%
\fi\endgroup%
\begin{picture}(7643,5307)(95,-4498)
\thinlines
\put(301,-3511){\framebox(900,750){}}
\put(751,-3061){\makebox(0,0)[b]{\smash{\SetFigFont{8}{14.4}{\rmdefault}{\mddefault}{\updefault}$1$ Step}}}
\put(751,-3286){\makebox(0,0)[b]{\smash{\SetFigFont{8}{14.4}{\rmdefault}{\mddefault}{\updefault}Delay}}}
\put(6150,-2169){\oval(1342,1342)}
\put(6151,-2311){\makebox(0,0)[b]{\smash{\SetFigFont{8}{14.4}{\rmdefault}{\mddefault}{\updefault}Channel}}}
\put(6151,-2086){\makebox(0,0)[b]{\smash{\SetFigFont{8}{14.4}{\rmdefault}{\mddefault}{\updefault}Noisy}}}
\put(4351,-361){\makebox(0,0)[b]{\smash{\SetFigFont{8}{14.4}{\rmdefault}{\mddefault}{\updefault}Designed}}}
\put(4351,-586){\makebox(0,0)[b]{\smash{\SetFigFont{8}{14.4}{\rmdefault}{\mddefault}{\updefault}Observer}}}
\put(4351,-3661){\makebox(0,0)[b]{\smash{\SetFigFont{8}{14.4}{\rmdefault}{\mddefault}{\updefault}Designed}}}
\put(4351,-3886){\makebox(0,0)[b]{\smash{\SetFigFont{8}{14.4}{\rmdefault}{\mddefault}{\updefault}Controller}}}
\put(2251,-586){\oval(1342,1342)}
\put(3751,-1186){\framebox(1200,1200){}}
\put(3751,-4486){\framebox(1200,1200){}}
\put(2626,-586){\vector( 1, 0){1125}}
\put(2251,539){\line( 0,-1){525}}
\put(2251, 14){\vector( 0,-1){225}}
\put(4951,-586){\line( 1, 0){1200}}
\put(6151,-586){\vector( 0,-1){1200}}
\put(3751,-3886){\line(-1, 0){3000}}
\put(751,-3886){\vector( 0, 1){375}}
\put(751,-2761){\line( 0, 1){2175}}
\put(751,-586){\vector( 1, 0){1050}}
{\color[gray]{0.5}
\put(3251,-1711){\makebox(0,0)[b]{\smash{\SetFigFont{8}{14.4}{\rmdefault}{\mddefault}{\updefault}Possible Control Knowledge}}}
\put(751,-1411){\line( 1, 0){3600}}
\put(4351,-1411){\vector( 0, 1){225}}
}
\put(2776,-3736){\makebox(0,0)[b]{\smash{\SetFigFont{8}{14.4}{\rmdefault}{\mddefault}{\updefault}$U_t$}}}
\put(2776,-4186){\makebox(0,0)[b]{\smash{\SetFigFont{8}{14.4}{\rmdefault}{\mddefault}{\updefault}Control Signals}}}
\put(4351,-886){\makebox(0,0)[b]{\smash{\SetFigFont{8}{14.4}{\rmdefault}{\mddefault}{\updefault}$\cal{O}$}}}
\put(4351,-4186){\makebox(0,0)[b]{\smash{\SetFigFont{8}{14.4}{\rmdefault}{\mddefault}{\updefault}$\cal{C}$}}}
\put(2251,-886){\makebox(0,0)[b]{\smash{\SetFigFont{8}{14.4}{\rmdefault}{\mddefault}{\updefault}$X_t$}}}
\put(2251,614){\makebox(0,0)[b]{\smash{\SetFigFont{8}{14.4}{\rmdefault}{\mddefault}{\updefault}$W_{t-1}$}}}
\put(426,-1486){\makebox(0,0)[b]{\smash{\SetFigFont{8}{14.4}{\rmdefault}{\mddefault}{\updefault}$U_{t-1}$}}}
\put(2251,-451){\makebox(0,0)[b]{\smash{\SetFigFont{8}{14.4}{\rmdefault}{\mddefault}{\updefault}Scalar}}}
\put(2251,-661){\makebox(0,0)[b]{\smash{\SetFigFont{8}{14.4}{\rmdefault}{\mddefault}{\updefault}System}}}
\put(6151,-3886){\vector(-1, 0){1200}}
\put(6151,-2536){\line( 0,-1){1350}}
\end{picture}
\end{center}
\caption{Control over a noisy communication channel without explicit
feedback of channel outputs.}
\label{fig:nofeedbackproblem}
\end{figure}

This section restricts the information pattern in stages. First, we
consider the problem of Figure~\ref{fig:nofeedbackproblem} in which
the observer can see the controls but not the channel outputs. Then,
we consider the problem of Figure~\ref{fig:nofeedbackatallproblem}
that restricts the observer to only see the states $X_t$. This section
is divided based on the approach rather than the problem.

In Section~\ref{sec:anytimenofeedback}, the solutions are based on
anytime codes without feedback. These give rise to sufficient
conditions that are more restrictive than the necessary conditions of
Theorem~\ref{thm:scalar_control_necessity}. The main result is
Theorem~\ref{thm:anytimenofeedbacknocontrolsDMC} --- a random
construction that shows it is possible, in the case of DMCs, to have
nearly memoryless time-varying observers and still achieve stability
without any feedback. All the complexity can in principle be shifted
to the controller side.

In Section~\ref{sec:cheating}, the solutions are based on explicitly
communicating the channel outputs back to the observer through either
the control signals or by making the plant itself ``dance'' in a
stable way that communicates limited information noiselessly with no
delay. Such solutions give rise to tight sufficient conditions. These
are not as constructive, but serve to establish the fundamental
connection between stabilization and communication with noiseless
feedback.

\subsection{Using anytime codes without feedback} \label{sec:anytimenofeedback}
Noisefree access to the control signals is not problematic in the case
of Corollary~\ref{cor:moment_scalarsufficiency_with_noise} since the
control signals are calculated from the perfect channel
feedback. Without such perfect feedback, it is more realistic to
consider only noisy access to the control signals. Furthermore,
observe that in Section~\ref{sec:observerwithmemory}, knowledge of the
actual applied controls is used to calculate $W_t$ from the observed
$X_{t+1}, X_{t-1}, U_t$. Thus, any bounded observation noise on the
control signals $U_t$ just translates into an effectively larger
$\Gamma$ bound on the state observation noise. By
Corollary~\ref{cor:moment_scalarsufficiency_with_noise}, any finite
$\Gamma$ can be dealt with and thus:

\begin{corollary} \label{cor:moment_scalar_control_sufficiency_weak}
It is possible to control an unstable scalar process driven by a
bounded disturbance over a noisy channel so that the $\eta$-moment of
$|X_t|$ stays finite for all time if the channel without feedback has
$C_{\mbox{any}}(\alpha) > \log_2 \lambda$ for some $\alpha > \eta \log_2
\lambda$ and the observer is allowed noisy access to the control signals and
the state process as long as the noise on both is bounded.
\end{corollary}
\vspace{0.1in}

As discussed in \cite{OurUpperBoundPaper}, without noiseless feedback
the anytime capacity will tend to be considerably lower for a given
$\alpha$, and so there will be a gap between the necessary condition
established in Theorem~\ref{thm:scalar_control_necessity} and the
sufficient condition in
Corollary~\ref{cor:moment_scalar_control_sufficiency_weak}. 

\begin{figure}
\begin{center}
\setlength{\unitlength}{2500sp}%
\begingroup\makeatletter\ifx\SetFigFont\undefined%
\gdef\SetFigFont#1#2#3#4#5{%
  \reset@font\fontsize{#1}{#2pt}%
  \fontfamily{#3}\fontseries{#4}\fontshape{#5}%
  \selectfont}%
\fi\endgroup%
\begin{picture}(7643,5307)(95,-4498)
\thinlines
\put(301,-3511){\framebox(900,750){}}
\put(751,-3061){\makebox(0,0)[b]{\smash{\SetFigFont{8}{14.4}{\rmdefault}{\mddefault}{\updefault}$1$ Step}}}
\put(751,-3286){\makebox(0,0)[b]{\smash{\SetFigFont{8}{14.4}{\rmdefault}{\mddefault}{\updefault}Delay}}}
\put(6150,-2169){\oval(1342,1342)}
\put(6151,-2311){\makebox(0,0)[b]{\smash{\SetFigFont{8}{14.4}{\rmdefault}{\mddefault}{\updefault}Channel}}}
\put(6151,-2086){\makebox(0,0)[b]{\smash{\SetFigFont{8}{14.4}{\rmdefault}{\mddefault}{\updefault}Noisy}}}
\put(4351,-361){\makebox(0,0)[b]{\smash{\SetFigFont{8}{14.4}{\rmdefault}{\mddefault}{\updefault}Designed}}}
\put(4351,-586){\makebox(0,0)[b]{\smash{\SetFigFont{8}{14.4}{\rmdefault}{\mddefault}{\updefault}Observer}}}
\put(4351,-3661){\makebox(0,0)[b]{\smash{\SetFigFont{8}{14.4}{\rmdefault}{\mddefault}{\updefault}Designed}}}
\put(4351,-3886){\makebox(0,0)[b]{\smash{\SetFigFont{8}{14.4}{\rmdefault}{\mddefault}{\updefault}Controller}}}
\put(2251,-586){\oval(1342,1342)}
\put(3751,-1186){\framebox(1200,1200){}}
\put(3751,-4486){\framebox(1200,1200){}}
\put(2626,-586){\vector( 1, 0){1125}}
\put(2251,539){\line( 0,-1){525}}
\put(2251, 14){\vector( 0,-1){225}}
\put(4951,-586){\line( 1, 0){1200}}
\put(6151,-586){\vector( 0,-1){1200}}
\put(3751,-3886){\line(-1, 0){3000}}
\put(751,-3886){\vector( 0, 1){375}}
\put(751,-2761){\line( 0, 1){2175}}
\put(751,-586){\vector( 1, 0){1050}}
\put(2776,-3736){\makebox(0,0)[b]{\smash{\SetFigFont{8}{14.4}{\rmdefault}{\mddefault}{\updefault}$U_t$}}}
\put(2776,-4186){\makebox(0,0)[b]{\smash{\SetFigFont{8}{14.4}{\rmdefault}{\mddefault}{\updefault}Control Signals}}}
\put(4351,-886){\makebox(0,0)[b]{\smash{\SetFigFont{8}{14.4}{\rmdefault}{\mddefault}{\updefault}$\cal{O}$}}}
\put(4351,-4186){\makebox(0,0)[b]{\smash{\SetFigFont{8}{14.4}{\rmdefault}{\mddefault}{\updefault}$\cal{C}$}}}
\put(2251,-886){\makebox(0,0)[b]{\smash{\SetFigFont{8}{14.4}{\rmdefault}{\mddefault}{\updefault}$X_t$}}}
\put(2251,614){\makebox(0,0)[b]{\smash{\SetFigFont{8}{14.4}{\rmdefault}{\mddefault}{\updefault}$W_{t-1}$}}}
\put(426,-1486){\makebox(0,0)[b]{\smash{\SetFigFont{8}{14.4}{\rmdefault}{\mddefault}{\updefault}$U_{t-1}$}}}
\put(2251,-451){\makebox(0,0)[b]{\smash{\SetFigFont{8}{14.4}{\rmdefault}{\mddefault}{\updefault}Scalar}}}
\put(2251,-661){\makebox(0,0)[b]{\smash{\SetFigFont{8}{14.4}{\rmdefault}{\mddefault}{\updefault}System}}}
\put(6151,-3886){\vector(-1, 0){1200}}
\put(6151,-2536){\line( 0,-1){1350}}
\end{picture}
\end{center}
\caption{Control over a noisy communication channel without any explicit
feedback path from controller to observer except through the plant.}
\label{fig:nofeedbackatallproblem}
\end{figure}
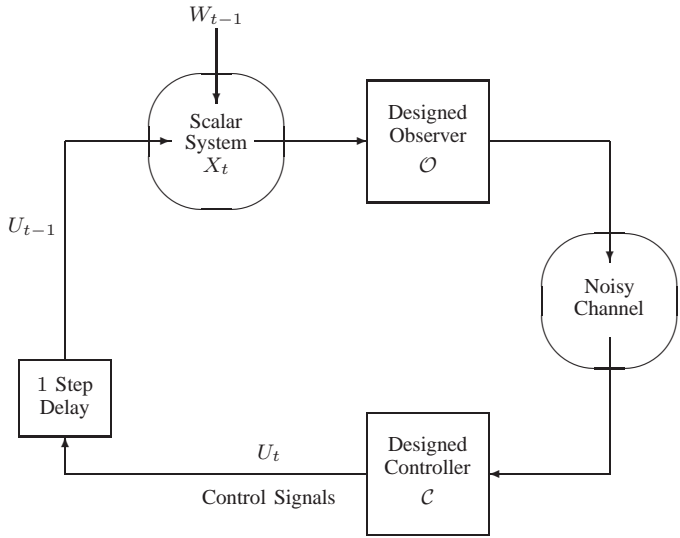
Next, consider the problem of figure \ref{fig:nofeedbackatallproblem}
that restricts the observer to only see the states $X_t$. The
challenge is that the observer of Section~\ref{sec:observerwithmemory}
needs to know the controls in order to remove their effect so as to
focus only on encoding the virtual process $\bar{X}_t$. As such, a new
type of observer is required:
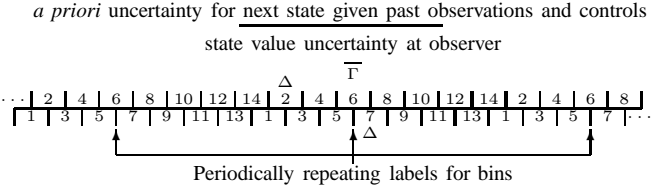
\begin{figure}
\begin{center}
\setlength{\unitlength}{1400sp}%
\begingroup\makeatletter\ifx\SetFigFont\undefined%
\gdef\SetFigFont#1#2#3#4#5{%
  \reset@font\fontsize{#1}{#2pt}%
  \fontfamily{#3}\fontseries{#4}\fontshape{#5}%
  \selectfont}%
\fi\endgroup%
\begin{picture}(12924,2898)(-2861,-2965)
\thinlines
{\color[rgb]{0,0,0}\put(3001,-2461){\line(-1, 0){4050}}
\put(-1049,-2461){\vector( 0, 1){450}}
}%
{\color[rgb]{0,0,0}\put(3001,-2461){\line( 1, 0){4350}}
\put(7351,-2461){\vector( 0, 1){450}}
}%
{\color[rgb]{0,0,0}\put(3151,-2611){\vector( 0, 1){600}}
}%
\put(3151,-2911){\makebox(0,0)[b]{\smash{\SetFigFont{8}{7.2}{\rmdefault}{\mddefault}{\updefault}{\color[rgb]{0,0,0}Periodically repeating labels for bins}%
}}}
\thicklines
{\color[rgb]{0,0,0}\put(1126,-161){\line( 1, 0){3600}}
}%
\put(2926,-11){\makebox(0,0)[b]{\smash{\SetFigFont{8}{7.2}{\rmdefault}{\mddefault}{\updefault}{\color[rgb]{0,0,0}{\em a priori} uncertainty for next state given past observations and controls}%
}}}
{\color[rgb]{0,0,0}\put(1351,-1636){\line( 1, 0){4500}}
}%
{\color[rgb]{0,0,0}\put(1651,-1336){\line( 0,-1){300}}
}%
{\color[rgb]{0,0,0}\put(2251,-1336){\line( 0,-1){300}}
}%
{\color[rgb]{0,0,0}\put(2851,-1336){\line( 0,-1){300}}
}%
{\color[rgb]{0,0,0}\put(3451,-1336){\line( 0,-1){300}}
}%
{\color[rgb]{0,0,0}\put(4051,-1336){\line( 0,-1){300}}
}%
{\color[rgb]{0,0,0}\put(4651,-1336){\line( 0,-1){300}}
}%
{\color[rgb]{0,0,0}\put(5251,-1336){\line( 0,-1){300}}
}%
{\color[rgb]{0,0,0}\put(1951,-1636){\line( 0,-1){300}}
}%
{\color[rgb]{0,0,0}\put(2551,-1636){\line( 0,-1){300}}
}%
{\color[rgb]{0,0,0}\put(3151,-1636){\line( 0,-1){300}}
}%
{\color[rgb]{0,0,0}\put(3751,-1636){\line( 0,-1){300}}
}%
{\color[rgb]{0,0,0}\put(4351,-1636){\line( 0,-1){300}}
}%
{\color[rgb]{0,0,0}\put(4951,-1636){\line( 0,-1){300}}
}%
{\color[rgb]{0,0,0}\put(5851,-1336){\line( 0,-1){300}}
}%
{\color[rgb]{0,0,0}\put(5551,-1636){\line( 0,-1){300}}
}%
{\color[rgb]{0,0,0}\put(1351,-1636){\line( 0,-1){300}}
}%
\thinlines
{\color[rgb]{0,0,0}\put(-2849,-1636){\line( 1, 0){4500}}
}%
{\color[rgb]{0,0,0}\put(-2549,-1336){\line( 0,-1){300}}
}%
{\color[rgb]{0,0,0}\put(-1949,-1336){\line( 0,-1){300}}
}%
{\color[rgb]{0,0,0}\put(-1349,-1336){\line( 0,-1){300}}
}%
{\color[rgb]{0,0,0}\put(-749,-1336){\line( 0,-1){300}}
}%
{\color[rgb]{0,0,0}\put(-149,-1336){\line( 0,-1){300}}
}%
{\color[rgb]{0,0,0}\put(451,-1336){\line( 0,-1){300}}
}%
{\color[rgb]{0,0,0}\put(1051,-1336){\line( 0,-1){300}}
}%
{\color[rgb]{0,0,0}\put(-2849,-1636){\line( 0,-1){300}}
}%
{\color[rgb]{0,0,0}\put(-2249,-1636){\line( 0,-1){300}}
}%
{\color[rgb]{0,0,0}\put(-1649,-1636){\line( 0,-1){300}}
}%
{\color[rgb]{0,0,0}\put(-1049,-1636){\line( 0,-1){300}}
}%
{\color[rgb]{0,0,0}\put(-449,-1636){\line( 0,-1){300}}
}%
{\color[rgb]{0,0,0}\put(151,-1636){\line( 0,-1){300}}
}%
{\color[rgb]{0,0,0}\put(751,-1636){\line( 0,-1){300}}
}%
{\color[rgb]{0,0,0}\put(1651,-1336){\line( 0,-1){300}}
}%
{\color[rgb]{0,0,0}\put(1351,-1636){\line( 0,-1){300}}
}%
{\color[rgb]{0,0,0}\put(5551,-1636){\line( 1, 0){2700}}
}%
{\color[rgb]{0,0,0}\put(5851,-1336){\line( 0,-1){300}}
}%
{\color[rgb]{0,0,0}\put(6451,-1336){\line( 0,-1){300}}
}%
{\color[rgb]{0,0,0}\put(7051,-1336){\line( 0,-1){300}}
}%
{\color[rgb]{0,0,0}\put(7651,-1336){\line( 0,-1){300}}
}%
{\color[rgb]{0,0,0}\put(8251,-1336){\line( 0,-1){300}}
}%
{\color[rgb]{0,0,0}\put(5551,-1636){\line( 0,-1){300}}
}%
{\color[rgb]{0,0,0}\put(6151,-1636){\line( 0,-1){300}}
}%
{\color[rgb]{0,0,0}\put(6751,-1636){\line( 0,-1){300}}
}%
{\color[rgb]{0,0,0}\put(7351,-1636){\line( 0,-1){300}}
}%
{\color[rgb]{0,0,0}\put(7951,-1636){\line( 0,-1){300}}
}%
{\color[rgb]{0,0,0}\put(3001,-831){\line( 1, 0){300}}
}%
\put(1951,-1261){\makebox(0,0)[b]{\smash{\SetFigFont{6}{7.2}{\rmdefault}{\mddefault}{\updefault}{\color[rgb]{0,0,0}$\Delta$}%
}}}
\put(1651,-1861){\makebox(0,0)[b]{\smash{\SetFigFont{6}{7.2}{\rmdefault}{\mddefault}{\updefault}{\color[rgb]{0,0,0}$1$}%
}}}
\put(1951,-1561){\makebox(0,0)[b]{\smash{\SetFigFont{6}{7.2}{\rmdefault}{\mddefault}{\updefault}{\color[rgb]{0,0,0}$2$}%
}}}
\put(2251,-1861){\makebox(0,0)[b]{\smash{\SetFigFont{6}{7.2}{\rmdefault}{\mddefault}{\updefault}{\color[rgb]{0,0,0}$3$}%
}}}
\put(2551,-1561){\makebox(0,0)[b]{\smash{\SetFigFont{6}{7.2}{\rmdefault}{\mddefault}{\updefault}{\color[rgb]{0,0,0}$4$}%
}}}
\put(2851,-1861){\makebox(0,0)[b]{\smash{\SetFigFont{6}{7.2}{\rmdefault}{\mddefault}{\updefault}{\color[rgb]{0,0,0}$5$}%
}}}
\put(3151,-1561){\makebox(0,0)[b]{\smash{\SetFigFont{6}{7.2}{\rmdefault}{\mddefault}{\updefault}{\color[rgb]{0,0,0}$6$}%
}}}
\put(3451,-1861){\makebox(0,0)[b]{\smash{\SetFigFont{6}{7.2}{\rmdefault}{\mddefault}{\updefault}{\color[rgb]{0,0,0}$7$}%
}}}
\put(3751,-1561){\makebox(0,0)[b]{\smash{\SetFigFont{6}{7.2}{\rmdefault}{\mddefault}{\updefault}{\color[rgb]{0,0,0}$8$}%
}}}
\put(4051,-1861){\makebox(0,0)[b]{\smash{\SetFigFont{6}{7.2}{\rmdefault}{\mddefault}{\updefault}{\color[rgb]{0,0,0}$9$}%
}}}
\put(4351,-1561){\makebox(0,0)[b]{\smash{\SetFigFont{6}{7.2}{\rmdefault}{\mddefault}{\updefault}{\color[rgb]{0,0,0}$10$}%
}}}
\put(4651,-1861){\makebox(0,0)[b]{\smash{\SetFigFont{6}{7.2}{\rmdefault}{\mddefault}{\updefault}{\color[rgb]{0,0,0}$11$}%
}}}
\put(4951,-1561){\makebox(0,0)[b]{\smash{\SetFigFont{6}{7.2}{\rmdefault}{\mddefault}{\updefault}{\color[rgb]{0,0,0}$12$}%
}}}
\put(5251,-1861){\makebox(0,0)[b]{\smash{\SetFigFont{6}{7.2}{\rmdefault}{\mddefault}{\updefault}{\color[rgb]{0,0,0}$13$}%
}}}
\put(5551,-1561){\makebox(0,0)[b]{\smash{\SetFigFont{6}{7.2}{\rmdefault}{\mddefault}{\updefault}{\color[rgb]{0,0,0}$14$}%
}}}
\put(-2549,-1861){\makebox(0,0)[b]{\smash{\SetFigFont{6}{7.2}{\rmdefault}{\mddefault}{\updefault}{\color[rgb]{0,0,0}$1$}%
}}}
\put(-2249,-1561){\makebox(0,0)[b]{\smash{\SetFigFont{6}{7.2}{\rmdefault}{\mddefault}{\updefault}{\color[rgb]{0,0,0}$2$}%
}}}
\put(-1949,-1861){\makebox(0,0)[b]{\smash{\SetFigFont{6}{7.2}{\rmdefault}{\mddefault}{\updefault}{\color[rgb]{0,0,0}$3$}%
}}}
\put(-1649,-1561){\makebox(0,0)[b]{\smash{\SetFigFont{6}{7.2}{\rmdefault}{\mddefault}{\updefault}{\color[rgb]{0,0,0}$4$}%
}}}
\put(-1349,-1861){\makebox(0,0)[b]{\smash{\SetFigFont{6}{7.2}{\rmdefault}{\mddefault}{\updefault}{\color[rgb]{0,0,0}$5$}%
}}}
\put(-1049,-1561){\makebox(0,0)[b]{\smash{\SetFigFont{6}{7.2}{\rmdefault}{\mddefault}{\updefault}{\color[rgb]{0,0,0}$6$}%
}}}
\put(-749,-1861){\makebox(0,0)[b]{\smash{\SetFigFont{6}{7.2}{\rmdefault}{\mddefault}{\updefault}{\color[rgb]{0,0,0}$7$}%
}}}
\put(-449,-1561){\makebox(0,0)[b]{\smash{\SetFigFont{6}{7.2}{\rmdefault}{\mddefault}{\updefault}{\color[rgb]{0,0,0}$8$}%
}}}
\put(-149,-1861){\makebox(0,0)[b]{\smash{\SetFigFont{6}{7.2}{\rmdefault}{\mddefault}{\updefault}{\color[rgb]{0,0,0}$9$}%
}}}
\put(151,-1561){\makebox(0,0)[b]{\smash{\SetFigFont{6}{7.2}{\rmdefault}{\mddefault}{\updefault}{\color[rgb]{0,0,0}$10$}%
}}}
\put(451,-1861){\makebox(0,0)[b]{\smash{\SetFigFont{6}{7.2}{\rmdefault}{\mddefault}{\updefault}{\color[rgb]{0,0,0}$11$}%
}}}
\put(751,-1561){\makebox(0,0)[b]{\smash{\SetFigFont{6}{7.2}{\rmdefault}{\mddefault}{\updefault}{\color[rgb]{0,0,0}$12$}%
}}}
\put(1051,-1861){\makebox(0,0)[b]{\smash{\SetFigFont{6}{7.2}{\rmdefault}{\mddefault}{\updefault}{\color[rgb]{0,0,0}$13$}%
}}}
\put(1351,-1561){\makebox(0,0)[b]{\smash{\SetFigFont{6}{7.2}{\rmdefault}{\mddefault}{\updefault}{\color[rgb]{0,0,0}$14$}%
}}}
\put(5851,-1861){\makebox(0,0)[b]{\smash{\SetFigFont{6}{7.2}{\rmdefault}{\mddefault}{\updefault}{\color[rgb]{0,0,0}$1$}%
}}}
\put(6151,-1561){\makebox(0,0)[b]{\smash{\SetFigFont{6}{7.2}{\rmdefault}{\mddefault}{\updefault}{\color[rgb]{0,0,0}$2$}%
}}}
\put(6451,-1861){\makebox(0,0)[b]{\smash{\SetFigFont{6}{7.2}{\rmdefault}{\mddefault}{\updefault}{\color[rgb]{0,0,0}$3$}%
}}}
\put(6751,-1561){\makebox(0,0)[b]{\smash{\SetFigFont{6}{7.2}{\rmdefault}{\mddefault}{\updefault}{\color[rgb]{0,0,0}$4$}%
}}}
\put(7051,-1861){\makebox(0,0)[b]{\smash{\SetFigFont{6}{7.2}{\rmdefault}{\mddefault}{\updefault}{\color[rgb]{0,0,0}$5$}%
}}}
\put(7351,-1561){\makebox(0,0)[b]{\smash{\SetFigFont{6}{7.2}{\rmdefault}{\mddefault}{\updefault}{\color[rgb]{0,0,0}$6$}%
}}}
\put(7651,-1861){\makebox(0,0)[b]{\smash{\SetFigFont{6}{7.2}{\rmdefault}{\mddefault}{\updefault}{\color[rgb]{0,0,0}$7$}%
}}}
\put(7951,-1561){\makebox(0,0)[b]{\smash{\SetFigFont{6}{7.2}{\rmdefault}{\mddefault}{\updefault}{\color[rgb]{0,0,0}$8$}%
}}}
\put(8251,-1861){\makebox(0,0)[b]{\smash{\SetFigFont{6}{7.2}{\rmdefault}{\mddefault}{\updefault}{\color[rgb]{0,0,0}$\cdots$}%
}}}
\put(-2849,-1561){\makebox(0,0)[b]{\smash{\SetFigFont{6}{7.2}{\rmdefault}{\mddefault}{\updefault}{\color[rgb]{0,0,0}$\cdots$}%
}}}
\put(3151,-611){\makebox(0,0)[b]{\smash{\SetFigFont{8}{7.2}{\rmdefault}{\mddefault}{\updefault}{\color[rgb]{0,0,0}state value uncertainty at observer}%
}}}
\put(3151,-1111){\makebox(0,0)[b]{\smash{\SetFigFont{6}{7.2}{\rmdefault}{\mddefault}{\updefault}{\color[rgb]{0,0,0}$\Gamma$}%
}}}
\put(3451,-2161){\makebox(0,0)[b]{\smash{\SetFigFont{6}{7.2}{\rmdefault}{\mddefault}{\updefault}{\color[rgb]{0,0,0}$\Delta$}%
}}}
\end{picture}
\end{center}
\caption{A 15-regularly-labeled lattice based quantizer. If the
  observer had known the controls, it would have centered the lattice
  to cover the top bar exactly. Because it does not, one additional
  quantization bin must be added at the end so that the uncertainty
  never covers two bins bearing the same label.}
\label{fig:latticeoverlaps}
\end{figure}
\begin{definition}
A {\em $\Delta$-lattice based quantizer} is a map (depicted in
Figure~\ref{fig:latticeoverlaps} that maps inputs $X$ to integer bins
$j$. The $j$-th bin spans $(\Delta\frac{j}{2}, \Delta(\frac{j}{2} +
1)]$ and is assigned to $X \in (\Delta(\frac{j}{2} + \frac{1}{4}),
\Delta(\frac{j}{2} + \frac{3}{4})]$ near the center of the bin. 

A {\em $L$-regularly-labeled} $\Delta$-lattice based quantizer is one
which outputs $j \bmod L$ when the input is assigned to bin $j$ ---
one for which the $L$ bin labels repeat periodically.

A {\em randomly-labeled} $\Delta$-lattice based quantizer is one
which outputs $A_j$ when the input it assigned to bin $j$ where the
$A_j$ are drawn iid from a specified distribution.
\end{definition}
\vspace{0.1in}
Lattice based quantizers have some nice properties:
\begin{lemma} \label{lem:latticeproperties}
\begin{itemize}
 \item[a.] If $X_{noisy}(t) = X_t + N_t$ with observation noise $N_t
           \in (\frac{-\Gamma}{2},\frac{+\Gamma}{2})$, then as long as
           $\Delta > 2\Gamma$, the bin $j$ selected by a
           $\Delta$-lattice based quantizer facing input
           $X_{noisy}(t)$ is guaranteed to contain $X_t$.

 \item[b.] There exists a constant $K$ depending only on $\lambda,
           \Delta, \Omega$ so that if $X_t$ is within a single
           particular bin, then $X_{t+n}$ can be in no more than $K
           \lambda^n$ possible adjacent bins whose positions are a
           function of the control inputs applied during those $n$
           time periods as well as the original bin index for $X_t$.

 \item[c.] If $L > K \lambda^n$ then knowing the $L$-regular label
           assigned to $X_{noisy}(t+n)$ is enough to determine a bin
           guaranteed to contain $X_{t+n}$ assuming knowledge of a bin
           containing $X_t$ as well as the control inputs applied
           during those $n$ time periods.
\end{itemize}
\end{lemma}
{\em Proof of [a]}: $X_{noisy}(t) \in (\Delta(\frac{j}{2} + \frac{1}{4}),
\Delta(\frac{j}{2} + \frac{3}{4})]$ implies 
$X_t \in (\Delta(\frac{j}{2} + \frac{1}{4}) - \frac{\Gamma}{2},
\Delta(\frac{j}{2} + \frac{3}{4})+\frac{\Gamma}{2}]$. But
$\frac{\Gamma}{2} < \frac{\Delta}{4}$ by assumption and hence 
$X_t \in (\Delta(\frac{j}{2} + \frac{1}{4}) - \frac{\Delta}{4},
\Delta(\frac{j}{2} + \frac{3}{4})+\frac{\Delta}{4}] = 
(\Delta\frac{j}{2}, \Delta(\frac{j}{2} + 1)]$ which is the extent of
the bin $j$.

{\em Proof of [b]}: First, suppose that the control actions were all
zero during the interval in question. Because the system is linear,
without loss of generality, assume that we start in the $j=0$ bin,
$[0, \Delta]$. After $n$ time-steps, this can reach at most $[0,
\lambda^n \Delta]$ without disturbances. The bounded disturbances can
contribute at most
\begin{eqnarray*}
\sum_{i=0}^{n-1} \lambda^i \frac{\Omega}{2}
& < & \lambda^n \frac{\Omega}{2} \sum_{i=1}^{\infty} \lambda^{-i} \\
& = & \lambda^n \frac{\Omega}{2 (\lambda - 1)}
\end{eqnarray*}
to each side, resulting in an interval of with total length
$\lambda^n( \Delta + \frac{\Omega}{\lambda - 1})$. 

By linearity, the effect of any control inputs is a simple translation
and is therefore just translates the interval by some positive or
negative amount. Because of the overlapping nature of the bins, a
single interval can overlap with at most 2 additional partial bins at
the boundaries.

Since the bins are spaced by $\frac{\Delta}{2}$, the number of
possible bins the state can be in is bounded by $2 + \lambda^n (2 +
\frac{2 \Omega}{\Delta(\lambda - 1)})$ and so $K = 4 + \frac{2
\Omega}{\Delta(\lambda - 1)}$ makes property [b] true.

{\em Proof that [a],[b] $\Rightarrow$ [c]}: [a] guarantees that the bin
corresponding to $X_{noisy}(t+n)$ is guaranteed to contain $X_{t+n}$.
[b] guarantees there are only at most $K \lambda^n < L$ adjacent bins
that the state could be in. Since the modulo operation used to assign
regular labels only assigns the same label to a bin $L$ positions away
or further, all of the $K \lambda^n$ positions have distinct labels
and hence the labeling of $X_{noisy}(t+n)$ picks out the unique
correct bin. \hfill $\Box$ \vspace{0.15in}

Lemma~\ref{lem:latticeproperties} allows the observer to just use
regular $\Delta$-lattice quantizer to translate the state positions
into bins since the control actions are side-information that is known
perfectly at the intended recipient (the controller). The overhead
implied by the constant $K$ can be amortized by looking at time in
blocks of $n$ and so does not asymptotically cost any rate. This can
be used to extend
Corollary~\ref{cor:moment_scalar_control_sufficiency_weak} to cases
without any access to the control. Every $n$ time-units, the observer
can just apply the appropriate regular $\Delta$-lattice quantizer and
send the bin labels through an anytime code that operates without
feedback. However, anytime codes without feedback have a natural tree
structure since the impact of the distant past must never die out.  In
the stabilization context, this tree structure forces the
observer/encoder to remember the bin sequence corresponding to all the
past states. This seems wasteful since closed-loop stability implies
that the plant state will keep returning to the bins in the
neighborhood of the origin. This suggests that this memory at the
observer is not necessary.

\begin{theorem} \label{thm:anytimenofeedbacknocontrolsDMC}
It is possible to control an unstable scalar process driven by a
bounded disturbance over a DMC so that the $\eta$-moment of $|X_t|$
stays finite for all time if the channel without feedback has random
coding error exponent $E_r(R) > \eta \log_2 \lambda$ for some $R >
\log_2 \lambda$ and the observer is allowed boundedly noisy access to
the state process.

Furthermore, there exists an $n > 0$ so this is possible by using an
observer consisting of a time-varying randomly-labeled
$\Delta$-lattice based quantizer that samples the state every $n$ time
steps and outputs a random label for the bin index. The random labels
are chosen iid from ${\cal A}^n$ according to the distribution that
maximizes the random coding error exponent at $R$. The controller must
have access to the common randomness used to choose the random bin
labels.
\end{theorem}
{\em Proof: } Fix a rate $R > \log_2 \lambda$ for which $E_r(R) > \eta
\log_2 \lambda$. Lemma~\ref{lem:latticeproperties} applies to our
quantizer. Pick $n, \Delta$ large enough so that $2^{nR} > K
\lambda^n$ where the $K$ comes from property [b] above. This gives:
\begin{itemize}
 \item[d.] Conditioned on actual past controls applied, the set of
          possible paths that the states $X_0, X_n, X_{2n}, \ldots $
          could have taken through the quantization bins is a subset
          of a trellis that has a maximum branching factor of $2^{nR}$
          Furthermore, the total length covered by the $d$-stage
          descendants of any particular bin is bounded above by $K
          \lambda^{dn}$. 
\end{itemize}

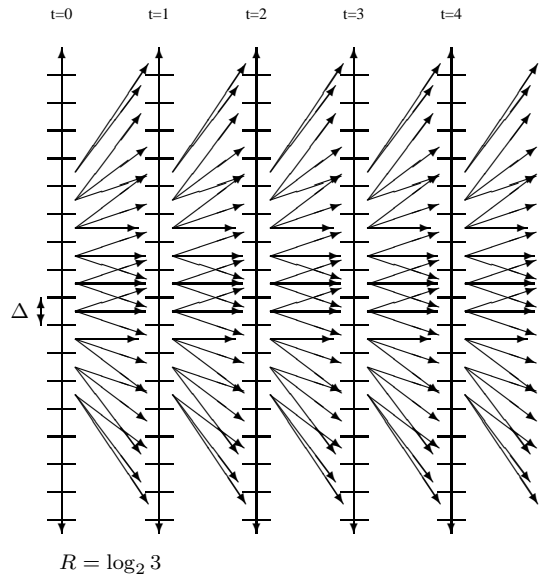
\begin{figure}
\begin{center}
\setlength{\unitlength}{2300sp}%
\begingroup\makeatletter\ifx\SetFigFont\undefined%
\gdef\SetFigFont#1#2#3#4#5{%
  \reset@font\fontsize{#1}{#2pt}%
  \fontfamily{#3}\fontseries{#4}\fontshape{#5}%
  \selectfont}%
\fi\endgroup%
\begin{picture}(5562,6118)(151,-6044)
\thinlines
{\color[rgb]{0,0,0}\put(601,-361){\vector( 0, 1){  0}}
\put(601,-361){\vector( 0,-1){5250}}
}%
{\color[rgb]{0,0,0}\put(451,-2761){\line( 1, 0){300}}
}%
{\color[rgb]{0,0,0}\put(451,-3061){\line( 1, 0){300}}
}%
{\color[rgb]{0,0,0}\put(451,-3361){\line( 1, 0){300}}
}%
{\color[rgb]{0,0,0}\put(451,-3661){\line( 1, 0){300}}
}%
{\color[rgb]{0,0,0}\put(451,-3961){\line( 1, 0){300}}
}%
{\color[rgb]{0,0,0}\put(451,-4261){\line( 1, 0){300}}
}%
{\color[rgb]{0,0,0}\put(451,-4561){\line( 1, 0){300}}
}%
{\color[rgb]{0,0,0}\put(451,-661){\line( 1, 0){300}}
}%
{\color[rgb]{0,0,0}\put(451,-961){\line( 1, 0){300}}
}%
{\color[rgb]{0,0,0}\put(451,-1261){\line( 1, 0){300}}
}%
{\color[rgb]{0,0,0}\put(451,-1561){\line( 1, 0){300}}
}%
{\color[rgb]{0,0,0}\put(451,-1861){\line( 1, 0){300}}
}%
{\color[rgb]{0,0,0}\put(451,-2161){\line( 1, 0){300}}
}%
{\color[rgb]{0,0,0}\put(451,-2461){\line( 1, 0){300}}
}%
{\color[rgb]{0,0,0}\put(451,-4861){\line( 1, 0){300}}
}%
{\color[rgb]{0,0,0}\put(451,-5161){\line( 1, 0){300}}
}%
{\color[rgb]{0,0,0}\put(451,-5461){\line( 1, 0){300}}
}%
{\color[rgb]{0,0,0}\put(1651,-361){\vector( 0, 1){  0}}
\put(1651,-361){\vector( 0,-1){5250}}
}%
{\color[rgb]{0,0,0}\put(1501,-2761){\line( 1, 0){300}}
}%
{\color[rgb]{0,0,0}\put(1501,-3061){\line( 1, 0){300}}
}%
{\color[rgb]{0,0,0}\put(1501,-3361){\line( 1, 0){300}}
}%
{\color[rgb]{0,0,0}\put(1501,-3661){\line( 1, 0){300}}
}%
{\color[rgb]{0,0,0}\put(1501,-3961){\line( 1, 0){300}}
}%
{\color[rgb]{0,0,0}\put(1501,-4261){\line( 1, 0){300}}
}%
{\color[rgb]{0,0,0}\put(1501,-4561){\line( 1, 0){300}}
}%
{\color[rgb]{0,0,0}\put(1501,-661){\line( 1, 0){300}}
}%
{\color[rgb]{0,0,0}\put(1501,-961){\line( 1, 0){300}}
}%
{\color[rgb]{0,0,0}\put(1501,-1261){\line( 1, 0){300}}
}%
{\color[rgb]{0,0,0}\put(1501,-1561){\line( 1, 0){300}}
}%
{\color[rgb]{0,0,0}\put(1501,-1861){\line( 1, 0){300}}
}%
{\color[rgb]{0,0,0}\put(1501,-2161){\line( 1, 0){300}}
}%
{\color[rgb]{0,0,0}\put(1501,-2461){\line( 1, 0){300}}
}%
{\color[rgb]{0,0,0}\put(1501,-4861){\line( 1, 0){300}}
}%
{\color[rgb]{0,0,0}\put(1501,-5161){\line( 1, 0){300}}
}%
{\color[rgb]{0,0,0}\put(1501,-5461){\line( 1, 0){300}}
}%
{\color[rgb]{0,0,0}\put(751,-2611){\vector( 3,-1){765}}
}%
{\color[rgb]{0,0,0}\put(751,-2611){\vector( 1, 0){750}}
}%
{\color[rgb]{0,0,0}\put(751,-2611){\vector( 3, 1){765}}
}%
{\color[rgb]{0,0,0}\put(751,-3211){\vector( 3,-1){765}}
}%
{\color[rgb]{0,0,0}\put(751,-3211){\vector( 1, 0){750}}
}%
{\color[rgb]{0,0,0}\put(751,-3211){\vector( 3, 1){765}}
}%
{\color[rgb]{0,0,0}\put(751,-2911){\vector( 1, 0){750}}
}%
{\color[rgb]{0,0,0}\put(751,-2911){\vector( 3, 1){765}}
}%
{\color[rgb]{0,0,0}\put(751,-2911){\vector( 3,-1){765}}
}%
{\color[rgb]{0,0,0}\put(751,-2311){\vector( 1, 0){675}}
}%
{\color[rgb]{0,0,0}\put(751,-2311){\vector( 3, 1){765}}
}%
{\color[rgb]{0,0,0}\put(751,-2311){\vector( 4, 3){768}}
}%
{\color[rgb]{0,0,0}\put(751,-2011){\vector( 3, 1){765}}
}%
{\color[rgb]{0,0,0}\put(751,-2011){\vector( 4, 3){768}}
}%
{\color[rgb]{0,0,0}\put(751,-2011){\vector( 3, 4){702}}
}%
{\color[rgb]{0,0,0}\put(751,-1711){\vector( 3, 4){702}}
}%
{\color[rgb]{0,0,0}\put(751,-1711){\vector( 2, 3){784.615}}
}%
{\color[rgb]{0,0,0}\put(751,-3511){\vector( 1, 0){675}}
}%
{\color[rgb]{0,0,0}\put(751,-3511){\vector( 3,-1){765}}
}%
{\color[rgb]{0,0,0}\put(751,-3511){\vector( 4,-3){768}}
}%
{\color[rgb]{0,0,0}\put(751,-3811){\vector( 3,-1){765}}
}%
{\color[rgb]{0,0,0}\put(751,-3811){\vector( 4,-3){768}}
}%
{\color[rgb]{0,0,0}\put(751,-3811){\vector( 3,-4){702}}
}%
{\color[rgb]{0,0,0}\put(751,-4111){\vector( 4,-3){768}}
}%
{\color[rgb]{0,0,0}\put(751,-4111){\vector( 3,-4){702}}
}%
{\color[rgb]{0,0,0}\put(751,-4111){\vector( 2,-3){784.615}}
}%
{\color[rgb]{0,0,0}\put(2701,-361){\vector( 0, 1){  0}}
\put(2701,-361){\vector( 0,-1){5250}}
}%
{\color[rgb]{0,0,0}\put(2551,-2761){\line( 1, 0){300}}
}%
{\color[rgb]{0,0,0}\put(2551,-3061){\line( 1, 0){300}}
}%
{\color[rgb]{0,0,0}\put(2551,-3361){\line( 1, 0){300}}
}%
{\color[rgb]{0,0,0}\put(2551,-3661){\line( 1, 0){300}}
}%
{\color[rgb]{0,0,0}\put(2551,-3961){\line( 1, 0){300}}
}%
{\color[rgb]{0,0,0}\put(2551,-4261){\line( 1, 0){300}}
}%
{\color[rgb]{0,0,0}\put(2551,-4561){\line( 1, 0){300}}
}%
{\color[rgb]{0,0,0}\put(2551,-661){\line( 1, 0){300}}
}%
{\color[rgb]{0,0,0}\put(2551,-961){\line( 1, 0){300}}
}%
{\color[rgb]{0,0,0}\put(2551,-1261){\line( 1, 0){300}}
}%
{\color[rgb]{0,0,0}\put(2551,-1561){\line( 1, 0){300}}
}%
{\color[rgb]{0,0,0}\put(2551,-1861){\line( 1, 0){300}}
}%
{\color[rgb]{0,0,0}\put(2551,-2161){\line( 1, 0){300}}
}%
{\color[rgb]{0,0,0}\put(2551,-2461){\line( 1, 0){300}}
}%
{\color[rgb]{0,0,0}\put(2551,-4861){\line( 1, 0){300}}
}%
{\color[rgb]{0,0,0}\put(2551,-5161){\line( 1, 0){300}}
}%
{\color[rgb]{0,0,0}\put(2551,-5461){\line( 1, 0){300}}
}%
{\color[rgb]{0,0,0}\put(1801,-2611){\vector( 3,-1){765}}
}%
{\color[rgb]{0,0,0}\put(1801,-2611){\vector( 1, 0){750}}
}%
{\color[rgb]{0,0,0}\put(1801,-2611){\vector( 3, 1){765}}
}%
{\color[rgb]{0,0,0}\put(1801,-3211){\vector( 3,-1){765}}
}%
{\color[rgb]{0,0,0}\put(1801,-3211){\vector( 1, 0){750}}
}%
{\color[rgb]{0,0,0}\put(1801,-3211){\vector( 3, 1){765}}
}%
{\color[rgb]{0,0,0}\put(1801,-2911){\vector( 1, 0){750}}
}%
{\color[rgb]{0,0,0}\put(1801,-2911){\vector( 3, 1){765}}
}%
{\color[rgb]{0,0,0}\put(1801,-2911){\vector( 3,-1){765}}
}%
{\color[rgb]{0,0,0}\put(1801,-2311){\vector( 1, 0){675}}
}%
{\color[rgb]{0,0,0}\put(1801,-2311){\vector( 3, 1){765}}
}%
{\color[rgb]{0,0,0}\put(1801,-2311){\vector( 4, 3){768}}
}%
{\color[rgb]{0,0,0}\put(1801,-2011){\vector( 3, 1){765}}
}%
{\color[rgb]{0,0,0}\put(1801,-2011){\vector( 4, 3){768}}
}%
{\color[rgb]{0,0,0}\put(1801,-2011){\vector( 3, 4){702}}
}%
{\color[rgb]{0,0,0}\put(1801,-1711){\vector( 3, 4){702}}
}%
{\color[rgb]{0,0,0}\put(1801,-1711){\vector( 2, 3){784.615}}
}%
{\color[rgb]{0,0,0}\put(1801,-3511){\vector( 1, 0){675}}
}%
{\color[rgb]{0,0,0}\put(1801,-3511){\vector( 3,-1){765}}
}%
{\color[rgb]{0,0,0}\put(1801,-3511){\vector( 4,-3){768}}
}%
{\color[rgb]{0,0,0}\put(1801,-3811){\vector( 3,-1){765}}
}%
{\color[rgb]{0,0,0}\put(1801,-3811){\vector( 4,-3){768}}
}%
{\color[rgb]{0,0,0}\put(1801,-3811){\vector( 3,-4){702}}
}%
{\color[rgb]{0,0,0}\put(1801,-4111){\vector( 4,-3){768}}
}%
{\color[rgb]{0,0,0}\put(1801,-4111){\vector( 3,-4){702}}
}%
{\color[rgb]{0,0,0}\put(1801,-4111){\vector( 2,-3){784.615}}
}%
{\color[rgb]{0,0,0}\put(3751,-361){\vector( 0, 1){  0}}
\put(3751,-361){\vector( 0,-1){5250}}
}%
{\color[rgb]{0,0,0}\put(3601,-2761){\line( 1, 0){300}}
}%
{\color[rgb]{0,0,0}\put(3601,-3061){\line( 1, 0){300}}
}%
{\color[rgb]{0,0,0}\put(3601,-3361){\line( 1, 0){300}}
}%
{\color[rgb]{0,0,0}\put(3601,-3661){\line( 1, 0){300}}
}%
{\color[rgb]{0,0,0}\put(3601,-3961){\line( 1, 0){300}}
}%
{\color[rgb]{0,0,0}\put(3601,-4261){\line( 1, 0){300}}
}%
{\color[rgb]{0,0,0}\put(3601,-4561){\line( 1, 0){300}}
}%
{\color[rgb]{0,0,0}\put(3601,-661){\line( 1, 0){300}}
}%
{\color[rgb]{0,0,0}\put(3601,-961){\line( 1, 0){300}}
}%
{\color[rgb]{0,0,0}\put(3601,-1261){\line( 1, 0){300}}
}%
{\color[rgb]{0,0,0}\put(3601,-1561){\line( 1, 0){300}}
}%
{\color[rgb]{0,0,0}\put(3601,-1861){\line( 1, 0){300}}
}%
{\color[rgb]{0,0,0}\put(3601,-2161){\line( 1, 0){300}}
}%
{\color[rgb]{0,0,0}\put(3601,-2461){\line( 1, 0){300}}
}%
{\color[rgb]{0,0,0}\put(3601,-4861){\line( 1, 0){300}}
}%
{\color[rgb]{0,0,0}\put(3601,-5161){\line( 1, 0){300}}
}%
{\color[rgb]{0,0,0}\put(3601,-5461){\line( 1, 0){300}}
}%
{\color[rgb]{0,0,0}\put(2851,-2611){\vector( 3,-1){765}}
}%
{\color[rgb]{0,0,0}\put(2851,-2611){\vector( 1, 0){750}}
}%
{\color[rgb]{0,0,0}\put(2851,-2611){\vector( 3, 1){765}}
}%
{\color[rgb]{0,0,0}\put(2851,-3211){\vector( 3,-1){765}}
}%
{\color[rgb]{0,0,0}\put(2851,-3211){\vector( 1, 0){750}}
}%
{\color[rgb]{0,0,0}\put(2851,-3211){\vector( 3, 1){765}}
}%
{\color[rgb]{0,0,0}\put(2851,-2911){\vector( 1, 0){750}}
}%
{\color[rgb]{0,0,0}\put(2851,-2911){\vector( 3, 1){765}}
}%
{\color[rgb]{0,0,0}\put(2851,-2911){\vector( 3,-1){765}}
}%
{\color[rgb]{0,0,0}\put(2851,-2311){\vector( 1, 0){675}}
}%
{\color[rgb]{0,0,0}\put(2851,-2311){\vector( 3, 1){765}}
}%
{\color[rgb]{0,0,0}\put(2851,-2311){\vector( 4, 3){768}}
}%
{\color[rgb]{0,0,0}\put(2851,-2011){\vector( 3, 1){765}}
}%
{\color[rgb]{0,0,0}\put(2851,-2011){\vector( 4, 3){768}}
}%
{\color[rgb]{0,0,0}\put(2851,-2011){\vector( 3, 4){702}}
}%
{\color[rgb]{0,0,0}\put(2851,-1711){\vector( 3, 4){702}}
}%
{\color[rgb]{0,0,0}\put(2851,-1711){\vector( 2, 3){784.615}}
}%
{\color[rgb]{0,0,0}\put(2851,-3511){\vector( 1, 0){675}}
}%
{\color[rgb]{0,0,0}\put(2851,-3511){\vector( 3,-1){765}}
}%
{\color[rgb]{0,0,0}\put(2851,-3511){\vector( 4,-3){768}}
}%
{\color[rgb]{0,0,0}\put(2851,-3811){\vector( 3,-1){765}}
}%
{\color[rgb]{0,0,0}\put(2851,-3811){\vector( 4,-3){768}}
}%
{\color[rgb]{0,0,0}\put(2851,-3811){\vector( 3,-4){702}}
}%
{\color[rgb]{0,0,0}\put(2851,-4111){\vector( 4,-3){768}}
}%
{\color[rgb]{0,0,0}\put(2851,-4111){\vector( 3,-4){702}}
}%
{\color[rgb]{0,0,0}\put(2851,-4111){\vector( 2,-3){784.615}}
}%
{\color[rgb]{0,0,0}\put(4801,-361){\vector( 0, 1){  0}}
\put(4801,-361){\vector( 0,-1){5250}}
}%
{\color[rgb]{0,0,0}\put(4651,-2761){\line( 1, 0){300}}
}%
{\color[rgb]{0,0,0}\put(4651,-3061){\line( 1, 0){300}}
}%
{\color[rgb]{0,0,0}\put(4651,-3361){\line( 1, 0){300}}
}%
{\color[rgb]{0,0,0}\put(4651,-3661){\line( 1, 0){300}}
}%
{\color[rgb]{0,0,0}\put(4651,-3961){\line( 1, 0){300}}
}%
{\color[rgb]{0,0,0}\put(4651,-4261){\line( 1, 0){300}}
}%
{\color[rgb]{0,0,0}\put(4651,-4561){\line( 1, 0){300}}
}%
{\color[rgb]{0,0,0}\put(4651,-661){\line( 1, 0){300}}
}%
{\color[rgb]{0,0,0}\put(4651,-961){\line( 1, 0){300}}
}%
{\color[rgb]{0,0,0}\put(4651,-1261){\line( 1, 0){300}}
}%
{\color[rgb]{0,0,0}\put(4651,-1561){\line( 1, 0){300}}
}%
{\color[rgb]{0,0,0}\put(4651,-1861){\line( 1, 0){300}}
}%
{\color[rgb]{0,0,0}\put(4651,-2161){\line( 1, 0){300}}
}%
{\color[rgb]{0,0,0}\put(4651,-2461){\line( 1, 0){300}}
}%
{\color[rgb]{0,0,0}\put(4651,-4861){\line( 1, 0){300}}
}%
{\color[rgb]{0,0,0}\put(4651,-5161){\line( 1, 0){300}}
}%
{\color[rgb]{0,0,0}\put(4651,-5461){\line( 1, 0){300}}
}%
{\color[rgb]{0,0,0}\put(3901,-2611){\vector( 3,-1){765}}
}%
{\color[rgb]{0,0,0}\put(3901,-2611){\vector( 1, 0){750}}
}%
{\color[rgb]{0,0,0}\put(3901,-2611){\vector( 3, 1){765}}
}%
{\color[rgb]{0,0,0}\put(3901,-3211){\vector( 3,-1){765}}
}%
{\color[rgb]{0,0,0}\put(3901,-3211){\vector( 1, 0){750}}
}%
{\color[rgb]{0,0,0}\put(3901,-3211){\vector( 3, 1){765}}
}%
{\color[rgb]{0,0,0}\put(3901,-2911){\vector( 1, 0){750}}
}%
{\color[rgb]{0,0,0}\put(3901,-2911){\vector( 3, 1){765}}
}%
{\color[rgb]{0,0,0}\put(3901,-2911){\vector( 3,-1){765}}
}%
{\color[rgb]{0,0,0}\put(3901,-2311){\vector( 1, 0){675}}
}%
{\color[rgb]{0,0,0}\put(3901,-2311){\vector( 3, 1){765}}
}%
{\color[rgb]{0,0,0}\put(3901,-2311){\vector( 4, 3){768}}
}%
{\color[rgb]{0,0,0}\put(3901,-2011){\vector( 3, 1){765}}
}%
{\color[rgb]{0,0,0}\put(3901,-2011){\vector( 4, 3){768}}
}%
{\color[rgb]{0,0,0}\put(3901,-2011){\vector( 3, 4){702}}
}%
{\color[rgb]{0,0,0}\put(3901,-1711){\vector( 3, 4){702}}
}%
{\color[rgb]{0,0,0}\put(3901,-1711){\vector( 2, 3){784.615}}
}%
{\color[rgb]{0,0,0}\put(3901,-3511){\vector( 1, 0){675}}
}%
{\color[rgb]{0,0,0}\put(3901,-3511){\vector( 3,-1){765}}
}%
{\color[rgb]{0,0,0}\put(3901,-3511){\vector( 4,-3){768}}
}%
{\color[rgb]{0,0,0}\put(3901,-3811){\vector( 3,-1){765}}
}%
{\color[rgb]{0,0,0}\put(3901,-3811){\vector( 4,-3){768}}
}%
{\color[rgb]{0,0,0}\put(3901,-3811){\vector( 3,-4){702}}
}%
{\color[rgb]{0,0,0}\put(3901,-4111){\vector( 4,-3){768}}
}%
{\color[rgb]{0,0,0}\put(3901,-4111){\vector( 3,-4){702}}
}%
{\color[rgb]{0,0,0}\put(3901,-4111){\vector( 2,-3){784.615}}
}%
{\color[rgb]{0,0,0}\put(4951,-2611){\vector( 3,-1){765}}
}%
{\color[rgb]{0,0,0}\put(4951,-2611){\vector( 1, 0){750}}
}%
{\color[rgb]{0,0,0}\put(4951,-2611){\vector( 3, 1){765}}
}%
{\color[rgb]{0,0,0}\put(4951,-3211){\vector( 3,-1){765}}
}%
{\color[rgb]{0,0,0}\put(4951,-3211){\vector( 1, 0){750}}
}%
{\color[rgb]{0,0,0}\put(4951,-3211){\vector( 3, 1){765}}
}%
{\color[rgb]{0,0,0}\put(4951,-2911){\vector( 1, 0){750}}
}%
{\color[rgb]{0,0,0}\put(4951,-2911){\vector( 3, 1){765}}
}%
{\color[rgb]{0,0,0}\put(4951,-2911){\vector( 3,-1){765}}
}%
{\color[rgb]{0,0,0}\put(4951,-2311){\vector( 1, 0){675}}
}%
{\color[rgb]{0,0,0}\put(4951,-2311){\vector( 3, 1){765}}
}%
{\color[rgb]{0,0,0}\put(4951,-2311){\vector( 4, 3){768}}
}%
{\color[rgb]{0,0,0}\put(4951,-2011){\vector( 3, 1){765}}
}%
{\color[rgb]{0,0,0}\put(4951,-2011){\vector( 4, 3){768}}
}%
{\color[rgb]{0,0,0}\put(4951,-2011){\vector( 3, 4){702}}
}%
{\color[rgb]{0,0,0}\put(4951,-1711){\vector( 3, 4){702}}
}%
{\color[rgb]{0,0,0}\put(4951,-1711){\vector( 2, 3){784.615}}
}%
{\color[rgb]{0,0,0}\put(4951,-3511){\vector( 1, 0){675}}
}%
{\color[rgb]{0,0,0}\put(4951,-3511){\vector( 3,-1){765}}
}%
{\color[rgb]{0,0,0}\put(4951,-3511){\vector( 4,-3){768}}
}%
{\color[rgb]{0,0,0}\put(4951,-3811){\vector( 3,-1){765}}
}%
{\color[rgb]{0,0,0}\put(4951,-3811){\vector( 4,-3){768}}
}%
{\color[rgb]{0,0,0}\put(4951,-3811){\vector( 3,-4){702}}
}%
{\color[rgb]{0,0,0}\put(4951,-4111){\vector( 4,-3){768}}
}%
{\color[rgb]{0,0,0}\put(4951,-4111){\vector( 3,-4){702}}
}%
{\color[rgb]{0,0,0}\put(4951,-4111){\vector( 2,-3){784.615}}
}%
{\color[rgb]{0,0,0}\put(376,-3361){\vector( 0,-1){  0}}
\put(376,-3361){\vector( 0, 1){300}}
}%
\put(601,-61){\makebox(0,0)[b]{\smash{\SetFigFont{6}{7.2}{\rmdefault}{\mddefault}{\updefault}{\color[rgb]{0,0,0}t=0}%
}}}
\put(1651,-61){\makebox(0,0)[b]{\smash{\SetFigFont{6}{7.2}{\rmdefault}{\mddefault}{\updefault}{\color[rgb]{0,0,0}t=1}%
}}}
\put(2701,-61){\makebox(0,0)[b]{\smash{\SetFigFont{6}{7.2}{\rmdefault}{\mddefault}{\updefault}{\color[rgb]{0,0,0}t=2}%
}}}
\put(3751,-61){\makebox(0,0)[b]{\smash{\SetFigFont{6}{7.2}{\rmdefault}{\mddefault}{\updefault}{\color[rgb]{0,0,0}t=3}%
}}}
\put(4801,-61){\makebox(0,0)[b]{\smash{\SetFigFont{6}{7.2}{\rmdefault}{\mddefault}{\updefault}{\color[rgb]{0,0,0}t=4}%
}}}
\put(151,-3286){\makebox(0,0)[b]{\smash{\SetFigFont{8}{7.2}{\rmdefault}{\mddefault}{\updefault}{\color[rgb]{0,0,0}$\Delta$}%
}}}
\put(1126,-5986){\makebox(0,0)[b]{\smash{\SetFigFont{8}{7.2}{\rmdefault}{\mddefault}{\updefault}{\color[rgb]{0,0,0}$R = \log_2 3$}%
}}}
\end{picture}
\end{center}
\caption{A short segment of the randomly labeled regular trellis from
the point of view of the controller that knows the actual control
signals applied in the past. The example has $R = \log_2 3$ and $\lambda
\approx 2.4$ with $\Delta$ large.}
\label{fig:encoding_trellis}
\end{figure}

\begin{figure}
\begin{center}
\mbox{\epsfxsize=3in \epsfysize=2in \epsfbox{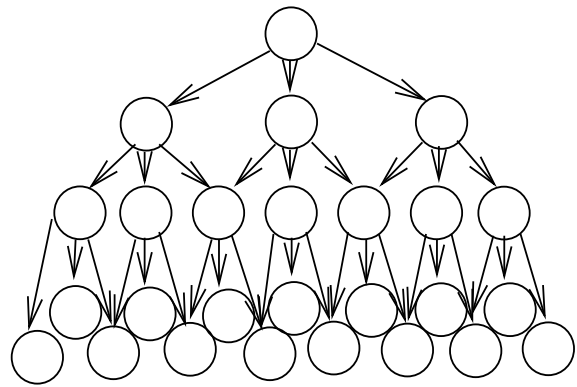}}
\end{center}
\caption{Locally, the trellis looks like a tree with the nodes
  corresponding to the intervals where the state might have been and
  the levels of the tree correspond to the time. It is not a tree
  because paths can remerge, but all labels on disjoint paths are
  chosen so that they are independent of each other.}
\label{fig:encoding_trellis_tree}
\end{figure}

Not all such paths through the trellis are necessarily possible, but
all possible paths do lie within the trellis. Figure
\ref{fig:encoding_trellis} shows what such a trellis looks like and
Figure~\ref{fig:encoding_trellis_tree} shows its tree like local
property. Furthermore, the labels on each bin are iid through both
time and across bins.

Call two paths of length $t$ through the trellis disjoint with depth
$d$ if their last common node was at depth $t-d$ and the paths are
disjoint after that. Consequently:
\begin{itemize}
 \item[e.] If two paths are disjoint in the trellis at a depth of $d$,
          then the channel inputs corresponding to the past $dn$
          channel uses are independent of each other.
\end{itemize}

The suboptimal controller just searches for the ML path through the
trellis. The trellis itself is constructed based on the controller's
memory of all past applied controls. Once an ML path has been
identified, a control signal is applied based on the bin estimate at
the end of the ML path. The control signal just attempts to drive the
center of that bin to zero.

Consider an error event at depth $d$. This represents the case that
the maximum likelihood path last intersected with the true path $dn$
time steps ago. By property [d] above, the control will be based on a
state estimate that can be at most $K \lambda^{dn}$ bins away from the
true state. Thus:
\begin{itemize}
 \item[f.] If an error event at depth $d$ occurs at time $t$, the
          state $|X_{t+n}|$ can be no larger than $K'
          \lambda^{(d+1)n}$ for some constant $K' = 2 \Delta K$ that
          does not depend on $d$ or $t$.
\end{itemize}
Property [f] plays the role of (\ref{eqn:controlledstatebound})
in this proof. 

By property [d], there are no more than $2^{dnR}$ possible false paths
that last intersected the true path $d$ stages ago. By the
memorylessness of the channel, the log-likelihood of each path is the
sum of the likelihood of the ``prefix'' of the path leading up to $d$
stages ago and the ``suffix'' of the path from that point onward. For
a path that is disjoint from the true path at a depth of $d$ to beat
all paths that end up at the true final state, the false path must
have a suffix log-likelihood that beats the suffix log-likelihood of
at least the true path. Property [e] guarantees that the channel
inputs corresponding to the false paths are pairwise independent of
the true inputs for the past $dn$ channel uses.

All that is required to apply Gallager's random block-coding analysis
of Chapter 5 in \cite{Gallager} is such a pairwise
independence\footnote{Notice that pairwise independence is also
obtained if the random labels were assigned using an appropriate
random time-varying infinite constraint-length convolutional code (with
the symbol-merging tricks of Figure~6.2.1 of \cite{Gallager} to match
the desired channel input-distribution) applied to the binary
expansion of the integer $j$ corresponding to the selected bin at each
stage. Since the closed-loop system is stable, the state is presumably
small and the bin is close to $0$. As such, all of the higher-order
bits in the binary expansion of the bin label are zeros and do not
cause any computational burden when operating the convolutional
code. This is related to the feedback convolutional codes with
variable constraint-lengths discussed further in
\cite{OurUpperBoundPaper}. Because of this, the computational burden
of running this observer is non-increasing with time.} between the
true and false codewords for a code of length $dn$.
\begin{itemize}
 \item[g.] The probability that the ML path diverges from the true path
          at depth $d$ is no more than $2^{-dnE_r(R)}$.
\end{itemize}

All that remains is to analyze the $\eta$-moment by combining [g] and
[f] and using the union bound to compute the expectation.
\begin{eqnarray*}
 E[|X_{t+n}|^\eta] 
 & \leq & 
 \sum_{d=0}^{\frac{t}{n}} 2^{-dnE_r(R)} (K' \lambda^{(d+1)n})^\eta \\
 & < & 
  (K'\lambda^n)^\eta \sum_{d=0}^{\infty}
 2^{-dnE_r(R)})\lambda^{\eta d n} \\
 & = & 
  (K'\lambda^n)^\eta \sum_{d=0}^{\infty} 
 2^{-dn(E_r(R) - \eta \log_2 \lambda)} \\
 & = & K'' < \infty
\end{eqnarray*}
where the final geometric sum converges since $E_r(R) > \eta \log_2
\lambda$. \hfill $\Box$ \vspace{0.15in}

Although the condition in
Theorem~\ref{thm:anytimenofeedbacknocontrolsDMC} is not tight, the
result has several nice features. First, it allows easy verification
of sufficiency for a good channel since $E_r(R)$ is easy to
calculate. Structurally, it demonstrates that there is no need to use
very complex observers. The intrinsic memory in the plant can play the
role of the memory that would otherwise need to be implemented in a
channel code. The complexity can be shifted to the controller, and
even that complexity is not too bad. Sequential decoding can be used
at the controller since it is known to have the same asymptotic
performance with respect to delay as the ML decoder\cite{ForneySeq,
JelinekSequential}. Because the closed-loop system is stable and
thereby renews itself constantly, the computational burden of running
sequential decoding (and hence the controller) does not grow
unboundedly with time \cite{Allerton05Control}.

Since $E_r(R,Q) > 0$ for all $R < C$ and the capacity-achieving
distribution $Q$, Theorem~\ref{thm:anytimenofeedbacknocontrolsDMC} can
also be recast in a weaker Shannon capacity-centric form:
\begin{corollary} \label{cor:some_moment_sufficiency}
If the observer is allowed boundedly noisy access to the plant state,
and the noisy channel is a DMC with Shannon capacity $C > \log_2
\lambda$, then there exists some $\eta > 0$ and an observer/controller
pair that stabilizes the system in closed loop so that the
$\eta$-moment of $|X_t|$ stays finite for all time.

Furthermore, there exists an $n > 0$ so this is possible by using an
observer consisting of a time-varying randomly-labeled
$\Delta$-lattice based quantizer that samples the state every $n$ time
steps and outputs a random label for the bin index. This random labels
are chosen iid from the ${\cal A}^n$ according to the
capacity-achieving input distribution. The controller must have access
to the common randomness used to choose the random bin labels.
\end{corollary}
\vspace{0.1in}

Applying Theorem~\ref{thm:almostsureresult} to
Corollary~\ref{cor:some_moment_sufficiency} immediately results in the
following new corollary:

\begin{corollary} \label{cor:almost_sure_sufficiency}
If the observer is allowed perfect access to the plant state, and the
noisy channel is a DMC with Shannon capacity $C > \log_2 \lambda$,
then there exists an observer/controller pair that stabilizes the
system (\ref{eqn:discretesystem}) in closed loop so that:
$$\lim_{t \rightarrow \infty} X_t = 0 \mbox{ almost surely}$$
as long as the initial condition $|X_0| \leq \frac{\Omega}{2}$ and the
disturbances $W_t = 0$. 

Furthermore, there exists an $n > 0$ so this is possible by using an
observer consisting of a time-varying randomly-labeled
$\Delta_t$-lattice based quantizer that samples the state every $n$
time steps and outputs a random label for the bin index. The
$\Delta_t$ shrink geometrically with time, and the random labels are
chosen iid from the ${\cal A}^n$ according to the capacity-achieving
input distribution. The controller must have access to the common
randomness used to choose the random bin labels.
\end{corollary}

\subsection{Communicating the channel outputs back to the observer} \label{sec:cheating}
In this section, the goal is to recover the tight condition on the
channel from Theorem~\ref{thm:moment_scalar_control_sufficiency}. To
do this, we construct a controller that explicitly communicates the
noisy channel outputs to the observer using whatever ``channels'' are
available to it. First we consider using a noiseless control signal to
embed the feedback information. This motivates the technique used
to communicate the feedback information by making the plant itself
dance in a stable way that tells the observer the channel output.

\subsubsection{Using the controls to communicate the channel outputs}
The idea is to ``cheat''\footnote{We call this ``cheating'' since it
violates the spirit of the requirement against access to the channel
outputs. However, it is important to establish this result because it
points out the need for a serious future study where the communication
constraints back from the controller to the observer are modeled more
carefully. A more realistic model for the problem should have a sensor
observing the plant connected via a communication channel to the
controller. The controller is then connected to an actuator through
another communication channel. The actuator finally acts upon the
plant itself. With no complexity constraints, this reduces to the case
studied here with the controller merely playing the role of a relay
bridging together two communication channels. The relay anytime
reliability will become the relevant quantity to study.}  and
communicate the channel outputs through the controls. The control
signal is thus serving dual purposes --- stabilization of the system
and the communication of channel outputs. Suppose the
observer had noiseless access to the control signals. The controller
can choose to quantize its real-valued controls to some suitable level
and then use the infinite bits remaining in the fractional part to
communicate the channel outputs to the observer. The observer can then
extract these bits noiselessly and give them to the anytime encoder as
noiseless channel feedback.

Of course, this additional fractional part will introduce an added
disturbance to the plant. One approach is to just consider the
quantization and channel output communication terms together as a
bounded noise on the control signals considered in
Section~\ref{sec:quantizedcontrols}. This immediately yields:
\begin{corollary} \label{cor:moment_scalar_control_sufficiency_stronger}
It is possible to control an unstable scalar process driven by a
bounded disturbance over a noisy channel so that the $\eta$-moment of
$|X_t|$ stays finite for all time if the channel has feedback anytime
capacity $C_{\mbox{any}}(\alpha) > \log_2 \lambda$ for some $\alpha >
\eta \log_2 \lambda$ and the observer is allowed to observe the
control signals perfectly.
\end{corollary}
\vspace{0.1in}

However, the additional disturbance introduced by the quantization of
the original control signal and the introduction of the new fractional
part representing the channel output is known perfectly at the
controller end. Meanwhile, the output of the virtual-process based
observer does not depend on the actual applied controls anyway since
it subtracts them off. So rather than compensating for this
quantization+signaling by expanding the uncertainty $\Omega$ and thus
changing the $\Delta$ at the observer, the controller can just clean
up after itself. This idea allows us to eliminate all access to the
control signals at the observer and generalizes to many cases of
countably large channel output output alphabets. 

\subsubsection{Removing noiseless access to the controls at the observer} \label{sec:nocontrols}

There are two tricks involved. The first is the idea of making the
plant ``dance'' appropriately and using the moves in the dance to
communicate the channel outputs. The second idea is to introduce an
artificial delay of $1$ time step in the determination of the
``non-dance'' component of the control signals. This makes the
non-dance component completely predictable by the observer and allows
the observer to clearly see the dance move corrupted only by the
bounded process disturbance. Putting it together gives:
\begin{figure}
\begin{center}
\setlength{\unitlength}{1800sp}%
\begingroup\makeatletter\ifx\SetFigFont\undefined%
\gdef\SetFigFont#1#2#3#4#5{%
  \reset@font\fontsize{#1}{#2pt}%
  \fontfamily{#3}\fontseries{#4}\fontshape{#5}%
  \selectfont}%
\fi\endgroup%
\begin{picture}(8640,8124)(439,-7423)
\thinlines
{\color[rgb]{0,0,0}\put(556,-3706){\oval(210,210)[bl]}
\put(556,-3166){\oval(210,210)[tl]}
\put(1246,-3706){\oval(210,210)[br]}
\put(1246,-3166){\oval(210,210)[tr]}
\put(556,-3811){\line( 1, 0){690}}
\put(556,-3061){\line( 1, 0){690}}
\put(451,-3706){\line( 0, 1){540}}
\put(1351,-3706){\line( 0, 1){540}}
}%
\put(901,-3361){\makebox(0,0)[b]{\smash{\SetFigFont{5}{14.4}{\rmdefault}{\mddefault}{\updefault}{\color[rgb]{0,0,0}$1$ Step}%
}}}
\put(901,-3586){\makebox(0,0)[b]{\smash{\SetFigFont{5}{14.4}{\rmdefault}{\mddefault}{\updefault}{\color[rgb]{0,0,0}Delay}%
}}}
{\color[rgb]{0,0,0}\put(2401,-2086){\oval(1342,1342)}
}%
{\color[rgb]{0,0,0}\put(8400,-3744){\oval(1342,1342)}
}%
{\color[rgb]{0,0,0}\put(3901,-5986){\framebox(1200,1200){}}
}%
{\color[rgb]{0,0,0}\put(2776,-2086){\vector( 1, 0){2925}}
}%
{\color[rgb]{0,0,0}\put(2401,-961){\line( 0,-1){525}}
\put(2401,-1486){\vector( 0,-1){225}}
}%
{\color[rgb]{0,0,0}\put(2059,-5386){\line(-1, 0){1158}}
\put(901,-5386){\vector( 0, 1){1575}}
}%
{\color[rgb]{0,0,0}\put(901,-3061){\line( 0, 1){975}}
\put(901,-2086){\vector( 1, 0){1050}}
}%
{\color[rgb]{0,0,0}\put(6001,-5986){\framebox(1200,1200){}}
}%
{\color[rgb]{0,0,0}\put(1576,-7261){\framebox(1200,1200){}}
}%
{\color[rgb]{0,0,0}\put(8401,-2686){\vector( 0,-1){675}}
}%
{\color[rgb]{0,0,0}\put(7801,-2686){\framebox(1200,1200){}}
}%
{\color[rgb]{0,0,0}\put(5701,-2686){\framebox(1200,1200){}}
}%
{\color[rgb]{0,0,0}\put(3601,-811){\framebox(1200,1200){}}
}%
\put(2401,-1951){\makebox(0,0)[b]{\smash{\SetFigFont{5}{14.4}{\rmdefault}{\mddefault}{\updefault}{\color[rgb]{0,0,0}Scalar }%
}}}
\put(2401,-2161){\makebox(0,0)[b]{\smash{\SetFigFont{5}{14.4}{\rmdefault}{\mddefault}{\updefault}{\color[rgb]{0,0,0}System}%
}}}
\put(2401,-2386){\makebox(0,0)[b]{\smash{\SetFigFont{5}{14.4}{\rmdefault}{\mddefault}{\updefault}{\color[rgb]{0,0,0}$X_t$}%
}}}
\put(2401,-886){\makebox(0,0)[b]{\smash{\SetFigFont{5}{14.4}{\rmdefault}{\mddefault}{\updefault}{\color[rgb]{0,0,0}$W_{t-1}$}%
}}}
\put(4501,-5386){\makebox(0,0)[b]{\smash{\SetFigFont{5}{14.4}{\rmdefault}{\mddefault}{\updefault}{\color[rgb]{0,0,0}Controller}%
}}}
\put(8401,-3886){\makebox(0,0)[b]{\smash{\SetFigFont{5}{14.4}{\rmdefault}{\mddefault}{\updefault}{\color[rgb]{0,0,0}Channel}%
}}}
\put(8401,-3661){\makebox(0,0)[b]{\smash{\SetFigFont{5}{14.4}{\rmdefault}{\mddefault}{\updefault}{\color[rgb]{0,0,0}Noisy}%
}}}
\put(4501,-5686){\makebox(0,0)[b]{\smash{\SetFigFont{5}{14.4}{\rmdefault}{\mddefault}{\updefault}{\color[rgb]{0,0,0}${\cal C}$}%
}}}
\put(901,-5761){\makebox(0,0)[b]{\smash{\SetFigFont{5}{14.4}{\rmdefault}{\mddefault}{\updefault}{\color[rgb]{0,0,0}$U_t$}%
}}}
\put(6301,-2086){\makebox(0,0)[b]{\smash{\SetFigFont{5}{14.4}{\rmdefault}{\mddefault}{\updefault}{\color[rgb]{0,0,0}Process}%
}}}
\put(6301,-2386){\makebox(0,0)[b]{\smash{\SetFigFont{5}{14.4}{\rmdefault}{\mddefault}{\updefault}{\color[rgb]{0,0,0}Simulator}%
}}}
\put(7351,-1861){\makebox(0,0)[b]{\smash{\SetFigFont{5}{14.4}{\rmdefault}{\mddefault}{\updefault}{\color[rgb]{0,0,0}$\bar{U}_t$}%
}}}
\put(601,-2536){\makebox(0,0)[b]{\smash{\SetFigFont{5}{14.4}{\rmdefault}{\mddefault}{\updefault}{\color[rgb]{0,0,0}$U_{t-1}$}%
}}}
\put(6001,-1186){\makebox(0,0)[b]{\smash{\SetFigFont{5}{14.4}{\rmdefault}{\mddefault}{\updefault}{\color[rgb]{0,0,0}$U_{t-1}$}%
}}}
\put(8401,-5761){\makebox(0,0)[b]{\smash{\SetFigFont{5}{14.4}{\rmdefault}{\mddefault}{\updefault}{\color[rgb]{0,0,0}$Z_t$}%
}}}
\put(5251,-511){\makebox(0,0)[b]{\smash{\SetFigFont{5}{14.4}{\rmdefault}{\mddefault}{\updefault}{\color[rgb]{0,0,0}$Z_{t-1}$}%
}}}
\put(4501,-5161){\makebox(0,0)[b]{\smash{\SetFigFont{5}{14.4}{\rmdefault}{\mddefault}{\updefault}{\color[rgb]{0,0,0}Tracking}%
}}}
\put(5551,-4636){\makebox(0,0)[b]{\smash{\SetFigFont{5}{14.4}{\rmdefault}{\mddefault}{\updefault}{\color[rgb]{0,0,0}$\widehat{U}_1^{t-1}(t-1)$}%
}}}
\put(6301,-1861){\makebox(0,0)[b]{\smash{\SetFigFont{5}{14.4}{\rmdefault}{\mddefault}{\updefault}{\color[rgb]{0,0,0}Virtual $\bar{X}$}%
}}}
{\color[rgb]{0,0,0}\put(2171,-5394){\circle{220}}
}%
{\color[rgb]{0,0,0}\put(8401,-4111){\line( 0,-1){1275}}
\put(8401,-5386){\vector(-1, 0){1200}}
}%
{\color[rgb]{0,0,0}\put(6001,-5386){\vector(-1, 0){900}}
}%
{\color[rgb]{0,0,0}\put(3901,-5386){\vector(-1, 0){1613}}
}%
{\color[rgb]{0,0,0}\put(7501,-5386){\line( 0,-1){1275}}
\put(7501,-6661){\vector(-1, 0){4725}}
}%
{\color[rgb]{0,0,0}\put(2176,-6061){\vector( 0, 1){555}}
}%
{\color[rgb]{0,0,0}\put(6901,-2086){\vector( 1, 0){900}}
}%
{\color[rgb]{0,0,0}\put(901,-4411){\line( 1, 0){3600}}
\put(4501,-4411){\vector( 0,-1){375}}
}%
{\color[rgb]{0,0,0}\multiput(451,-7411)(9.00000,0.00000){801}{\makebox(1.6667,11.6667){\SetFigFont{5}{6}{\rmdefault}{\mddefault}{\updefault}.}}
\multiput(451,-4111)(9.00000,0.00000){801}{\makebox(1.6667,11.6667){\SetFigFont{5}{6}{\rmdefault}{\mddefault}{\updefault}.}}
\multiput(451,-7411)(0.00000,8.99183){368}{\makebox(1.6667,11.6667){\SetFigFont{5}{6}{\rmdefault}{\mddefault}{\updefault}.}}
\multiput(7651,-7411)(0.00000,8.99183){368}{\makebox(1.6667,11.6667){\SetFigFont{5}{6}{\rmdefault}{\mddefault}{\updefault}.}}
}%
{\color[rgb]{0,0,0}\put(4201,-2086){\vector( 0, 1){1275}}
}%
{\color[rgb]{0,0,0}\put(4801,-211){\vector( 1, 0){900}}
}%
{\color[rgb]{0,0,0}\put(6301,-811){\vector( 0,-1){675}}
}%
{\color[rgb]{0,0,0}\multiput(5701,-811)(9.02256,0.00000){134}{\makebox(1.6667,11.6667){\SetFigFont{5}{6}{\rmdefault}{\mddefault}{\updefault}.}}
\multiput(5701,389)(9.02256,0.00000){134}{\makebox(1.6667,11.6667){\SetFigFont{5}{6}{\rmdefault}{\mddefault}{\updefault}.}}
\multiput(5701,-811)(0.00000,9.02256){134}{\makebox(1.6667,11.6667){\SetFigFont{5}{6}{\rmdefault}{\mddefault}{\updefault}.}}
\multiput(6901,-811)(0.00000,9.02256){134}{\makebox(1.6667,11.6667){\SetFigFont{5}{6}{\rmdefault}{\mddefault}{\updefault}.}}
}%
{\color[rgb]{0,0,0}\put(5251,-211){\line( 0, 1){900}}
\put(5251,689){\line( 1, 0){3150}}
\put(8401,689){\vector( 0,-1){2175}}
}%
\put(6601,-5161){\makebox(0,0)[b]{\smash{\SetFigFont{5}{14.4}{\rmdefault}{\mddefault}{\updefault}{\color[rgb]{0,0,0}Feedback}%
}}}
\put(6601,-5386){\makebox(0,0)[b]{\smash{\SetFigFont{5}{14.4}{\rmdefault}{\mddefault}{\updefault}{\color[rgb]{0,0,0}Anytime}%
}}}
\put(6601,-5611){\makebox(0,0)[b]{\smash{\SetFigFont{5}{14.4}{\rmdefault}{\mddefault}{\updefault}{\color[rgb]{0,0,0}Decoder}%
}}}
\put(2175,-5428){\makebox(0,0)[b]{\smash{\SetFigFont{5}{14.4}{\rmdefault}{\mddefault}{\updefault}{\color[rgb]{0,0,0}+}%
}}}
\put(2192,-6527){\makebox(0,0)[b]{\smash{\SetFigFont{5}{14.4}{\rmdefault}{\mddefault}{\updefault}{\color[rgb]{0,0,0}Channel }%
}}}
\put(2192,-6752){\makebox(0,0)[b]{\smash{\SetFigFont{5}{14.4}{\rmdefault}{\mddefault}{\updefault}{\color[rgb]{0,0,0}Output}%
}}}
\put(2192,-6977){\makebox(0,0)[b]{\smash{\SetFigFont{5}{14.4}{\rmdefault}{\mddefault}{\updefault}{\color[rgb]{0,0,0}Encoder}%
}}}
\put(8401,-1861){\makebox(0,0)[b]{\smash{\SetFigFont{5}{14.4}{\rmdefault}{\mddefault}{\updefault}{\color[rgb]{0,0,0}Feedback}%
}}}
\put(8401,-2086){\makebox(0,0)[b]{\smash{\SetFigFont{5}{14.4}{\rmdefault}{\mddefault}{\updefault}{\color[rgb]{0,0,0}Anytime}%
}}}
\put(8401,-2311){\makebox(0,0)[b]{\smash{\SetFigFont{5}{14.4}{\rmdefault}{\mddefault}{\updefault}{\color[rgb]{0,0,0}Encoder}%
}}}
\put(6001,-7261){\makebox(0,0)[b]{\smash{\SetFigFont{5}{14.4}{\rmdefault}{\mddefault}{\updefault}{\color[rgb]{0,0,0}Joint Decoder/Controller}%
}}}
\put(4201,-61){\makebox(0,0)[b]{\smash{\SetFigFont{5}{14.4}{\rmdefault}{\mddefault}{\updefault}{\color[rgb]{0,0,0}Channel}%
}}}
\put(4201,-286){\makebox(0,0)[b]{\smash{\SetFigFont{5}{14.4}{\rmdefault}{\mddefault}{\updefault}{\color[rgb]{0,0,0}Output}%
}}}
\put(4201,-511){\makebox(0,0)[b]{\smash{\SetFigFont{5}{14.4}{\rmdefault}{\mddefault}{\updefault}{\color[rgb]{0,0,0}Extractor}%
}}}
\put(6301, 14){\makebox(0,0)[b]{\smash{\SetFigFont{5}{14.4}{\rmdefault}{\mddefault}{\updefault}{\color[rgb]{0,0,0}Joint}%
}}}
\put(6301,-211){\makebox(0,0)[b]{\smash{\SetFigFont{5}{14.4}{\rmdefault}{\mddefault}{\updefault}{\color[rgb]{0,0,0}Decoder}%
}}}
\put(6301,-436){\makebox(0,0)[b]{\smash{\SetFigFont{5}{14.4}{\rmdefault}{\mddefault}{\updefault}{\color[rgb]{0,0,0}Controller}%
}}}
\put(6301,-661){\makebox(0,0)[b]{\smash{\SetFigFont{5}{14.4}{\rmdefault}{\mddefault}{\updefault}{\color[rgb]{0,0,0}Copy}%
}}}
\end{picture}
\end{center}
\caption{Overlaying messages onto the control signal and recovering
the messages at the observer. The control signal is generated based on
unit-delayed channel outputs with the current output being
communicated back.}
\label{fig:plantdancing}
\end{figure}
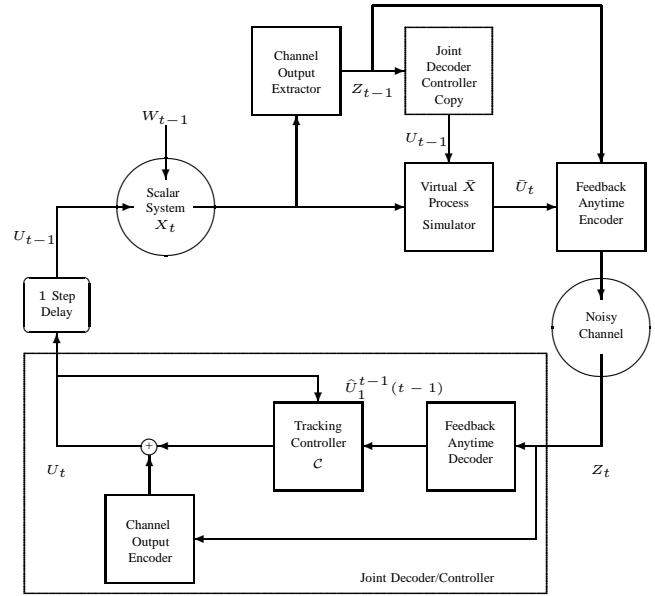

\begin{figure}
\begin{center}
\setlength{\unitlength}{1700sp}%
\begingroup\makeatletter\ifx\SetFigFont\undefined%
\gdef\SetFigFont#1#2#3#4#5{%
  \reset@font\fontsize{#1}{#2pt}%
  \fontfamily{#3}\fontseries{#4}\fontshape{#5}%
  \selectfont}%
\fi\endgroup%
\begin{picture}(3900,4110)(2101,-3640)
\thinlines
{\color[rgb]{0,0,0}\put(5851,-2611){\vector(-1, 0){1050}}
}%
{\color[rgb]{0,0,0}\put(4801,-2161){\vector(-1, 0){450}}
}%
{\color[rgb]{0,0,0}\put(4201,-211){\line( 1, 0){300}}
}%
{\color[rgb]{0,0,0}\put(4426,314){\vector( 0,-1){525}}
}%
{\color[rgb]{0,0,0}\put(2251,-811){\line( 0,-1){2550}}
}%
{\color[rgb]{0,0,0}\put(2251,-2161){\vector( 1, 0){2100}}
}%
{\color[rgb]{0,0,0}\put(2251,-3361){\vector( 1, 0){3600}}
}%
{\color[rgb]{0,0,0}\put(2251,-811){\vector( 1, 0){2550}}
}%
\thicklines
{\color[rgb]{0,0,0}\put(3676,-961){\line( 1, 0){2250}}
}%
{\color[rgb]{0,0,0}\put(3676,-1261){\line( 1, 0){2250}}
}%
{\color[rgb]{0,0,0}\put(4126,-961){\line( 0,-1){300}}
}%
{\color[rgb]{0,0,0}\put(4576,-961){\line( 0,-1){300}}
}%
{\color[rgb]{0,0,0}\put(5026,-961){\line( 0,-1){300}}
}%
{\color[rgb]{0,0,0}\put(3676,-961){\line( 0,-1){300}}
}%
{\color[rgb]{0,0,0}\put(5476,-961){\line( 0,-1){300}}
}%
{\color[rgb]{0,0,0}\put(5926,-961){\line( 0,-1){300}}
}%
\put(4801,-1711){\makebox(0,0)[b]{\smash{\SetFigFont{6}{7.2}{\rmdefault}{\mddefault}{\updefault}{\color[rgb]{0,0,0}based on past }%
}}}
\put(4801,-1861){\makebox(0,0)[b]{\smash{\SetFigFont{6}{7.2}{\rmdefault}{\mddefault}{\updefault}{\color[rgb]{0,0,0}channel outputs}%
}}}
\put(4351,-436){\makebox(0,0)[b]{\smash{\SetFigFont{6}{7.2}{\rmdefault}{\mddefault}{\updefault}{\color[rgb]{0,0,0}$\Omega$}%
}}}
\put(4351,-661){\makebox(0,0)[b]{\smash{\SetFigFont{6}{7.2}{\rmdefault}{\mddefault}{\updefault}{\color[rgb]{0,0,0}width of bin}%
}}}
\put(4801,-2236){\makebox(0,0)[lb]{\smash{\SetFigFont{6}{7.2}{\rmdefault}{\mddefault}{\updefault}{\color[rgb]{0,0,0}letter encoded by $F(b_t)$}%
}}}
\put(4351,-2686){\makebox(0,0)[rb]{\smash{\SetFigFont{6}{7.2}{\rmdefault}{\mddefault}{\updefault}{\color[rgb]{0,0,0}applied by controller}%
}}}
\put(5851,-3586){\makebox(0,0)[rb]{\smash{\SetFigFont{6}{7.2}{\rmdefault}{\mddefault}{\updefault}{\color[rgb]{0,0,0}Control signal calculated from delayed control }%
}}}
\put(2101,-736){\makebox(0,0)[rb]{\smash{\SetFigFont{6}{7.2}{\rmdefault}{\mddefault}{\updefault}{\color[rgb]{0,0,0}Adjustment at decoder}%
}}}
\put(2101,-886){\makebox(0,0)[rb]{\smash{\SetFigFont{6}{7.2}{\rmdefault}{\mddefault}{\updefault}{\color[rgb]{0,0,0}computed based on past}%
}}}
\put(2101,-1036){\makebox(0,0)[rb]{\smash{\SetFigFont{6}{7.2}{\rmdefault}{\mddefault}{\updefault}{\color[rgb]{0,0,0}channel output feedback}%
}}}
\put(4801,-1186){\makebox(0,0)[b]{\smash{\SetFigFont{6}{7.2}{\rmdefault}{\mddefault}{\updefault}{\color[rgb]{0,0,0}0}%
}}}
\put(4801,-1561){\makebox(0,0)[b]{\smash{\SetFigFont{6}{7.2}{\rmdefault}{\mddefault}{\updefault}{\color[rgb]{0,0,0}Decoding regions}%
}}}
\put(4951,-286){\makebox(0,0)[lb]{\smash{\SetFigFont{6}{7.2}{\rmdefault}{\mddefault}{\updefault}{\color[rgb]{0,0,0}Correct answer: $b_t = -1$}%
}}}
\put(3901,-1186){\makebox(0,0)[b]{\smash{\SetFigFont{6}{7.2}{\rmdefault}{\mddefault}{\updefault}{\color[rgb]{0,0,0}-2}%
}}}
\put(4351,-1186){\makebox(0,0)[b]{\smash{\SetFigFont{6}{7.2}{\rmdefault}{\mddefault}{\updefault}{\color[rgb]{0,0,0}-1}%
}}}
\put(5251,-1186){\makebox(0,0)[b]{\smash{\SetFigFont{6}{7.2}{\rmdefault}{\mddefault}{\updefault}{\color[rgb]{0,0,0}1}%
}}}
\put(5701,-1186){\makebox(0,0)[b]{\smash{\SetFigFont{6}{7.2}{\rmdefault}{\mddefault}{\updefault}{\color[rgb]{0,0,0}2}%
}}}
\put(6001,-961){\makebox(0,0)[lb]{\smash{\SetFigFont{6}{7.2}{\rmdefault}{\mddefault}{\updefault}{\color[rgb]{0,0,0}$|{\cal B}|=5$}%
}}}
\put(6001,-1186){\makebox(0,0)[lb]{\smash{\SetFigFont{6}{7.2}{\rmdefault}{\mddefault}{\updefault}{\color[rgb]{0,0,0}five possible}%
}}}
\put(6001,-1411){\makebox(0,0)[lb]{\smash{\SetFigFont{6}{7.2}{\rmdefault}{\mddefault}{\updefault}{\color[rgb]{0,0,0}output letters}%
}}}
\put(5851,-2686){\makebox(0,0)[lb]{\smash{\SetFigFont{6}{7.2}{\rmdefault}{\mddefault}{\updefault}{\color[rgb]{0,0,0}Adjusted by $-\lambda(U_{t-1} - U'_{t-1})$}%
}}}
\put(4351,-2461){\makebox(0,0)[rb]{\smash{\SetFigFont{6}{7.2}{\rmdefault}{\mddefault}{\updefault}{\color[rgb]{0,0,0}Actual control $U'_t$}%
}}}
\put(4426,314){\makebox(0,0)[b]{\smash{\SetFigFont{6}{7.2}{\rmdefault}{\mddefault}{\updefault}{\color[rgb]{0,0,0}$X_{t+1} - \lambda X_{t} = U'_{t} + W_{t}$}%
}}}
\put(5851,-2911){\makebox(0,0)[lb]{\smash{\SetFigFont{6}{7.2}{\rmdefault}{\mddefault}{\updefault}{\color[rgb]{0,0,0}at controller to compensate}%
}}}
\end{picture}
\end{center}
\caption{How to communicate the channel outputs through the plant with
state observations only. The controller restricts its main control
signal to be calculated with an extra delay of $1$ time unit and then
adjusts it by $-\lambda (U_{t-1} - U'_{t-1})$ to eliminate the effect
of the past communication. The final control signal applied is shifted
slightly to encode which $b_t$ was received. The decoder uses the past
$b_0^{t-1}$ to align its decoding regions and then reads off $b_t$ by
using $X_{t+1} - \lambda X_t$.} \label{fig:latticedecoder} 
\end{figure}
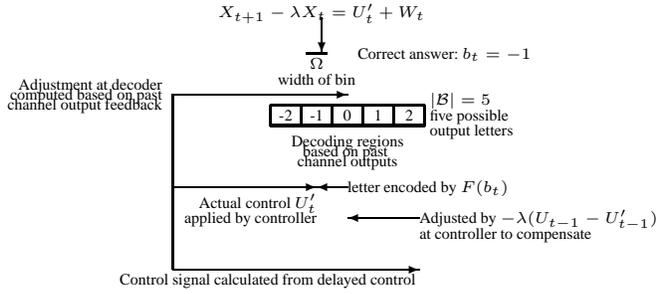

\begin{theorem} \label{thm:moment_scalar_control_sufficiency_without_feedback_with_noise}
Given a noisy channel with a countable alphabet, identify the channel
output alphabet with the integers and suppose that there
exist $K > 0, \beta > \eta$ so that the channel outputs $B_t$ satisfy:
${\cal P}(|B_t| \geq i) \leq K i^{-\beta}$ for all $t$ regardless of
the channel inputs.

Then, it is possible to control an unstable scalar plant driven by a
bounded disturbance over that channel so that the $\eta$-moment of
$|X_t|$ stays finite for all time if the channel has feedback anytime
capacity $C_{\mbox{any}}(\alpha) \geq \log_2 \lambda$ for some $\alpha
> \eta \log_2 \lambda$ even if the observer is only allowed to observe
the state $X_t$ corrupted by bounded noise.
\end{theorem}
{\em Proof: } The overall strategy is illustrated in
Figure~\ref{fig:plantdancing}. The channel output extraction at the
observer is illustrated in Figure~\ref{fig:latticedecoder} in the
context of a channels with output alphabet size $|{\cal B}| = 5$.

Let $U_t(b_0^{t-1})$ be the control that would be
applied from Theorem~\ref{thm:delayed_sufficiency} as transformed by
the action of Theorem~\ref{thm:control_noise_impact} if necessary. It
only depends on the strictly past channel outputs.

Let $b_t$ be the current channel output. The control applied is: 
\begin{equation} \label{eqn:countablecontrolcase}
U'_t(b_0^t) = U_t(b_0^{t-1}) + F(b_t)
-\lambda\left(U'_{t-1}(b_0^{t-1}) - U_{t-1}(b_0^{t-2}) \right)
\end{equation}
where the function $F(b_t)$ is the ``dance move'' corresponding to the
channel output. 

First consider the case that perfect state observations $X_t$ are
available at observer. At time $t$ the observer can see the control
signal only as it is corrupted by the process disturbance since
$U_{t-1} + W_{t-1} = X_{t} - \lambda X_{t-1}$. By observing $X$
perfectly, the observer has in effect gained boundedly noisy access to
the $U$ with $\Gamma_u = \Omega$. Now suppose that the observations of
$X$ were boundedly noisy with some $\Gamma$. In that case:
\begin{eqnarray*}
& &  |U_{t-1} - (X_{noisy}(t) - \lambda X_{noisy}(t-1))| \\
 & = &
 |U_{t-1} - (X_t - \lambda X_{t-1}) + (\lambda N_{t-1} - N_t)| \\
 & = &  
 |-W_{t-1} + (\lambda N_{t-1} - N_t)| \\
 & \leq & \Omega + (\lambda + 1)\Gamma
\end{eqnarray*}
In this case, the effective observation noise on the controls is
bounded by $\Gamma_u = \Omega + (\lambda + 1)\Gamma$. 

Just by looking at the state and its history, the observer has access
to $U^o_t$ with the property that $|U'_t - U^o_t| \leq \Gamma_u$. To
ensure decodability of $b_t$, set $F(b_t) = 3 \Gamma_u b_t$ so the
channel outputs are modulated to be integer multiples of $3\Gamma_u$.

At time $t=0$, the observer is unchanged since there is nothing for it
to learn and no applied controls. At time $t=1$, because of the
induced delay of $1$ extra time step, there are no delayed controls
ready to apply either and so the applied control only consists of $3
\Gamma_u b_0$. This is observed up to precision $\Gamma_u$ and so the
observer can uniquely recover $b_0$ and feed it to its anytime
encoder. 

Assume now that the observer was successful in learning $b_0^{t-1}$ in
the past. Then it can compute the $U_t(b_0^{t-1})$ term as well as the
$U'_{t-1}(b_0^{t-2}) - U_{t-1}(b_0^{t-1})$ using this knowledge and
can subtract both of them from its observed $U^o_t$. This leaves only
the $3 \Gamma_u b_t$ term which can be uniquely decoded given that the
observation noise is no more than $\Gamma_u$ in either direction. By
induction, the observer can effectively recover the past channel
outputs from its noiseless observations of the control signal and can
thereby operate the feedback anytime-encoder successfully.

The communication of each channel output $b_t$ only impacts the very
next state by shifting it by $3 \Gamma_u b_t$. At the next time, it is
canceled out by the correction term $-\lambda\left(U'_{t-1}(b_0^{t-1})
- U_{t-1}(b_0^{t-2})\right)$. The non-dancing controlled state
$X'_{t+1} = (X_{t+1} - 3 \Gamma_u B_t)$ has at least a power-law tail
${\cal P}(X'_{t+1} \geq x) \leq K' x^{-(\eta + \epsilon)}$ for some
$K'$ and $\epsilon > 0$. Then $E[|X|^\eta] = $
\begin{eqnarray*} 
& = &  \int_0^\infty 
{\cal P}(|X' + 3\Gamma_u B| \geq m^{\frac{1}{\eta}}) dm \\
& \leq &  \int_0^\infty 
{\cal P}\left(2\max(|X'|, 3\Gamma_u |B|) \geq m^{\frac{1}{\eta}}\right)
dm \\
& \leq &  \int_0^\infty 
{\cal P}\left(|X'| \geq \frac{1}{2}m^{\frac{1}{\eta}}\right) dm +  \\
& & + {\cal P}\left(|B| \geq \frac{1}{6
  \Gamma_u}m^{\frac{1}{\eta}}\right) dm \\
& \leq &  \int_0^\infty 
K' \left(\frac{1}{2}m^{\frac{1}{\eta}}\right)^{-(\eta + \epsilon)}
+ K \left(\frac{1}{6 \Gamma_u}m^{\frac{1}{\eta}}\right)^{-\beta} dm
\end{eqnarray*}
Since $\beta > \eta$, this converges and so the $\eta$-moment of $X$
also exists. \hfill $\Box$

The channel output condition in
\ref{thm:moment_scalar_control_sufficiency_without_feedback_with_noise}
is clearly satisfied whenever the channel has a finite output
alphabet. Beyond that case, it is satisfied in generic situations when
the input alphabet is finite and the transition probabilities $p(b|a)$
individually have an light enough tail for each one of the finite $a$
values.\footnote{For example, an AWGN channel with a hard-input
constraint and quantized outputs.} When the channel input alphabet is
itself countable, the condition is harder to check.

If information must flow noiselessly from the controller to the
observer, the key question is to quantify the instantaneous zero-error
capacity of the effective channel {\em through the plant.} Here, the
bounded support of $W$ and the unconstrained nature of $U$ are
critical since they allow the instantaneous zero-error capacity of
that effective channel to be infinite. Of course, there remains the
problem of the dual-nature of the control signal --- it is
simultaneously being asked to stabilize the plant as well as to
feedback information about the channel outputs. The theorem shows that
the ability of the controller to move the plant provides enough
feedback to the encoder in the case of finite channel output alphabets
or channels with uniformly exponentially bounded output statistics.

At an abstract level, the controller is faced with the problem of
causal ``writing on dirty paper''\cite{costadirtypaper} where the
information it wishes to convey in one time step is the channel output
and the dirty paper consists of the control signals it must apply to
keep the system stable and to counteract the effect of the writing it
did in previous time steps. Here, the problem is finessed by
introducing the artificial delay at the controller to ensure that the
``dirt'' is side-information known both to the transmitter and the
receiver. For finite output alphabets, it is also possible to take a
direct ``precoding'' approach to do this by encoding the channel
outputs by placing the control to the appropriate value modulo $3
\Gamma_u (|{\cal B}| + 1)$.  This is a bounded perturbation of the
control inputs and Theorem~\ref{thm:control_noise_impact} tells us
that this does not break stability if the $\Delta$ is adjusted
appropriately. 

Finally, it might seem that this particular ``dance'' by the plant
will be a disaster for performance metrics beyond stabilization. This
is probably true, but we conjecture that such implicit feedback
through the plant will be usable without much loss of performance. If
it has memory, the observer can notice when and how the channel has
misbehaved since the plant's state will start growing rather than
staying near $0$. The $\Delta$-lattice based quantizer used in the
observer for Theorem~\ref{thm:anytimenofeedbacknocontrolsDMC} could
not exploit this because it was memoryless and used uniformly sized
bins regardless of whether the state was large or small.

\section{Continuous time systems} \label{sec:continuous}
\subsection{Overview}
So far, we have considered a discrete-time model
(\ref{eqn:discretesystem}) for the dynamic system that must be
stabilized over the communication link. This has simplified the
discussion by having a common clock that drives both the system and
the uses of the noisy channel. In general, there will be a $\tau_c$
that represents the time between channel uses. This allows translating
everything into absolute time units.

\begin{equation} \label{eqn:continuoussystem} 
\dot{X}(t) = \lambda X(t) + U(t) + W(t), \ \ t \geq 0
\end{equation} 
where the bounded disturbance $|W(t)| \leq \frac{\Omega}{2}$ and there
is a known initial condition $X(0)=0$. If the open-loop system is
unstable, then $\lambda > 0$.

Sampling can be used to extend both the necessity and sufficiency
results to the continuous time case. The basic result is that
stability requires an anytime capacity greater than $\lambda$ nats per
second.

\subsection{Necessity}
For necessity, we are free to choose the disturbance signal $W(t)$ and
consequently can restrict ourselves to piecewise constant
signals\footnote{zero order hold} that stay constant for time
$\tau$. By sampling at the rate $\frac{1}{\tau}$, the sampled state
evolves as
$X(\tau(i+1)) =$
\begin{equation}  \label{eqn:sampledcontinuous} 
e^{\lambda\tau}X(\tau i) + (\frac{e^{\lambda\tau} - 1}{\lambda} W_i) +
\int_{i\tau}^{(i+1)\tau} U(s) e^{\lambda(\tau(i+1) - s)} ds
\end{equation}
Notice that (\ref{eqn:sampledcontinuous}) is just a discrete time
system with $\lambda' = e^{\lambda\tau}$ taking the role of $\lambda$ in
(\ref{eqn:discretesystem}), and the disturbance is bounded by
$\Omega' = \frac{\Omega (e^{\lambda\tau} - 1)}{\lambda}$. All that
remains is to reinterpret the earlier theorem. 

By setting $\tau = \tau_c$ to match up the sampling times to the
channel use times, it is clear that the appropriate anytime capacity
must exceed $\log_2 \lambda' = \tau_c \lambda \log_2 e$ bits per
channel use. By converting units to nats per second\footnote{Assuming
that $\dot{X}$ is in per second units.}, we get the intuitively
appealing result that the anytime capacity must be greater that
$\lambda$ nats/sec.\footnote{This truly justifies nats as the
``natural'' unit of information!} Similarly, to hold the $\eta$-th
moment constant, the probability of error must drop with delay faster
than $K2^{-(\eta \lambda \log_2 e)d\tau_c}$ where $d$ is in units of
channel uses and thus $d\tau_c$ has units of seconds. Thus, we get the
following pair of theorems:

\begin{theorem} \label{thm:continuous_control_necessity}
For a given noisy channel and $\eta > 0$, if there exists an observer
${\cal O}$ and controller ${\cal C}$ for the unstable scalar
continuous time system that achieves $E[|X(t)|^\eta] < K$ for all $t$
and bounded driving noise signals $|W(t)| \leq \frac{\Omega}{2}$, then
the channel's feedback anytime capacity $C_{\mbox{any}}(\eta \lambda
\log_2 e) \geq \lambda$ nats per second.
\end{theorem}

\begin{theorem} \label{thm:general_continuous_control_necessity}
For a given noisy channel and decreasing function $f(m)$, if
there exists an observer ${\cal O}$ and controller ${\cal C}$ for the
unstable continuous-time scalar system that achieves ${\cal P}(|X(t)|
> m) < f(m)$ for all $t$ and all bounded driving noise signals $|W(t)|
\leq \frac{\Omega}{2}$, then $C_{\mbox{g-any}}(g) \geq \lambda$ nats per
second for the noisy channel considered with the encoder having access
to noiseless feedback and $g(d)$ having the form $g(d) = f(K e^{\lambda d})$
for some constant $K$.
\end{theorem}

\subsection{Sufficiency}

For sufficiency, the disturbance is arbitrary but we are free to
sample the signal as desired at the observer and apply piecewise
constant control signals. Sampling every $\tau$ units of time gives
rise to (\ref{eqn:sampledcontinuous}) only with the roles of $W$ and
$U$ reversed. It is clear that $W_i = \int_{i\tau}^{(i+1)\tau} W(s)
e^{\lambda(\tau(i+1) - s)} ds$ is still bounded by substituting in the
upper and lower bounds and then noticing that $|W_i| \leq \frac{\Omega
(e^{\lambda\tau} -1)}{2\lambda}$.

Thus, the same argument above holds and the sufficiency
Theorems~\ref{thm:general_scalar_control_sufficiency},
\ref{thm:moment_scalar_control_sufficiency}, and
\ref{thm:moment_scalar_control_sufficiency_without_feedback_with_noise}
as well as
Corollaries~\ref{cor:moment_scalar_control_sufficiency_stronger} and
\ref{cor:moment_scalar_control_sufficiency_weak} translate cleanly
into continuous time. In each, the relevant anytime capacity must be
greater than $\lambda$ nats per second. Since the necessary and
sufficient conditions are right next to each other, it is clear that
the choice of sampling time does not impact the sense of stability
that can be achieved. Of course, this need not be optimal in terms of
performance.

Finally, if the channel we face is an input power-constrained
$\infty$-bandwidth AWGN channel, more can be
said. Section~\ref{sec:awgncasewithfeedback} makes it clear that
nothing special is required in this case: using linear controllers and
observers is good enough if the average power constraint is high
enough. But what if the channel had a hard amplitude constraint
that allowed the encoder no more than $P$ power per unit time?  In
this case, it is possible to generalize
Theorem~\ref{thm:anytimenofeedbacknocontrolsDMC} in an interesting
way.

In \cite{SahaiSeqPPM} we give an explicit construction of a
feedback-free anytime code for the infinite bandwidth AWGN channel
that uses a sequential form of orthogonal signaling. In the
$\infty$-bandwidth AWGN channel, pairwise orthogonality between
codewords plays the role that pairwise independence does for
DMCs. Applying that principle through the proof of
Theorem~\ref{thm:anytimenofeedbacknocontrolsDMC}, the observer/encoder
can simply be a {\em time-invariant} regular partition of the state
space with the bins being labeled with orthogonal pulses, each with an
energy equal to the hard limit for the channel.\footnote{In
particular, the following sequence of pulses work with an appropriate
scaling. For $0 \leq t \leq \tau$, set $g_{i,\tau}(t) =
\frac{1}{\tau} \mbox{sgn}\left(\sin(\frac{4\pi i}{\tau} t)\right)$
and $g_{-i,\tau}(t) = \frac{1}{\tau} \mbox{sgn}\left(\sin(\frac{2\pi
(2i-1)}{\tau} t)\right)$ and zero everywhere else. Here $\tau$ is the
time between taking samples of the state. The $g_{i,\tau}$ functions
are orthogonal, and the $i$-th function is the channel input
corresponding to the $i$-th lattice bin for the plant state
observation.} The encoder just pieces together pulses with shapes
corresponding to where the state is at the sampling times. The
controller then searches for the most likely path based on the channel
output signal as well as the past control values, and then applies a
control based on the current estimate. This approach allows the use of
occasional bandwidth expansion to deal with unlucky streaks of channel
noise while keeping the channel input power constant. The details of
this approach are given in \cite{SahaiMemorylessControl}.

\section{A Hierarchy Of Communication Problems}
\label{sec:hierarchy}

In this final section, we interpret some of the results in a different
way inspired by the approach used in computational complexity
theory. There, the scarce resource is the time and space available for
computation and the asymptotic question is whether or not a certain
family of problems (indexed by $n$) can be solved using the limited
amount of resource available. While explicit algorithms for solving
problems do play a role, ``reductions'' from one problem to another
also feature prominently in relating the resource requirements among
related problems \cite{Hopcraft}.

In communication, the scarce resource can be thought of as being the
available channel.\footnote{This might in turn be related to other
more primitive scarce resources like power or bandwidth available for
communication.} Problems should be ordered by what channels are good
enough for them. We begin with some simple definitions and then see
how they apply to classical results from information theory. Finally,
we interpret our current results in this framework.

\begin{definition}
A {\em communication problem} is a partially specified random system
together with an information pattern and a performance objective. This
is specified by a triple: $({\cal S}, {\cal I}, {\cal V})$. The
partially specified random system ${\cal S} = (S_0, S_1, \ldots)$ in
which $S_i$ are real valued functions on $[0,1]^{i+1} \times
\real^i$. The output of the $S_i$ function is denoted $X_i$. The
information pattern ${\cal I}$ identifies what variables each of the
$i$-th encoders and decoders has access to. The performance objective
${\cal V}$ is a statement that must evaluate to either true or false
once the entire random system is specified. 
\end{definition}
\vspace{0.1in}

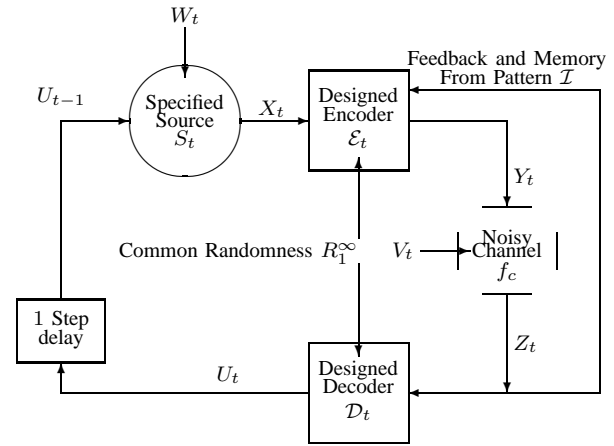
\begin{figure}
\begin{center}
\setlength{\unitlength}{2047sp}%
\begingroup\makeatletter\ifx\SetFigFont\undefined%
\gdef\SetFigFont#1#2#3#4#5{%
  \reset@font\fontsize{#1}{#2pt}%
  \fontfamily{#3}\fontseries{#4}\fontshape{#5}%
  \selectfont}%
\fi\endgroup%
\begin{picture}(7074,5268)(214,-4498)
{\color[rgb]{0,0,0}\thinlines
\put(2251,-586){\oval(1342,1342)}
}%
{\color[rgb]{0,0,0}\put(3751,-1186){\framebox(1200,1200){}}
}%
{\color[rgb]{0,0,0}\put(3751,-4486){\framebox(1200,1200){}}
}%
{\color[rgb]{0,0,0}\put(2926,-586){\vector( 1, 0){825}}
}%
{\color[rgb]{0,0,0}\put(2251,539){\line( 0,-1){525}}
\put(2251, 14){\vector( 0,-1){ 75}}
}%
{\color[rgb]{0,0,0}\put(4951,-586){\line( 1, 0){1200}}
\put(6151,-586){\vector( 0,-1){1050}}
}%
{\color[rgb]{0,0,0}\put(3751,-3886){\line(-1, 0){3000}}
\put(751,-3886){\vector( 0, 1){375}}
}%
{\color[rgb]{0,0,0}\put(751,-2761){\line( 0, 1){2175}}
\put(751,-586){\vector( 1, 0){825}}
}%
{\color[rgb]{0,0,0}\put(7276,-3886){\line( 0, 1){3675}}
\put(7276,-211){\vector(-1, 0){2325}}
}%
{\color[rgb]{0,0,0}\put(5101,-2161){\vector( 1, 0){600}}
}%
\put(2251,-661){\makebox(0,0)[b]{\smash{\SetFigFont{8}{14.4}{\rmdefault}{\mddefault}{\updefault}{\color[rgb]{0,0,0}Source}%
}}}
\put(2776,-3736){\makebox(0,0)[b]{\smash{\SetFigFont{8}{14.4}{\rmdefault}{\mddefault}{\updefault}{\color[rgb]{0,0,0}$U_t$}%
}}}
\put(2251,614){\makebox(0,0)[b]{\smash{\SetFigFont{8}{14.4}{\rmdefault}{\mddefault}{\updefault}{\color[rgb]{0,0,0}$W_{t}$}%
}}}
\put(4351,-3661){\makebox(0,0)[b]{\smash{\SetFigFont{8}{14.4}{\rmdefault}{\mddefault}{\updefault}{\color[rgb]{0,0,0}Designed}%
}}}
\put(4351,-361){\makebox(0,0)[b]{\smash{\SetFigFont{8}{14.4}{\rmdefault}{\mddefault}{\updefault}{\color[rgb]{0,0,0}Designed}%
}}}
\put(2251,-451){\makebox(0,0)[b]{\smash{\SetFigFont{8}{14.4}{\rmdefault}{\mddefault}{\updefault}{\color[rgb]{0,0,0}Specified}%
}}}
\put(2251,-886){\makebox(0,0)[b]{\smash{\SetFigFont{8}{14.4}{\rmdefault}{\mddefault}{\updefault}{\color[rgb]{0,0,0}$S_t$}%
}}}
\put(4351,-586){\makebox(0,0)[b]{\smash{\SetFigFont{8}{14.4}{\rmdefault}{\mddefault}{\updefault}{\color[rgb]{0,0,0}Encoder}%
}}}
\put(4351,-886){\makebox(0,0)[b]{\smash{\SetFigFont{8}{14.4}{\rmdefault}{\mddefault}{\updefault}{\color[rgb]{0,0,0}${\cal E}_t$}%
}}}
\put(6151, 89){\makebox(0,0)[b]{\smash{\SetFigFont{8}{14.4}{\rmdefault}{\mddefault}{\updefault}{\color[rgb]{0,0,0}Feedback and Memory}%
}}}
\put(4351,-3886){\makebox(0,0)[b]{\smash{\SetFigFont{8}{14.4}{\rmdefault}{\mddefault}{\updefault}{\color[rgb]{0,0,0}Decoder}%
}}}
\put(751,-3061){\makebox(0,0)[b]{\smash{\SetFigFont{8}{14.4}{\rmdefault}{\mddefault}{\updefault}{\color[rgb]{0,0,0}$1$ Step}%
}}}
\put(751,-361){\makebox(0,0)[b]{\smash{\SetFigFont{8}{14.4}{\rmdefault}{\mddefault}{\updefault}{\color[rgb]{0,0,0}$U_{t-1}$}%
}}}
\put(4351,-4186){\makebox(0,0)[b]{\smash{\SetFigFont{8}{14.4}{\rmdefault}{\mddefault}{\updefault}{\color[rgb]{0,0,0}${\cal D}_t$}%
}}}
\put(6151,-2236){\makebox(0,0)[b]{\smash{\SetFigFont{8}{14.4}{\rmdefault}{\mddefault}{\updefault}{\color[rgb]{0,0,0}Channel}%
}}}
\put(6151,-2086){\makebox(0,0)[b]{\smash{\SetFigFont{8}{14.4}{\rmdefault}{\mddefault}{\updefault}{\color[rgb]{0,0,0}Noisy}%
}}}
{\color[rgb]{0,0,0}\put(7276,-3886){\vector(-1, 0){2325}}
}%
{\color[rgb]{0,0,0}\put(6151,-2686){\vector( 0,-1){1200}}
}%
{\color[rgb]{0,0,0}\put(226,-3511){\framebox(1050,750){}}
}%
{\color[rgb]{0,0,0}\put(4351,-2011){\vector( 0, 1){975}}
}%
{\color[rgb]{0,0,0}\put(4351,-2311){\vector( 0,-1){1125}}
}%
{\color[rgb]{0,0,0}\put(5851,-1636){\line( 1, 0){600}}
\put(6451,-1636){\line( 1,-1){300}}
\put(6751,-1936){\line( 0,-1){450}}
\put(6751,-2386){\line(-1,-1){300}}
\put(6451,-2686){\line(-1, 0){600}}
\put(5851,-2686){\line(-1, 1){300}}
\put(5551,-2386){\line( 0, 1){450}}
\put(5551,-1936){\line( 1, 1){300}}
}%
\put(3301,-511){\makebox(0,0)[b]{\smash{\SetFigFont{8}{14.4}{\rmdefault}{\mddefault}{\updefault}{\color[rgb]{0,0,0}$X_t$}%
}}}
\put(6151,-136){\makebox(0,0)[b]{\smash{\SetFigFont{8}{14.4}{\rmdefault}{\mddefault}{\updefault}{\color[rgb]{0,0,0}From Pattern ${\cal I}$}%
}}}
\put(751,-3286){\makebox(0,0)[b]{\smash{\SetFigFont{8}{14.4}{\rmdefault}{\mddefault}{\updefault}{\color[rgb]{0,0,0}delay}%
}}}
\put(5026,-2236){\makebox(0,0)[rb]{\smash{\SetFigFont{8}{14.4}{\rmdefault}{\mddefault}{\updefault}{\color[rgb]{0,0,0}$V_t$}%
}}}
\put(4351,-2236){\makebox(0,0)[rb]{\smash{\SetFigFont{8}{14.4}{\rmdefault}{\mddefault}{\updefault}{\color[rgb]{0,0,0}Common Randomness $R_1^\infty$}%
}}}
\put(6226,-1336){\makebox(0,0)[lb]{\smash{\SetFigFont{8}{14.4}{\rmdefault}{\mddefault}{\updefault}{\color[rgb]{0,0,0}$Y_t$}%
}}}
\put(6226,-3361){\makebox(0,0)[lb]{\smash{\SetFigFont{8}{14.4}{\rmdefault}{\mddefault}{\updefault}{\color[rgb]{0,0,0}$Z_t$}%
}}}
\put(6151,-2486){\makebox(0,0)[b]{\smash{\SetFigFont{8}{14.4}{\rmdefault}{\mddefault}{\updefault}{\color[rgb]{0,0,0}$f_c$}%
}}}
\end{picture}
\end{center}
\caption{Abstractly, a communication problem consists of a partially
specified random system consisting of a known and possibly interactive
source together with an information pattern. The noisy channel and
encoder/decoders need to be specified before all the random variables
become properly defined.}
\label{fig:abstract_problem}
\end{figure}

As depicted in Figure~\ref{fig:abstract_problem}, the communication
problem is thus an open system that awaits interconnection with
encoder, channel, and decoder maps. The channel is a 
measurable map $f_c$ from $[0,1] \times \real$ into
$\real$. The encoder and decoder are both represented by a possibly
time-varying sequence of real valued functions compatible with the
information pattern ${\cal I}$.

Once all the maps are specified, the random system becomes completely
specified by tying them to an underlying probability space consisting
of three iid sequences $(W_i, V_i, R_i)$ of continuous uniform random
variables on $[0,1]$. The $W_1^i$ are connected to the first input of
$S_i$ while $V_i$ is connected to the first input of the memoryless
channel. As is usual, the output of the encoder is connected to the
remaining input of the channel, and all the past outputs of the
channel are connected to the decoding functions as per the information
patterns.  Finally, assume that common randomness $R_i$ is made
available to both the encoder and decoder so that they may do random
coding if desired. Once everything is connected, it is possible to
evaluate the truth or falsehood of ${\cal V}$.

\begin{definition}
A channel is said to {\em solve} the problem if there exist suitable
encoder and decoder maps compatible with the given information pattern
so that the combined random system satisfies the performance objective
${\cal V}$. 

Communication problem $A$ is {\em harder than} problem $B$ if any
channel $f_c$ that solves $A$ also solves $B$.
\end{definition}
\vspace{0.1in}

Each particular communication problem therefore divides channels into
two classes: those that solve it and those that do not. Suitable
families of communication problems, ordered by hardness, can then be
used to sort channels as well. Channels that solve harder problems are
better than ones that do not. The equivalence of certain families of
communication problems means that they induce the same orderings on
communication channels. This will become clearer by the examples of
the next few sections.

\subsection{Classical Examples}

\subsubsection{The Shannon communication problem}

Shannon identified the problem of communicating bits reliably as one
of the core problems of communication. In our framework, this problem 
is formalized as follows:
\begin{itemize}
 \item $X_i = 1$ if $W_i > \frac{1}{2}$ and $X_i = 0$ otherwise. The
 functions $S_i$ ignore all other inputs.

 \item The information pattern ${\cal I}$ specifies that ${\cal D}_i$
 has access to $Z_1^i$. The encoder information pattern is complete in
 the case of communication with feedback: ${\cal E}_i$ has access to
 $X_1^i$ as well as $Z_1^{i-1}$. Without feedback, ${\cal E}_i$ has
 access only to $X_1^i$. 

 \item The performance objective ${\cal V}(\epsilon, d)$ is satisfied
 if ${\cal P}(X_i \neq U_{i+d}) \leq \epsilon$ for every $i \geq 0$. 
\end{itemize}

The Shannon communication problem naturally comes in a pair of
families $A^f_{\epsilon, d}$ with feedback and $A^{nf}_{\epsilon, d}$
without feedback. These families are indexed by the tolerable
probability of bit error $\epsilon$ and end-to-end delay $d$.

To obtain other rates $R > 0$, adjust the source functions as
follows:
\begin{itemize}
 \item $X_i = \frac{j}{2^{\lfloor Ri \rfloor - \lfloor
 R(i-1)\rfloor}}$ if $W_i \in [\frac{j}{2^{\lfloor Ri \rfloor - \lfloor
 R(i-1)\rfloor}}, \frac{j+1}{2^{\lfloor Ri \rfloor - \lfloor
 R(i-1)\rfloor}})$ for integer $j \geq 0$. The possibly time-varying
 functions $S_i$ ignore all other inputs.
\end{itemize}

These naturally result in families $A^f_{R, \epsilon, d}$ and
$A^{nf}_{R, \epsilon, d}$ for the feedback and feedback-free cases
respectively. It is immediately clear that $A^{nf}_{R, \epsilon, d}$
is harder than $A^{f}_{R, \epsilon, d}$ and furthermore problems with
smaller $\epsilon$ or $d$ are harder than those with larger ones. It
is also true that $A^{f}_{R, \epsilon, d}$ is harder than $A^{f}_{R',
  \epsilon, d}$ whenever $R \leq R'$ in that it is more challenging to
communicate reliably at a high rate rather than a low one. 

The set of channels with classical Shannon feedback capacity of at
least $R$ is therefore:
\begin{equation} \label{eqn:channel_capacity_constrained}
{\cal C}^f_R =  \bigcap_{\epsilon > 0} \bigcap_{R' < R} \bigcup_{d > 0} \{f_c | f_c \mbox{
  solves } A^f_{R',\epsilon, d} \}
\end{equation}
and similarly for ${\cal C}^{nf}_R$. The classical result that
feedback does not increase capacity tells us that ${\cal C}^f_R =
{\cal C}^{nf}_R$. Because of this, we just call them both ${\cal
  C}_R$. 

\subsubsection{The zero-error communication problem}
A second problem is the one of zero error communication. It is defined
exactly the same as the Shannon communication problem above, except
that $\epsilon = 0$.

The channels that have feedback zero-error capacity of at least $R$
with feedback are therefore:
\begin{equation} \label{eqn:zero_capacity_constrained}
{\cal C}^f_{0,R} = \bigcap_{R' < R} \bigcup_{d > 0} \{f_c | f_c \mbox{ solves } A^f_{R',0, d} \}
\end{equation}
and similarly for ${\cal C}^{nf}_{0,R}$. In this case, the result with
and without feedback can be different and furthermore, ${\cal
C}^{nf}_{0,R} \subset {\cal C}^f_{0,R} \subset {\cal C}_R$
\cite{ShannonZeroError}. In this sense, zero-error communication is
fundamentally a harder problem than $\epsilon$-error communication.

\subsubsection{Estimation problems with distortion constraints}
Consider iid real valued sources with cumulative distribution
functions $F_X(t) = {\cal P}(X \leq t)$.

\begin{itemize}
 \item $X_i = F^{-1}_X(W_i)$ ignoring all the other inputs. This gives
 the desired source statistics.

 \item The information patterns remain as in the Shannon problem.

 \item The performance objective ${\cal V}(\rho, D, d)$ is satisfied
 if $\lim_{n \rightarrow \infty} \frac{1}{n} E[\sum_{i=1}^n
 \rho(X_i,U_{i+d})] \leq D$. 
\end{itemize}

Call these estimation problems $A^f_{(F_X, \rho, D, d)}$ and
$A^{nf}_{(F_X, \rho, D, d)}$ (for the cases with/without feedback) and
once again associate them with the set of channels that solve them in
the limit of large delays:

\begin{equation} \label{eqn:channel_distortion_constrained}
{\cal C}^f_{e,(F_X, \rho, D)} =  \bigcap_{D' > D} \bigcup_{d > 0} \{f_c | f_c \mbox{
  solves } A^f_{(F_X, \rho, D', d)} \}
\end{equation}
and similarly for ${\cal C}^{nf}_{e,(F_X, \rho, D)}$. For cases where
the distortion $\rho$ is bounded, the existing separation result can
be interpreted as follows:
\begin{equation} \label{eqn:classical_separation_result}
{\cal C}_{R(D)} = {\cal C}^{nf}_{e,(F_X, \rho, D)} = {\cal
  C}^{f}_{e,(F_X, \rho, D)}
\end{equation}
where $R(D)$ is the information-theoretic rate-distortion curve. 

The interpretation of this separation theorem is that in the limit of
large delays, estimation problems with a fidelity constraint are no
harder or easier than Shannon communication problems dealing with
bits. Both families of problems induce essentially the same partial
order on channels.

\subsection{Anytime communication problems}
The anytime communication problems are natural generalizations of the
binary data communication problems above. Everything remains as in the
Shannon communication problem, only the performance measure
changes. Let $U_t = 0.\widehat{X}_0(t),\widehat{X}_1(t),\widehat{X}_2(t),\ldots$
when written out in binary notation. This can always be done and the
parsing of the string is unique no matter what the rate is. 

\begin{itemize}
 \item ${\cal V}_{(K, \alpha)}$ is satisfied if ${\cal P}(X_i \neq
 \widehat{X}_i(i+d)) \leq K 2^{-\alpha d}$ for every $i \geq 0, d \geq
 0$. 
\end{itemize}

Call these problems $A^f_{(R, \alpha, K)}$ when feedback is allowed
and $A^{nf}_{(R, \alpha, K)}$ when it is not permitted. Once again, it
is clear that the non-feedback problems are harder than the
corresponding feedback problems. Furthermore, $A_{(R, \alpha, K)}$ is
harder than $A_{(R, \alpha', K)}$ if $\alpha' \leq \alpha$ in addition
to the usual fact of $A_{(R, \alpha, K)}$ being harder than $A_{(R',
  \alpha, K)}$ if $R' \leq R$.  Similarly, smaller $K$ values are
harder than larger ones. 

The channels with $\alpha$-anytime feedback capacity of at least $R$
are then given by:
\begin{equation} \label{eqn:anytime_capacity_constrained}
{\cal C}^f_{a,(R,\alpha)} =  \bigcap_{R' < R} \bigcap_{\alpha' < \alpha}
  \bigcup_{K > 0} \{f_c | f_c \mbox{ solves } A^f_{(R',\alpha', K)} \}
\end{equation}
with a similar definition for ${\cal C}^{nf}_{a,(R,\alpha)}$. It is
immediately clear that 
$${\cal C}^f_{0,R} \subseteq {\cal C}^f_{a,(R,\alpha)} \subseteq {\cal
  C}_R$$
The case of $\alpha = 0$ is defined as the limit:
\begin{equation} \label{eqn:anytime_limit__capacity_constrained}
{\cal C}^f_{a,(R,0)} =  \bigcup_{\alpha > 0} {\cal C}^f_{a,(R,\alpha)}
\end{equation}
It turns out in this case that ${\cal C}^f_{a,(R,0)} = {\cal
C}^{nf}_{a,(R,0)} = {\cal C}_R$ since infinite random tree codes can
be used to communicate reliably at all rates below the Shannon
capacity \cite{OurSourceCodingPaper}.

However, for other $\alpha > 0$, 
$${\cal C}^{nf}_{0,R} \subset {\cal C}^{nf}_{a,(R,\alpha)} \subset
{\cal C}^{f}_{a,(R,\alpha)} \subset {\cal C}_R$$ 
and
$${\cal C}^{nf}_{0,R} \subset {\cal C}^{f}_{0,R} \subset {\cal
C}^{f}_{a,(R,\alpha)} \subset {\cal C}_R$$ with all of these being
strict inclusion relations. ${\cal C}^f_{0,R}$ and ${\cal
C}^{nf}_{a,(R,\alpha)}$ are not subsets of each other in general.

In this sense, there is a non-trivial hierarchy of problems with
Shannon communication as the easiest example and zero-error
communication as the hardest.

\subsection{Control and the relation to anytime communication}

The stabilization problems considered in this paper are different in
that they are interactive. The formulation should be apparent by
comparing Figure~\ref{fig:abstract_problem} with
Figure~\ref{fig:problem}.
\begin{itemize}
 \item $X_i$ represents the state of the scalar control problem with
 unstable system dynamics given by $\lambda > 1$. The $W_t$ is the
 bounded disturbance and $U_i$ represents the control signal used to
 generate $X_{i+1}$. 

 \item The information pattern with and without feedback is as before.

 \item The performance objective ${\cal V}_{(\eta, K)}$ is satisfied
 if $E[|X_i|^\eta] \leq K$ for all $i \geq 0$. 
\end{itemize}

Call this problem $A^{f}_{(\lambda, \eta, K)}$ for cases with feedback
and $A^{nf}_{(\lambda, \eta, K)}$ for cases without feedback available
at the encoder. The problem without feedback is harder than the
problem with feedback. It is also clear that $A^{f}_{(\lambda, \eta,
K)}$ is harder than $A^{f}_{(\lambda', \eta, K)}$ whenever $\lambda
\geq \lambda'$ and similarly for $A^{nf}$. The same holds if $\eta$ is
made larger or $K$ is made smaller.

\begin{equation} \label{eqn:control_capacity_constrained}
{\cal C}^f_{c,(\lambda,\eta)} =  \bigcap_{\lambda' < \lambda} 
                                 \bigcap_{\eta' < \eta} 
                                 \bigcup_{K > 0} 
 \{f_c | f_c \mbox{ solves } A^f_{(\lambda',\eta', K)} \}
\end{equation}
with a similar definition for ${\cal C}^{nf}_{c,(\lambda, \eta)}$. 
The necessity result of Theorem~\ref{thm:scalar_control_necessity}
establishes that 
$$
{\cal C}^{nf}_{c,(\lambda, \eta)} \subseteq 
{\cal C}^f_{c,(\lambda,  \eta)} \subseteq 
{\cal C}^f_{a,(\log_2 \lambda, \eta \log_2 \lambda)}
$$
while Theorem~\ref{thm:moment_scalar_control_sufficiency} establishes
the other direction for the case of feedback:
\begin{equation} \label{eqn:equivalence_in_channel_terms}
{\cal C}^{nf}_{c,(\lambda, \eta)} \subseteq 
{\cal C}^f_{c,(\lambda,  \eta)}  = 
{\cal C}^f_{a,(\log_2 \lambda, \eta \log_2 \lambda)}
\end{equation}

Meanwhile without feedback and restricting to the set of finite output
alphabet channels (ie.~where the range of $f_c$ has finite
cardinality.) denoted ${\cal C}_{\mbox{fin}}$,
Theorem~\ref{thm:moment_scalar_control_sufficiency_without_feedback_with_noise}
implies:
$$ {\cal C}^{f}_{a,(\log_2 \lambda, \eta \log_2 \lambda)} 
              \cap {\cal C}_{\mbox{fin}} 
   \subseteq 
   {\cal C}^{nf}_{c,(\lambda, \eta)}
              \cap {\cal C}_{\mbox{fin}}
$$ 
Combining with (\ref{eqn:equivalence_in_channel_terms}) gives
the following result for finite output alphabet channels:
\begin{equation} \label{eqn:finite_equivalence_in_channel_terms}
{\cal C}^{nf}_{c,(\lambda, \eta)} \cap {\cal C}_{\mbox{fin}} 
=
{\cal C}^f_{c,(\lambda, \eta)}  \cap {\cal C}_{\mbox{fin}} 
= 
{\cal C}^f_{a,(\log_2 \lambda, \eta \log_2 \lambda)}
                                \cap {\cal C}_{\mbox{fin}} 
\end{equation}

Finally, notice how the mapping from $(\lambda,\eta)$ to $(R,\alpha)$
is one-to-one and onto. By setting $\lambda = 2^R$ and $\eta =
\frac{\alpha}{R}$ it is possible to translate in the opposite
direction and this does provide some additional insight. For example,
in the anytime communication problem, it is clear that increasing $R$
from $2$ to $3$ while keeping $\alpha$ constant at $6$ results in a
harder problem. When translated to stabilization, without the results
established here, it is far from obvious that the equivalent move from
$\lambda = 4$ to $\lambda = 8$ with a simultaneous drop in the
required $\eta$ from $3$ to $2$ is also a move in a fundamentally
harder direction.

\subsection{Discussion}
Traditionally, this hierarchy of communication problems had not been
explored since there were apparently only two interesting levels:
problems equivalent to classical Shannon communication and those
equivalent to zero-error communication. Anytime communication problems
are intermediate between the two. Though feedback anytime
communication problems are interesting on their own, the equivalence
with feedback stabilization makes them even more fundamental.

It is interesting to consider where Schulman's interactive computation
problems fit in this sort of hierarchy. Because a constant factor
slowdown is permitted by the asymptotics, such problems of interactive
computation do not distinguish between channels of different Shannon
capacity. In the language of this section, this means that Shannon
communication problems are harder than those of interactive
computation considered in \cite{Schulman}.

Furthermore, the noisy channel definition given here can be extended
to include channels with memory. Simply make the current channel
output depend on all the current and past $V_t$ and $Y_t$. In that
case, (\ref{eqn:equivalence_in_channel_terms}) will continue to
hold. Since the finite-output alphabet constructions never needed
memorylessness, (\ref{eqn:finite_equivalence_in_channel_terms}) will
also hold.

The constructive nature of the proofs for the underlying theorems
makes them akin to the ``reductions'' used in theoretical computer
science to show that two problems belong to the same complexity
class. They are direct translations at the level of problems and
solutions. In contrast, the classical separation results go through
the mutual information characterization of $R(D)$ and $C$.  It would
be interesting to study a suitable analog of
(\ref{eqn:classical_separation_result}) for channels with
memory. Feedback can now increase the capacity so the with-feedback
and feedback-free problems are no longer equivalent. However, it would
be nice to see a direct reduction of Shannon's communication problem
to an estimation problem that encompasses such cases as well. The
asymptotic equivalence situation is likely even richer in the
multiuser setting where traditional separation theorems do not hold.

\section*{Acknowledgments}
The authors would like to thank Mukul Agarwal, Shashibhushan Borade,
Devavrat Shah, and Lav Varshney for comments on earlier versions of
this paper. We thank Nicola Elia for several constructive discussions
about the subject matter of this paper and Sekhar Tatikonda for many
discussions over a long period of time which have influenced this work
in important ways. Finally, we thank the anonymous reviewers for
a careful reading of the paper and helpful feedback.

\bibliographystyle{./IEEEtran}
\bibliography{./IEEEabrv,./MyMainBibliography}

\end{document}